\documentclass[12pt,fleqn]{article}
\usepackage{verbatim,amssymb,amsfonts,amsmath,graphics,amsthm,cite}
\usepackage{graphicx}
\usepackage{caption}
\usepackage{subcaption}
\numberwithin{equation}{section}

\newcommand{\calP}{\mathcal{P}}

\newcommand{\ms}{\mathsf{s}}
\newcommand{\rms}{\sqrt{\mathsf{s}}}
\newcommand{\rrms}{\sqrt{\mathsf{s}'}}

\newcommand{\HS}{{\mathcal{H}}}

\newcommand{\bs}{\boldsymbol}
\newcommand{\ka}{\kappa}
\newcommand{\intq}{\int\frac{d^3\bs{q}}{2\sqrt{1+\bs{q}^2}}}
\newcommand{\intqq}{\int\frac{d^3\bs{q}'}{2\sqrt{1+{\bs{q}'}^2}}}
\newcommand{\ha}{\hat{a}}
\newcommand{\hphi}{\hat{\phi}}
\newcommand{\hB}{\hat{B}}
\newcommand{\hc}{\hat{c}}
\newcommand{\hV}{\hat{V}}

\usepackage[colorlinks=true,linkcolor=blue,citecolor=blue,urlcolor=blue]{hyperref}

\begin{document}
\title{\textbf{Point-form dynamics of quasistable states}}
\author{M.~Gadella$^a$, F.~G\'omez-Cubillo$^b$, L.~Rodriguez$^c$ and\\ S.~Wickramasekara$^c$\vspace{.3cm}\\
\textit{$^a$Department of Theoretical Physics,} \\\textit{Atomic Physics and Optics}\\
\textit{University of Valladolid}\\
\textit{Valladolid, Spain}\vspace{.3cm}\\
\textit{$^b$Department of Mathematical Analysis}\\
\textit{University of Valladolid}\\
\textit{Valladolid, Spain}\vspace{.3cm}\\
\textit{$^c$Department of Physics}\\ \textit{Grinnell College}\\\textit{Grinnell, IA 50112}}
\date{\today}
\maketitle
\begin{abstract}
\noindent\rule{\linewidth}{0.5mm}
\noindent We present a field theoretical model of point-form dynamics which exhibits resonance scattering. In particular, 
we construct point-form Poincar\'e generators explicitly from field operators and show that in the vector spaces for the in-states and out-states (endowed with certain analyticity and topological properties suggested by the structure of the $S$-matrix) these operators integrate to furnish 
differentiable representations of the causal Poincar\'e semigroup, the semidirect product of the semigroup of spacetime translations into the forward 
lightcone and the group of Lorentz transformations. We also show that there exists a class of \emph{irreducible} representations of the Poincar\'e semigroup defined by a complex mass and a half-integer spin. The complex mass characterizing the representation naturally appears in the construction as the square root of the pole position of the propagator. These representations provide a  description of resonances in the same vein as Wigner's 
unitary irreducible representations of the Poincar\'e group provide a description of stable particles.
\noindent\rule{\linewidth}{0.5mm}
\end{abstract}
\tableofcontents

\section{Introduction}\label{sec1}
The purpose of this paper is to present a quantum field theoretical model leading to resonance scattering and construct the dynamics in the point-form where all four components of the momentum 
operator become interaction-dependent and the Lorentz generators remain interaction free. We study the integrability of the Poincar\'e algebra spanned by the point-form generators 
and show that under certain topological and analytic properties imposed on the spaces of in and out scattering states, a subset (more precisely, a cone) of the algebra 
integrates to a representation of what we call the  \emph{causal Poincar\'e semigroup},  
\begin{equation}
{\cal P}_+=\{(\Lambda,a): \Lambda\in SO(3,1),\ a^0\geq0,\ a_\mu a^\mu\geq0\}.\label{0.1}
\end{equation}
This set is clearly closed under the usual composition rule of the Poincar\'e group, $(\Lambda_2,a_2)(\Lambda_1,a_1)=(\Lambda_2\Lambda_1,a_2+\Lambda_2 a_1)$, 
but ${\cal P}_+$ is a semigroup, rather than a group,  because not every element of ${\cal P}_+$ is invertible in ${\cal P}_+$.  It is the 
semidirect product of the semigroup of spacetime translations into the closed forward light cone and the group of Lorentz transformations. We show that a certain class of representations of 
this semigroup provide the framework for a consistent theory of resonance scattering and decay phenomena.

Two different approaches to the characterization of resonances and decaying states by representations of 
the semigroup \eqref{0.1} have been recently pursued, each with its particular strengths and limitations. In \cite{bohm&harshman, bkw1,bkw2,sw1,sw2}, the emphasis has 
been on the point-form dynamics and the integrability of the Poincar\'e algebra to obtain the semigroup representations. Of particular 
importance is the manifest Lorentz invariance of the analyticity properties identified in \cite{bkw1,bkw2,sw1,sw2} as needed for 
semigroup integrability. While the formulation of 
\cite{sw1,sw2} shows that the interactions leading to semigroup representations can be encoded in the mass operator along the 
lines of the Bakamjian-Thomas construction, concrete realizations of such interactions, particularly in a manner that satisfies 
supplementary requirements such as cluster decomposition,  have not been given in \cite{bkw1,bkw2,sw1,sw2}. In \cite{appg1,appg2}, on the 
other hand, the emphasis has been on particular models in which interaction-incorporating Poincar\'e algebras have been explicitly constructed in terms 
of quantum fields.  However, Lorentz invariance of the analyticity properties used in \cite{appg1,appg2} is not manifest and therewith 
the integrability of the Poincar\'e algebra is unclear. 
The appearance of purely spacelike translations, in particular, show that the construction of \cite{appg1,appg2} does not have a well-defined semigroup representation consistent 
with special relativity.

This paper is a synthesis of the above two approaches, building on the strengths of each. The key feature of  \cite{bkw1,bkw2,sw1,sw2} that enables 
a well-defined representation of the causal Poincar\'e semigroup is the use of point-form dynamics where, with the subscript $0$ denoting the interaction-free operators, 
the Poincar\'e generators have the form  $\hat{P}^\mu=\hat{P}_0^\mu+\Delta \hat{P}^\mu$ and $\hat{J}^{\mu\nu}=\hat{J}^{\mu\nu}_0$. 
In the Bakamjian-Thomas construction developed in \cite{sw1,sw2}, the perturbations to the four momentum operator 
$\Delta \hat{P}^\mu$ were  in turn  induced from perturbations to the mass operator, $\hat{M}=\hat{M}_0+\Delta\hat{M}$,  
such that the velocity operators were interaction free: $\hat{Q}^\mu=\frac{\hat{P}^\mu}{\hat{M}}=\frac{\hat{P}_0^\mu}{\hat{M}_0}=\hat{Q}_0^\mu$. The property that 
the velocity operators are interaction-free and transform as a four vector under Lorentz transformations, also generated by interaction-free 
operators,  has the great advantage that supplementary conditions on the mass wavefunctions 
can be readily imposed while preserving Lorentz invariance. Specifically, in  \cite{bkw1,bkw2,sw1,sw2} wavefunctions were taken to be a class of smooth Hardy functions 
in the square mass variable (i.e., boundary values of functions analytic in the open lower or upper half complex plane of the square mass) 
and smooth, rapidly decreasing functions of the velocity variables.    

In contrast, in most quantum field theoretical treatments of interacting systems, including \cite{appg1,appg2}, the 
dynamics appear in the instant-form, defined by interaction-free momentum  and angular momentum operators, $\hat{\bs{P}}=\hat{\bs{P}}_0$ and $\hat{\bs{J}}=\hat{\bs{J}}_0$, 
which generate a representation of the Euclidean group, and  interaction-incorporating  Hamiltonian $\hat{H}=\hat{H}_0+\hat{V}$ and Lorentz boost generators 
$\hat{\bs{K}}=\hat{\bs{K}}_0+\hat{\bs{W}}$. This form of dynamics is a direct consequence of choosing a constant-time hypersurface as 
 the quantization surface and integrating the energy-momentum  $\hat{T}^{\mu\nu}$ and Lorentz  
$\hat{M}^{\rho\mu\nu}$ tensor densities on this surface to obtain the Poincar\'e generators. 
 In order to obtain dynamics in the point-form, and therewith construct the representations of the causal 
 Poincar\'e semigroup, here we use a forward hyperboloid $\tau^2={x^0}^2-\bs{x}^2$, $x^0>0$, of spacetime as the spacelike hypersurface on which 
 the tensor densities $\hat{T}^{\mu\nu}$ and $\hat{M}^{\rho\mu\nu}$ are integrated. Once this is done, 
 many of the standard techniques of quantum field theory can be used to construct the vacuum state and 
 solve the eigenvalue problem for the observables of the interacting system.  The main technical results that we report are the 
 construction of the point-form Poincar\'e generators for a particular model and the 
 integration of these generators to obtain a representation of the causal Poincar\'e semigroup. We show that the physical significance of 
 these representations is that they provide the appropriate 
 mathematical structure for understanding relativistic resonances and decaying states. 
 
 The organization of the paper is as follows: In section \ref{sec2}, we will discuss different forms of dynamics, show that the integration of  tensor 
 densities on the forward hyperboloid leads to point-form dynamics and obtain the interaction-incorporating four momentum operators for our model. In section \ref{sec3}, we will solve the eigenvalue problem for the 
 interacting momentum operators by constructing  the creation and annihilation operators for the interacting system. We will also how the new vacuum state for the interacting system, which is 
 different from the initial Fock vacuum,  is 
 obtained. The construction of the rigged  Hilbert spaces for resonance scattering and the representations of the causal Poincar\'e semigroup in these spaces 
 is the subject of section \ref{sec4}.  We will make a few concluding remarks in section  \ref{sec5}. We gather details of some calculations in Appendices \ref{appA} and \ref{appB}.

 \section{Point-form dynamics}\label{sec2}
 In Galilean (non-relativistic) quantum physics, the interactions of a system of particles are  commonly understood as a perturbation that changes the Hamiltonian while leaving
 other Galilean generators invariant. Thus, 
 \begin{eqnarray}
\hat{H}&=&\hat{H}_0+\hat{V}\nonumber\\
 \hat{\bs{P}}&=&\hat{\bs{P}}_0\nonumber\\
\hat{\bs{J}}&=&\hat{\bs{J}}_0\nonumber\\
\hat{\bs{K}}&=&\hat{\bs{K}}_0\nonumber\\
 \hat{M}&=&\hat{M}_0\label{1.1}
 \end{eqnarray}
 where the subscript $0$ indicates the generators of the corresponding non-interacting system. We will use this notation 
throughout this paper  for all interaction-free generators and the finite symmetry transformations they generate.
 If the interaction-free operators fulfill the commutation relations of 
 the Galilean algebra, it follows 
 readily that the interaction-incorporating operators defined by \eqref{1.1} also fulfill these commutation relations 
 for any interaction term $\hat{V}$ that commutes with $\hat{\bs{P}}_0$, $\hat{\bs{J}}_0$, $\hat{\bs{K}}_0$ and $\hat{M}_0$.  
 
 However, in the general case, 
 the Lie algebra that governs the symmetry structure of a given system of particles puts more stringent constraints 
 on which of the generators may be modified to incorporate interactions. For instance,  
 interactions cannot be incorporated into a system of particles in Lorentzian (relativistic) quantum physics by modifying the Hamiltonian alone because 
 the commutation relations 
 \begin{equation}
 \left[{\hat{K}_0}^i,{\hat{P}_0}^j\right]=i\delta^{ij}\hat{H}_0\label{1.2}
 \end{equation}
 show that at least some of the momenta $\hat{\bs{P}}_0$ or boost generators $\hat{\bs{K}}_0$ must be modified so that interaction-incorporating 
 operators also furnish a representation of the Poincar\'e algebra. Hence, the problem is how many and which of the generators of the algebra may remain 
unaffected when interactions are introduced into a system of particles. 
 
 Dirac was perhaps the first to systematically study this problem for the Poincar\'e algebra \cite{dirac}. In this study, he identified three forms of dynamics: 
the instant-form, point-form and front-form. Each form is characterized by a subgroup of the Poincar\'e group that remains interaction-free. 
These kinematic subgroups in turn  can be defined as stability groups of certain spacelike surfaces of spacetime.  For instance, 
the kinematic subgroup of instant-form dynamics is the Euclidean group generated by $\hat{\bs{P}}$ and $\hat{\bs{J}}$ and it leaves any surface 
defined by $t=constant$ invariant. The kinematic subgroup of point-form dynamics is the full Lorentz group generated by $\hat{\bs{J}}$ and $\hat{\bs{K}}$. 
It leaves any hyperboloid defined by $x^\mu x_\mu={constant}$ invariant. The kinematic subgroup of front-form  dynamics is generated by $\hat{P}_1$, 
$\hat{P}_2$, $\hat{P}_-:=\hat{P}_0-\hat{P}_3$, $\hat{J}_3$, $\hat{K}_3$, $\hat{K}_{1-}=\hat{K}_1-\hat{J}_2$ and $\hat{K}_{2-}=\hat{K}_2-\hat{J}_1$ and it 
leaves 
a surface defined by $x^0-x^3={constant}$ invariant. Note that Dirac's characterization of dynamics relies on spacelike surfaces and their stability groups, 
rather than, for instance, the choice of a time parameter for the evolution. 

On the other hand, within the context of quantum field theory, early papers of Tomonaga \cite{tomonaga}  
and Schwinger \cite{schwinger} show that quantization can be done on \emph{any} spacelike 
hypersurface, even a self-intersecting one. Despite this early understanding of the capabilities of the formalism and despite 
quantization on curved surfaces being a well-investigated subject, 
most quantum field theoretical studies of particle physics  are predominantly formulated in instant-form dynamics 
and are therewith tied to the use of equal-time commutation relations for the field operators. In fact, some treatments of quantum field theory 
even give the impression that the instant-form is modal \cite{weinberg}, rather than the result of a choice 
for the kinematic subgroup. The geometry of the flat surfaces $t={constant}$ and $x^0-x^3={constant}$ that underlie the 
instant-form and front-form is clearly simpler than, say, that of the hyperboloid $x^\mu x_\mu={constant}$ that underlies the point-form, 
and one might surmise that this simplicity is one reason that the instant-form and, to some extent, the front-form are more favored. Nevertheless, 
it should also be noted that the simplicity of the quantization surface does not necessarily make all aspects of a theory conceptually 
or computationally simpler. For instance, the integrability of the interaction-incorporating  Lorentz generators or the unitarity of the resulting 
group representation are often intractable problems. 

In this paper, we use the forward hyperboloid $x^\mu x_\mu=\tau^2,\ \ x^0>0$ as the surface on which the tensor densities $\hat{T}^{\mu\nu}$ and $\hat{M}^{\rho\mu\nu}$, 
constructed out of field operators and their derivatives, are  
integrated to obtain Poincar\'e generators in the point-form. Such a surface may also be considered a \emph{quantization surface} and the theory may be developed 
by, for instance,  quantizing preexisting classical fields on this surface by imposing commutation relations (see \eqref{2.1.18} below) consistent with the geometry of the surface. 
There exist several studies that have taken this approach \cite{sommerfield, gromes,disessa,biernat}. 
However, the point of view that we seek to advance here is that the choice of the surface on which $\hat{T}^{\mu\nu}$ and $\hat{M}^{\rho\mu\nu}$ are integrated to obtain the Poincar\'e generators 
is quite independent of the construction of quantum fields.  In other words, once we have obtained the relevant quantum fields by whatever method we use, 
we may integrate the resulting tensor densities on the appropriate spacelike hupersurface to obtain the Poincar\'e generators for any form of dynamics 
that we seek. In this spirit, our study is grounded in Wigner's discovery that the state space of a 
relativistic quantum particle furnishes a unitary irreducible representation of the Poincar\'e group. The creation and annihilation operators can be introduced as operators 
that transform between the one-particle states and the vacuum. From these creation and annihilation operators, the field operators can be defined the usual way. 
We then use these field operator to construct $\hat{T}^{\mu\nu}$ and $\hat{M}^{\rho\mu\nu}$ and obtain the point-form generators. This way, the use of the hyperboloid 
$x^\mu x_\mu=\tau^2,\ x^0\geq0$ is intrinsically consistent with the Poincar\'e invariant definition of a particle, in contrast to some of the earlier studies on quantization on the 
forward hyperboloid \cite{sommerfield,gromes,disessa}. The treatment of \cite{biernat} does address the issue of the consistency of quantizing a classical field 
on the hyperboloid and the Poincar\'e invariant definition of a particle. In this regard, our approach is similar to \cite{biernat} although its starting point is a bit different from ours.   

As mentioned  in the introduction,  the point-form dynamics has the 
advantage that certain analyticity properties underlying representations of the causal Poincar\'e semigroup can be implemented 
in a Lorentz invariant manner. This point is developed in \cite{sw1,sw2} within the  Bakamjian-Thomas construction, where all interactions 
are induced from a perturbation to the mass operator alone. While this is clearly not going to be the case in a field theoretical model, at least one that uses 
the Lagrangian formalism, our study will show that all the essential properties of resonance scattering and its description by Poincar\'e semigroup representations are still encoded in the analytic 
structure and singularities of various functions of the interacting square mass variable. 

\subsection{Point-form dynamics: preliminaries}\label{sec2.1}
\subsubsection{Introduction}
In this section, we gather some elementary facts about quantum fields, set up the notation to be used throughout the paper and 
discuss the problem of integrating the energy-momentum tensor for a massive, spin zero particle 
on the hyperboloid $x^\mu x_\mu=\tau^2,\ \ x^0>0$.  We will then apply the formalism developed in this section to obtain the specific expressions of Poincar\'e generators for the model considered in section \ref{sec2.2}. 

Our starting point is that in relativistic quantum mechanics, the mathematical image of a particle is a unitary irreducible representation of the Poincar\'e group. For a particle of mass $m$ and spin zero, 
the group operators furnishing the representation can be defined by their action on the momentum eigenstates: 
\begin{equation}
\hat{U}(\Lambda,a)\left|\bs{q}, m\right.\rangle=e^{im\Lambda q\cdot a}\left|\bs{\Lambda q}, m\right.\rangle\label{2.1.1}
\end{equation}
The $\left|\bs{q}\right.\rangle$ are generalized eigenvectors of the four momentum operators $\hat{P}^\mu$ and mass operator $\hat{M}=\sqrt{\hat{P}^\mu\hat{P}_\mu}$:
\begin{eqnarray}
\hat{P}^\mu\left|\bs{q}, m\right.\rangle&=&mq^\mu\left|\bs{q}, m\right.\rangle\nonumber\\
\hat{M}\left|\bs{q}, m\right.\rangle&=&m\left|\bs{q}, m\right.\rangle\label{2.1.2}
\end{eqnarray}
However, we have labeled the eigenvectors of $\hat{P}^\mu$ by the spacial part $\bs{q}$ of the eigenvalues of the 
velocity operators $\hat{Q}_\mu=\frac{\hat{P}_\mu}{\hat{M}}$, rather than the more common choice of momentum eigenvalues. 
Clearly, $q^2=q_\mu q^\mu=1$ and $q^0=\sqrt{1+\bs{q}^2}$. This choice corresponds to the complete system of commuting operators (CSCO) involving $\hat{Q}^\mu$, as opposed to the more commonly used one 
including the momentum operators, and it will become quite useful later in the paper when we consider analytic extensions in the mass variable. 
The boldface notation $\bs{\Lambda q}$ is to indicate the spatial part of the four vector $\Lambda q$. 

The vectors $\left|\bs{q},m\right.\rangle$ are normalized as 
\begin{equation}
\langle\bs{q},m|\bs{q'},m\rangle=2q^0\delta(\bs{q}-\bs{q}')\label{2.1.3}  
\end{equation}

We define $\hat{a}^\dagger(\bs{q},m)$ as the operator that maps the vacuum state $\left|0\right.\rangle$ to the one-particle velocity eigenstate 
$\left|\bs{q},m\right.\rangle$:
\begin{equation}
\left|\bs{q},m\right.\rangle=\hat{a}^\dagger\left(\bs{q},m\right)\left|0\right.\rangle\label{2.1.4}
\end{equation}
It follows  that the annihilation operator $\hat{a}(\bs{q},m)$ acts on $|\bs{q},m\rangle$ as
\begin{equation}
\hat{a}(\bs{q},m)|\bs{q}', m\rangle=2q^0\delta(\bs{q}-\bs{q}')|0\rangle.\label{2.1.5}
\end{equation}
and $\hat{a}^\dagger(\bs{q},m)$ and $\hat{a}(\bs{q},m)$ fulfill the commutation relations 
\begin{equation}
\left[\hat{a}(\bs{q},m),\hat{a}^\dagger(\bs{q}',m)\right]=2q^0\delta(\bs{q}-\bs{q}')\label{2.1.6}
\end{equation}

The transformation formula \eqref{2.1.1} and the definitions \eqref{2.1.4} and \eqref{2.1.5} imply that under the Poincar\'e group, the $\ha^\dagger(\bs{q},m)$ and $\ha(\bs{q},m)$ transform as 
\begin{eqnarray}
{\hat{U}}(\Lambda,a)\ha^\dagger(\bs{q}) {\hat{U}}^{-1}(\Lambda,a)=e^{im\Lambda{q}\cdot a}\ha^\dagger(\bs{\Lambda q})\nonumber\\
{\hat{U}}(\Lambda,a)\ha(\bs{q}){\hat{U}}^{-1}(\Lambda,a)=e^{-im\Lambda{q}\cdot a}\ha(\bs{\Lambda q})\label{2.1.7}
\end{eqnarray}
Differentiating \eqref{2.1.7} with respect to $a^\mu$ and evaluating the derivative at the identity of the Poincar\'e group $(I,0)$, we obtain the commutation relations
\begin{eqnarray}
\left[\hat{P}^\mu,\ha^\dagger(\bs{q},m)\right]=mq^\mu\ha^\dagger(\bs{q},m)\nonumber\\
\left[\hat{P}^\mu,\ha(\bs{q},m)\right]=-mq^\mu\ha(\bs{q},m)\label{2.1.8}
\end{eqnarray}
That is, the $\ha^\dagger(\bs{q},m)$ and $\ha(\bs{q},m)$ are the raising and lowering operators for the momentum operators which have continuous spectra.

The operators $\ha^\dagger(\bs{q},m)$ and $\ha(\bs{q},m)$ furnish a basis for the algebra of observables in that any operator can be expanded as sums of products of 
$\ha^\dagger(\bs{q},m)$ and $\ha(\bs{q},m)$. The expansion of the momentum operators, in particular, gives
\begin{equation}
\hat{P}^\mu=\intq\,mq^\mu\ha^\dagger(\bs{q},m)\ha(\bs{q},m)\label{2.1.9}
\end{equation}

We can now define the field operators for our massive scalar particle and its antiparticle in terms of the creation and annihilation operators:  
\begin{eqnarray}
\hat{\phi}(x)&:=&\frac{1}{(2\pi)^{3/2}}\int \frac{d^3q}{2q^0}m\Bigl(\hat{a}(\bs{q},m)e^{-imq\cdot x}+\hat{b}^\dagger(\bs{q},m)e^{imq\cdot x}\Bigr)\nonumber\\
 \hat\phi^\dagger(x)&:=&\frac{1}{(2\pi)^{3/2}}\int \frac{d^3q}{2q^0}m\Bigl(\hat{a}^\dagger(\bs{q},m)e^{imq\cdot x}+\hat{b}(\bs{q},m)e^{-imq\cdot x}\Bigr)\label{2.1.10}
 \end{eqnarray}
Here, $\hat{b}^\dagger(\bs{q})$ and $\hat{b}(\bs{q},m)$ are the creation and annihilation operators for the antiparticle that corresponds to our particle. In the concrete model 
discussed in the next subsection, we consider a neutral particle so $\hat{b}^\dagger(\bs{q},m)$ and $\hat{b}(\bs{q},m)$ coincide with $\ha^\dagger(\bs{q},m)$ and $\ha(\bs{q},m)$, respectively. 
It follows from \eqref{2.1.7} that under the Poincar\'e group, the field operators \eqref{2.1.10} transform as 
\begin{eqnarray}
{\hat{U}}(\Lambda,a)\hphi(x){\hat{U}}^{-1}(\Lambda,a)&=&\hphi(\Lambda x+a)\nonumber\\
{\hat{U}}(\Lambda,a)\hphi^\dagger(x){\hat{U}}^{-1}(\Lambda,a)&=&\hphi^\dagger(\Lambda x+a).\label{2.1.11}
\end{eqnarray}

As a consequence of \eqref{2.1.1} or, equivalently, \eqref{2.1.7}, the field operators \eqref{2.1.10} fulfill the equations of motion, 
\begin{equation}
\left(\partial_\mu\partial^\mu+m^2\right)\hphi(x)=0,\quad \left(\partial_\mu\partial^\mu+m^2\right)\hphi^\dagger(x)=0.\label{2.1.12}
\end{equation}
These equations simply re-state the fact that the square mass operator $\hat{M}^2=\hat{P}_\mu\hat{P}^\mu$ is a Casimir operator of the Poincar\'e algebra and that 
in an irreducible representation such as \eqref{2.1.1}, it is proportional to the identity $\hat{M}^2=m^2\hat{I}$. 

A Lagrangian density can also be written down, 
\begin{equation}
\hat{\cal L}(x)=\partial_\mu\hphi^\dagger(x)\partial^\mu\hphi(x)-m^2\hphi^\dagger(x)\hphi(x),\label{2.1.13}
\end{equation}
so that \eqref{2.1.12} results from it by way of Euler-Lagrange equations.

Aside from the use of the velocity operators, all of the above results are quite familiar from the usual formulation of quantum field theory that uses the $x^0={constant}$ as the quantization surface and therewith
the equal time commutation relations. However, the point we wish to emphasize is that, as seen above, they are grounded in the definition of a particle as an entity that has representation by a unitary irreducible representation of the Poincar\'e group and, as such, should be independent of our choose of quantization surface or the integration of $\hat{T}^{\mu\nu}$ and $\hat{M}^{\rho\mu\nu}$ on that surface. Although we will use the forward hyperboloid as the integration surface, we can and will maintain all the relationships, such as \eqref{2.1.1}-\eqref{2.1.9},  that directly follow from the Poincar\'e invariant notion of a particle.  Only the commutation relations for the 
field operators, which depend on the geometry of the surface,  will be different from the instant-form case: 
\begin{eqnarray}
x^\mu\left[\hat{\phi}(x'),\partial_\mu\hat{\phi}^\dagger(x)\right]_{x^2={x'}^2=\tau^2}&=&ix^0\delta(\bs{x}'-\bs{x})\nonumber\\
\left[\hat{\phi}(x'),\hat{\phi}(x)\right]_{x^2={x'}^2=\tau^2}&=&0\nonumber\\
\left[\hat{\phi}^\dagger(x'),\hat{\phi}^\dagger(x)\right]_{x^2={x'}^2=\tau^2}&=&0\label{2.1.18}
\end{eqnarray}
These may be obtained from the commutation relations 
\eqref{2.1.6} for the creation and annihilation operators and the expansion of the field operators \eqref{2.1.10} in terms of creation and annihilation 
operators. The \eqref{2.1.18} are a manifestly covariant version of the equal-time commutation relations that could have been anticipated from the fact the hyperboloid $x^\mu x_\mu=\tau^2$ is a 
Lorentz invariant surface whereas the $x^0={constant}$ is not.

\subsubsection{Quantization of classical fields}\label{sec2.1.2}
Had our starting point been classical fields satisfying the equations of motion \eqref{2.1.12}, rather than quantum particles satisfying \eqref{2.1.1}, we could have extracted the creation and annihilation operators by 
inverting the defining identities  \eqref{2.1.10}.  The first step in this process is defining a Lorentz invariant 
inner product under which the fundamental solutions $\varphi(\bs{q},x):=e^{-imq\cdot x}$ and  $\varphi^*(\bs{q},x):=e^{imq\cdot x}$ to \eqref{2.1.12} are orthogonal. The most commonly used inner product 
for this purpose is 
\begin{equation}
\left( \psi,\phi\right):=i\int d^3x\Bigl(\psi^*(x)\partial_t\phi(x)-\phi(x)\partial_t\psi^*(x)\Bigr)\label{2.1.14}
\end{equation}
defined on a hypersurface $x^0=constant$. However, as pointed out in the introduction, this choice leads to instant-form dynamics. 

For point-form dynamics, we must choose a forward hyperboloid $x_\mu x^\mu=\tau^2,\ x^0\geq0$  define on this surface an inner-product that is both Lorentz invariant and $\tau$-independent. 
To this end,  we recall the general result \cite{schweber} that for any 
spacelike hypersurface $\sigma$, the inner product 
\begin{equation}
\left( \psi,\phi\right)_\sigma:=i\int d\sigma^\mu(x)\left(\psi^*(x)\partial_\mu\phi(x)-\phi(x)\partial_\mu\psi^*(x)\right)\label{2.1.15}
\end{equation}
is $\sigma$-independent and (obviously) Lorentz invariant. 
When $\sigma$ is the forward hyperboloid $x^\mu x_\mu=\tau^2,\ \ x^0>0$, then 
\begin{equation}
d\sigma^\mu=2d^4x\,\delta(x^2-\tau^2)\theta(x^0)x^\mu\label{2.1.16}
\end{equation}
and the defining relation \eqref{2.1.15} becomes 
\begin{equation}
\left( \psi,\phi\right)_\tau:=i\int 2d^4x\,\delta(x^2-\tau^2)\theta(x^0)x^\mu\Bigl(\psi^*(x)\partial_\mu\phi(x)-\phi(x)\partial_\mu\psi^*(x)\Bigr)\label{2.1.17}
\end{equation}
It is straightfoward to show using \eqref{2.1.10} that under this inner product, the plane wave solutions 
$\varphi(q,x)=e^{-imq\cdot x}$ and $\varphi(q,x)^*=e^{imq\cdot x}$ of the Klein-Gordon equation \eqref{2.1.12} are orthogonal on the hyperboloid $x^\mu x_\mu=\tau^2$. 
Therefore, starting with classical fields, one would be able to use the inner product \eqref{2.1.17} and the commutation relations \eqref{2.1.18}, now imposed \emph{a priori}, 
to develop quantum fields and the corresponding creation and annihilation operators. The commutation relations \eqref{2.1.6} for the latter can then be derived,  
\emph{a posteriori}, finally arriving at the same particle description as that given by \eqref{2.1.1}. 

In passing, we also note that if $\sigma$ is the surface defined by $x^0={constant}$, then $d\sigma^\mu=\eta^{0\mu}dx^3$ and \eqref{2.1.15} reduces to the familiar 
expression \eqref{2.1.14}.

\subsubsection{Integration of tensor densities on the hyperboloid}\label{sec2.1.3}
With the above preliminary results, we can readily show that, for interactions that involve no derivative coupling, the 
integration of $\hat{T}^{\mu\nu}$ and $\hat{M}^{\rho\mu\nu}$ on the forward hyperboloid leads to point-form dynamics. To that end, consider a Lagrangian density
of the form 
\begin{equation}
\hat{{\cal L}}(x)=\hat{{\cal L}}_1(x)+\hat{{\cal L}}_2(x)+\hat{{\cal L}}_{\rm int}(x)\label{2.1.19}
\end{equation}
where $\hat{{\cal L}}_1(x)$ and $\hat{{\cal L}}_2(x)$ are the free particle Lagrangian \eqref{2.1.13} with $(m_1,\hat{\phi}_1)$ and 
$(m_2,\hat{\phi}_2)$, respectively, and $\hat{{\cal L}}_{\rm int}(x)$ is the interaction Lagrangian.  Defining operators canonically conjugated to $\partial^\mu\hphi_1$ and $\partial^\mu\hphi_2$, 
\begin{eqnarray}
\hat{\pi}^\mu_1(x):=\frac{\partial{\hat{\cal L}}}{\partial(\partial^\mu\hat{\phi}_1)}\nonumber\\
\hat{\pi}^\mu_2(x):=\frac{\partial{\hat{\cal L}}}{\partial(\partial^\mu\hat{\phi}_2)}\label{2.1.20}
\end{eqnarray}
we obtain the energy-momentum tensor $\hat{T}^{\mu\nu}$:
\begin{equation}
\hat{T}^{\mu\nu}:=\sum_{i=1}^2\hat{\pi}^\mu_i\partial^\nu\hat{\phi}_i+\partial^\mu{\hat{\phi}}^\dagger_i {{\hat{\pi}_i}}^{\dagger^\nu}-\eta^{\mu\nu}\hat{\cal L}\label{2.1.21}
\end{equation}

Next, suppose that \emph{the interaction Lagrangian $\hat{{\cal L}}_{\rm int}$ does not contain derivatives of the fields $\hat{\phi}_1$ and $\hat{\phi}_2$}.  Then, from \eqref{2.1.20}, 
\begin{equation}
\hat{\pi}_i^\mu=\partial^\mu{\hat{\phi}^\dagger}_i,\quad i=1,2\label{2.1.22}
\end{equation}
and \eqref{2.1.21} becomes 
\begin{equation}
\hat{T}^{\mu\nu}=\hat{T}_1^{\mu\nu}+\hat{T}_2^{\mu\nu}-\eta^{\mu\nu}{\hat{\cal L}}_{\rm int},\label{2.1.23}
\end{equation}
where 
\begin{equation}
\hat{T}_i^{\mu\nu}=\partial^\mu\hat{\phi}_i^\dagger\partial^\nu\hat{\phi}_i+\partial^\nu\hat{\phi}_i^\dagger\partial^\mu\hat{\phi}_i-\eta^{\mu\nu}\hat{{\cal L}}_i, 
\quad i=1,2\label{2.1.24}
\end{equation}
\\

Similarly, we define the Lorentz tensor density operator by
\begin{equation}
\hat{M}^{\rho\mu\nu}=x^\mu\hat{T}^{\rho\nu}-x^\nu\hat{T}^{\rho\mu}\label{2.1.25}
\end{equation}
For an interaction Lagrangian without derivative coupling, we substitute \eqref{2.1.23} in \eqref{2.1.25} to obtain
\begin{equation}
\hat{M}^{\rho\mu\nu}=\hat{M}_1^{\rho\mu\nu}+\hat{M}_2^{\rho\mu\nu}+\left(x^\mu\eta^{\rho\nu}-x^\nu\eta^{\rho\mu}\right)\hat{\cal L}_{\rm int}\label{2.1.26}
\end{equation}

The Poincar\'e generators for the interacting system can be obtained by integrating $\hat{T}^{\mu\nu}$ and $\hat{M}^{\rho\mu\nu}$ on the forward hyperboloid. For the momentum operators, we have 
\begin{equation}
\hat{P}^\mu=\int 2d^4x\,\delta (x^2-\tau^2)\theta(x^0)x_\nu\hat{T}^{\mu\nu} \label{2.1.27}
\end{equation}
Substituting from \eqref{2.1.23}, we obtain
\begin{equation}
\hat{P}^\mu=\hat{P}_0^\mu+\hat{P}^\mu_{\rm int}\label{2.1.28}
\end{equation}
where 
\begin{eqnarray}
&&\hat{P}_0^\mu=\hat{P}_1^\mu+\hat{P}_2^\mu=\int 2d^4x\delta (x^2-\tau^2)\theta(x^0)x_\nu\left(\hat{T}_1^{\mu\nu}+\hat{T}_2^{\mu\nu}\right) \nonumber\\
&&\hat{P}^\mu_{\rm int}=-\int 2d^4x\delta (x^2-\tau^2)\theta(x^0)x_\nu \eta^{\mu\nu}\hat{{\cal L}}_{\rm int}=-\int 2d^4x\delta (x^2-\tau^2)\theta(x^0)x^\mu\hat{{\cal L}}_{\rm int}\nonumber\\
\label{2.1.29}
\end{eqnarray}
For the Lorentz group generators, using \eqref{2.1.26} we obtain 
\begin{eqnarray}
\hat{J}^{\mu\nu}&=&\int 2d^4x\delta (x^2-\tau^2)\theta(x^0)x_\rho \hat{M}^{\rho\mu\nu}\nonumber\\
&=&\int 2d^4x\delta (x^2-\tau^2)\theta(x^0)x_\rho\left(\hat{M}_1^{\rho\mu\nu}+\hat{M}_2^{\rho\mu\nu}\right)\nonumber\\
&&\quad +\int 2d^4x\delta (x^2-\tau^2)\theta(x^0)x_\rho \left(x^\mu\eta^{\rho\nu}-x^\nu\eta^{\rho\mu}\right)\hat{{\cal L}}_{\rm int}\nonumber\\
&=&\hat{J}_1^{\mu\nu}+\hat{J}_2^{\mu\nu}+\int 2d^4x\delta (x^2-\tau^2)\theta(x^0)\left(x^\mu x^\nu-x^\nu x^\mu\right)\hat{{\cal L}}_{\rm int}\nonumber\\
\label{2.1.30}
\end{eqnarray}
The last term clearly vanishes. Therefore, we have 
\begin{equation}
\hat{J}^{\mu\nu}=\hat{J}^{\mu\nu}_1+\hat{J}_2^{\mu\nu}\equiv \hat{J}_0^{\mu\nu}\label{2.1.31}
\end{equation}
If we start with a Poincar\'e scalar for the Lagrangian \eqref{2.1.19}, then the operators defined by \eqref{2.1.27} and \eqref{2.1.30} fulfill the characteristic 
commutation relations of the Poincar\'e algebra. Furthermore, we see from \eqref{2.1.29} and \eqref{2.1.31} that for an interaction Lagrangian without derivative coupling, 
all four components of the momentum vector acquire interactions 
while the Lorentz generators remain interaction-free. This is Dirac's point-form dynamics. Note further that  this result does not depend on the inner-product \eqref{2.1.17} 
on the hyperboloid as would be necessary if we had started with a classical field theory. 

\subsubsection{Expansion of generators in terms of creation and annihilation operators}
We noted earlier that the plane wave solutions to the Klein-Gordon equation are orthogonal with respect to the 
inner product \eqref{2.1.17} on the forward hyperboloid. This fact and the definition of $\hat{P}^\mu$ and 
$\hat{J}^{\mu\nu}$ as integrals of the respective tensor densities on the hyperboloid can be used to obtain expansions of these 
operators in terms of creation and annihilation operators. Since we have set up the formalism so as to ensure 
that the Poincar\'e invariant definition of a particle is meaningful, we expect that these integral expressions for the Poincar\'e 
generators have the same form as those that follow from the representations of the Poincar\'e group.
As an example, we show here that the momentum operators $\hat{P}^\mu$ for a free particle obtained by integrating $\hat{T}^{\mu\nu}$ 
has the same form as \eqref{2.1.9} when expressed in terms of creation and annihilation operators. Recall that we obtained \eqref{2.1.9} directly 
from the representation structure of the Poincar\'e group, without making use of fields. 
 
To that end, using the expansions \eqref{2.1.10} for the field operators in \eqref{2.1.24} for $\hat{T}^{\mu\nu}$, 
\begin{eqnarray}
\hat{T}^{\mu\nu}&=&\frac{1}{(2\pi)^3}\int \frac{d^3\bs{q}}{2\sqrt{1+\bs{q}^2}}\int \frac{d^3\bs{q}'}{2\sqrt{1+{\bs{q}'}^2}}m^4\nonumber\\
&&\left(q^\mu {q'}^\nu+q^\nu{q'}^\mu-\eta^{\mu\nu}(q\cdot q'+1)\right)e^{-im(q+q')\cdot x}\hat{a}(\bs{q},m)\hat{b}(\bs{q}',m)\nonumber\\
&&+\left(-q^\mu {q'}^\nu-q^\nu{q'}^\mu+\eta^{\mu\nu}(q\cdot q'-1)\right)e^{-im(q-q')\cdot x}\hat{a}(\bs{q'},m)\hat{a}^\dagger(\bs{q},m)\nonumber\\
&&+\left(-q^\mu {q'}^\nu-q^\nu{q'}^\mu+\eta^{\mu\nu}(q\cdot{q'}-1)\right)e^{-m(q'-q)\cdot x}\hat{b}^\dagger(\bs{q},m)\hat{b}(\bs{q'},m)\nonumber\\
&&+\left(q^\mu {q'}^\nu+q^\nu{q'}^\mu-\eta^{\mu\nu}(q\cdot q'+1)\right)e^{im(q+q')\cdot x}\hat{a}^\dagger(\bs{q}',m)\hat{b}^\dagger(\bs{q},m)\nonumber\\
\label{2.1.32}
\end{eqnarray}
Therefore, the momentum operators are 
\begin{eqnarray}
\hat{P}^\mu&=&\frac{1}{(2\pi)^3}\int 2d^4x\delta(x^2-\tau^2)\theta(x^0)x_\nu\hat{T}^{\mu\nu}\nonumber\\
&=&-\frac{1}{(2\pi)^3}\int 2d^4x\delta(x^2-\tau^2)\theta(x^0)m^4\int \frac{d^3\bs{q}}{2\sqrt{1+\bs{q}^2}}\int \frac{d^3\bs{q}'}{2\sqrt{1+{\bs{q}'}^2}}\nonumber\\
&&\left(q^\mu x\cdot{q'}+{q'}^\mu x\cdot{q}-x^\mu(q\cdot{q'}+1)\right)\nonumber\\
&&\qquad \times \left(e^{-im(q+q')\cdot x}\hat{a}(\bs{q},m)\hat{b}(\bs{q}',m)+e^{im(q+q')\cdot x}\hat{a}^\dagger(\bs{q}',m)\hat{b}^\dagger(\bs{q},m)\right)\nonumber\\
&&+\left(-q^\mu x\cdot{q'}-{q'}^\mu x\cdot q+x^\mu(q\cdot q'-1)\right)\nonumber\\
&&\qquad\times \left(e^{-im(q-q')\cdot x}\hat{a}(\bs{q}',m)\hat{a}^\dagger(\bs{q},m)+e^{im(q-q')\cdot x}\hat{b}^\dagger(\bs{q},m)\hat{b}(\bs{q}',m)\right)\nonumber\\
\label{2.1.33}
\end{eqnarray}
In order to evaluate the integrals, let us make the change of variables 
\begin{equation}
P=m(q+q')\quad \text{and}\quad {\sf p}=m(q-q'),\label{2.1.34}
\end{equation}
introduce the notation 
\begin{equation}
I(p,\tau):=\int 2d^4x\,\delta(x^2-\tau^2)\theta(x^0)e^{-ix\cdot p},\label{2.1.35}
\end{equation}
and define 
\begin{equation}
{\cal I}^\mu(p,\tau)\equiv i\frac{\partial}{\partial p_\mu}I(p,\tau)=\int 2d^4x\,\delta(x^2-\tau^2)\theta(x^0)x^\mu e^{-ix\cdot{p}}.\label{2.1.37}
\end{equation}
For notational simplicity, below we will suppress $\tau$ in $I(p,\tau)$ and ${\cal I}^\mu(p,\tau)$ because these quantities will be always evaluated on a fixed hyperboloid. 

\noindent Then, in terms of $P$ and $\sf p$, 
\begin{eqnarray}
\left(q^\mu x\cdot{q'}+{q'}^\mu x\cdot{q}-x^\mu(q\cdot{q'}+1)\right)=\frac{1}{2m^2}\Bigl(-P^\mu x\cdot{\sf p}+{\sf p}^\mu x\cdot{\sf p}+x^\mu P\cdot P\Bigr)\nonumber\\
\left(-q^\mu {q'}\cdot x-{q'}^\mu q\cdot x+x^\mu(q\cdot{q'}-1)\right)=\frac{1}{2m^2}\Bigl(-P^\mu x\cdot P+{\sf p}^\mu x\cdot{\sf p}-x^\mu{\sf p}\cdot{\sf p}\Bigr)\nonumber\\
\label{2.1.36}
\end{eqnarray}
Using \eqref{2.1.37} and \eqref{2.1.36} in \eqref{2.1.33}, we obtain
\begin{eqnarray}
\hat{P}^\mu&=&-\frac{1}{2(2\pi)^3}\int\frac{d^3\bs{q}}{2\sqrt{1+\bs{q}^2}}\int \frac{d^3\bs{q}'}{2\sqrt{1+{\bs{q}'}^2}}m^2\nonumber\\
&&\Bigl(P^\mu P\cdot{\cal I}(P)-{\sf p}^\mu{\sf p}\cdot{\cal I}(P)-P\cdot P{\cal I}^\mu(P)\Bigr)\hat{a}^\dagger(\bs{q}',m)\hat{b}^\dagger(\bs{q},m)\nonumber\\
&&\Bigl(P^\mu P\cdot{\cal I}(-P)-{\sf p}^\mu{\sf p}\cdot{\cal I}(-P)-P\cdot P{\cal I}^\mu(-P)\Bigr)\hat{a}(\bs{q},m)\hat{b}(\bs{q}',m)\nonumber\\
&&-\Bigl(P^\mu P\cdot{\cal I}({\sf{p}})-{\sf{p}}^\mu{\sf{p}}\cdot{\cal I}({\sf p})+{\sf{p}}\cdot{\sf{p}}{\cal I}^\mu({\sf{p}})\Bigr)\hat{a}(\bs{q}',m)\hat{a}^\dagger(\bs{q},m)\nonumber\\
&&-\Bigl(P^\mu P\cdot{\cal I}(-{\sf p})-{\sf p}^\mu{\sf p}\cdot{\cal I}(-{\sf p})+{\sf p}\cdot{\sf p}{\cal I}^\mu(-{\sf p})\Bigr)\hat{b}^\dagger(\bs{q},m)\hat{b}(\bs{q}',m)\nonumber\\
\label{2.1.38}
\end{eqnarray}
The evaluation of the integrals \eqref{2.1.35} and \eqref{2.1.37} are given in Appendix \ref{appA}. With these results, it can be seen that the first two terms of \eqref{2.1.38} vanish and the last two
terms, \emph{after normal ordering},  reduce to 
\begin{eqnarray}
\hat{P}^\mu&=&\frac{1}{2(2\pi)^3}\int\frac{d^3\bs{q}}{2\sqrt{1+\bs{q}^2}}\int \frac{d^3\bs{q}'}{2\sqrt{1+{\bs{q}'}^2}}m^2\nonumber\\
&&P^\mu \Bigl(P\cdot{\cal I}({\sf p})\hat{a}^\dagger(\bs{q}',m)\hat{a}(\bs{q},m)+P\cdot{\cal I}(-{\sf p})\hat{b}^\dagger(\bs{q},m)\hat{b}(\bs{q}',m)\Bigr)\nonumber\\
&=&\int\frac{d^3\bs{q}}{2\sqrt{1+\bs{q}^2}} mq^\mu\Bigl(\hat{a}^\dagger(\bs{q},m)\hat{a}(\bs{q},m)+\hat{b}^\dagger(\bs{q},m)\hat{b}(\bs{q},m)\Bigr)
\label{2.1.39}
\end{eqnarray}
This is the same as \eqref{2.1.9} and gives further legitimacy 
to integrating field operators on the forward hyperboloid to set up the quantum theory. A similar calculation can be carried out for the Lorentz group generators 
$\hat{J}^{\mu\nu}$. However, we shall never need that expression as the $\hat{J}^{\mu\nu}$ are  not modified by interactions. 

We now make use of the results of this section for the particular model developed below.

\subsection{Point-form dynamics: the model}\label{sec2.2}
We consider a system of three neutral  particles, one with mass $M$ and spin 
zero and each of the other two with mass $m$ and spin $s$. Let $\hat{U}_1(\Lambda,a)$ be the operators 
that furnish the unitary irreducible representation of the Poincar\'e group that describes the first particle. These operators act on the generalized eigenvectors 
of the momentum operators the same way as in \eqref{2.1.1}:
\begin{equation}
{\hat{U}}_1(\Lambda,a)|\bs{q},M\rangle=e^{iM\Lambda q\cdot a}|\bs{\Lambda q},M\rangle\label{2.2.1}
\end{equation}
Likewise, each of the other two particles is described by a unitary irreducible representation of the Poincar\'e group. The tensor product space 
that describes the system consisting of these two-paticles is not irreducible under the Poincar\'e transformations, 
but it has a direct sum decomposition into subspaces which furnish irreducible 
representations. 
This decomposition involves a direct integral  over the square mass variable and a sum over angular momentum:
 \begin{equation}
 {\cal H}^{(m,s)}\otimes{\cal H}^{(m,s)}=\sum_{j=0}^\infty\int_{4m^2}^\infty d{\rho}\oplus {\cal H}^{({\rho },j)}\label{2.2.3}
 \end{equation}
 where $\rho=4m^2(1+\ka^2)$. Consider the $j=0$ subspace of \eqref{2.2.3}. This subspace carries a 
 \emph{reducible} representation of the Poincar\'e group, which 
 we denote by the operators $\hat{U}_2(\Lambda,a)$, that describes a quantum system with zero spin and  a continuous 
 mass distribution $m(\ka)\equiv\sqrt{\rho}=2m\sqrt{1+\ka^2}$. The generalized eigenstates of the momentum operator transform under $\hat{U}_2(\Lambda,a)$ as
 \begin{equation}
{\hat{U}}_2(\Lambda,a)|\bs{q},m(\ka)\rangle=e^{im(\ka)\Lambda q\cdot a}|\bs{\Lambda q},m(\ka)\rangle\label{2.2.2}
\end{equation}
Our model consists of making the system with mass $M$ and spin zero interact with the system with the continuous mass distribution $m(\ka)=2m\sqrt{1+\ka^2}$ 
and spin zero.  Hence, it is a three particle interaction, subject to the constraint that only the $s$-wave of the two identical particles participates. In a sense, it is a relativistic version 
of the well-known Friedrichs model \cite{friedrichs} in that a system corresponding to a discrete eigenvalue $M$ interacts with a system corresponding to a continuum $m(\ka)$. 
 
Such interactions are easiest to construct within the Lagrangian formalism. Therefore, as 
in \eqref{2.1.4}-\eqref{2.1.5}, we first define creation and annihilation operators $\ha^\dagger(\bs{q},M)$ and $\ha(\bs{q},M)$ for the first system
and $\hB^\dagger(\bs{q},m(\ka))$ and $\hB(\bs{q},m(\ka))$ for the second system as operators that map between the state vectors 
of \eqref{2.2.1} and \eqref{2.2.2} and the Poincar\'e invariant vacuum, respectively. From these operators, we then construct field operators and the interacting 
Lagrangian density.

It follows that creation and annihilation operators for the model fulfill the following commutation relations: 
\begin{eqnarray}
\left[\hat{a}(\bs{q},M),\hat{a}^\dagger({\bs{q}}',M)\right]&=&2q^0\delta(\bs{q}-\bs{q}')\label{2.2.4}\\
\left[\hat{B}(\bs{q},\ka),\hat{B}^\dagger(\bs{q}',\ka')\right]&=&2q^0\delta(\bs{q}-\bs{q}')\delta(\ka-\ka')\label{2.2.5}
\end{eqnarray}
For notational simplicity, we will from now on suppress the mass variable $M$ in $\hat{a}$ and $\hat{a}^\dagger$. 

As in  \eqref{2.1.7}, the operators of \eqref{2.2.4} and  \eqref{2.2.5} transform under the Poincar\'e group as follows: 
\begin{eqnarray}
{\hat{U}}_1(\Lambda,a)\ha^\dagger(\bs{q}) {\hat{U}}_1^{-1}(\Lambda,a)=e^{iM\Lambda{q}\cdot a}\ha^\dagger(\bs{\Lambda q})\nonumber\\
{\hat{U}}_1(\Lambda,a)\ha(\bs{q}){\hat{U}}_1^{-1}(\Lambda,a)=e^{-iM\Lambda{q}\cdot a}\ha(\bs{\Lambda q})\nonumber\\
{\hat{U}}_2(\Lambda,a)\hB^\dagger(\bs{q},k){\hat{U}}_2^{-1}(\Lambda,a)=e^{im(\ka)\Lambda{q}\cdot a}\hB^\dagger(\bs{\Lambda q},\ka)\nonumber\\
{\hat{U}}_2(\Lambda,a)\hB(\bs{q},k){\hat{U}}_2^{-1}(\Lambda,a)=e^{-im(\ka)\Lambda{q}\cdot a}\hB(\bs{\Lambda q},\ka)\label{2.2.6}
\end{eqnarray}
In particular, the parameter $\ka$ that determines the mass spectrum of the system defined by $\hB$ and $\hB^\dagger$ is Poincar\'e invariant. 

By differentiating \eqref{2.2.6} with respect to $a^\mu$ and evaluating the derivatives at the identity $(I,0)$ of the Poincar\'e group, we obtain the following eigenvalue 
equations for the momentum operators: 
\begin{eqnarray}
\left[\hat{P}_1^\mu,\ha^\dagger(\bs{q})\right]&=&Mq^\mu\ha^\dagger(\bs{q})\nonumber\\
\left[\hat{P}_1^\mu,\ha(\bs{q})\right]&=&-Mq^\mu\ha(\bs{q})\nonumber\\
\left[\hat{P}_2^\mu,\hB^\dagger(\bs{q},\ka)\right]&=&m(\ka)q^\mu\hB^\dagger(\bs{q},\ka)\nonumber\\
\left[\hat{P}_2^\mu,\hB(\bs{q},\ka)\right]&=&-m(\ka)q^\mu\hB(\bs{q},\ka)\label{2.2.7}
\end{eqnarray}

Next, we define field operators $\hphi_1$ and $\hphi_2$ for the two systems by means of $(\ha,\ha^\dagger)$ and $(\hB, \hB^\dagger)$, respectively,  the usual way: 
\begin{eqnarray}
\hat{\phi}_1(x)&=&\frac{1}{(2\pi)^{3/2}}\intq M\Bigl(\hat{a}(\bs{q})e^{-iMq\cdot x}+\hat{a}^\dagger(\bs{q})e^{iMq\cdot x}\Bigr)\nonumber\\
\hat{\phi}_2(x,\lambda)&=&\frac{1}{(2\pi)^{3/2}}\int_{-\infty}^\infty d\ka\intq m(\ka)\cos(\lambda\ka)\nonumber\\
&&\quad\quad\quad\qquad \Bigl(\hat{B}(\bs{q},\ka)e^{-im(\ka)q\cdot x}+\hat{B}^\dagger(\bs{q},\ka)e^{im(\ka)q\cdot x}\Bigr)
\label{2.2.8}
\end{eqnarray}
Note that $\hat{\phi}_2$ is an even function of $\lambda$, the parameter conjugated to $\kappa$ of the variable mass $m(\kappa)$. 
It follows from \eqref{2.2.8} and the transformation formulas \eqref{2.2.6} that $\hphi_1$ and $\hphi_2$ transform under the Poincar\'e transformations as: 
\begin{eqnarray}
{\hat{U}}_1(\Lambda,a)\hphi_1(x){\hat{U}}_1^{-1}(\Lambda,a)&=&\hphi_1(\Lambda x+a)\nonumber\\
{\hat{U}}_2(\Lambda,a)\hphi_2(x,\lambda){\hat{U}}_2^{-1}(\Lambda,a)&=&\hphi_2(\Lambda x+a,\lambda)\label{2.2.9}
\end{eqnarray}
\\

On the hyperboloid $x^2=\tau^2$, the field $\hat{\phi}_1$ satisfies the commutation relations \eqref{2.1.18} (only, now we have $\hat{\phi}_1^\dagger=\hat{\phi}_1$). 
The field $\hat{\phi}_2$ fulfills the commutation relations
\begin{equation}
x^\mu\left[\hphi_2(x',\lambda'),\frac{\partial}{\partial x^\mu} \hphi_2(x,\lambda)\right]_{x^2={x'}^2=\tau^2}=
ix^0\delta(\bs{x'}-\bs{x})\frac{1}{2}\Bigl(\delta(\lambda+\lambda')+\delta(\lambda-\lambda')\Bigr)\label{2.2.10}
\end{equation} 
 
 The field $\hphi_1$ fulfills the usual Klein-Gordon equation \eqref{2.1.12}. The corresponding Lagrangian density is  
 \begin{equation}
 \hat{\cal L}_1:=\frac{1}{2}\left(\partial_\rho\hphi_1\partial^\rho\hphi_1-M^2\hphi_1^2\right)\label{2.2.11}
 \end{equation}
 and  the energy-momentum tensor density is 
 \begin{equation}
 \hat{T}_1^{\mu\nu}(x):=\frac{1}{2}\left(\partial^\mu\hphi_1(x)\partial^\nu\hphi_1(x)+\partial^\nu\hphi_1(x)\partial^\mu\hphi_1(x)-\eta^{\mu\nu}
 \left(\partial_\rho\hphi_1\partial^\rho\hphi_1-M^2\hphi_1^2\right)\right).\label{2.2.12}
 \end{equation}
 The field $\hphi_2$ with its continuous mass distribution is governed by the Lagrangian density 
 \begin{equation}
 \hat{{\cal L}}_2:=\frac{1}{2}\left(\partial_\rho\hphi_2\partial^\rho\hphi_2-4m^2\hphi_2^2-4m^2\left(\frac{\partial\hphi_2}{\partial\lambda}\right)^2\right)\label{2.2.13}
 \end{equation}
 with the equation of motion
 \begin{equation}
\left(\partial_\rho\partial^\rho+4m^2-4m^2\frac{\partial^2}{\partial\lambda^2}\right)\hphi_2(x,\lambda)=0\label{2.2.14}
\end{equation}  
 and the energy-momentum tensor density
 \begin{eqnarray}
 \hat{T}_2^{\mu\nu}(x)&:=&\frac{1}{2}\int_{-\infty}^\infty d\lambda\left(\partial^\mu\hphi_2(x,\lambda)\partial^\nu\hphi_2(x,\lambda)+\partial^\nu\hphi_2(x,\lambda)\partial^\mu\hphi_2(x,\lambda)\right)\nonumber\\
&& -\frac{1}{2}\eta^{\mu\nu}\int_{-\infty}^\infty d\lambda\,
 \left(\partial_\rho\hphi_2(x,\lambda)\partial^\rho\hphi_2(x,\lambda)-4m^2\hphi_2^2(x,\lambda)-4m^2\left(\frac{\partial\hphi_2(x,\lambda)}{\partial\lambda}\right)^2\right)\nonumber\\
 \label{2.2.15}
\end{eqnarray}
The same tedious integration on the hyperboloid leading to \eqref{2.1.39} can be repeated for $\hat{T}_1^{\mu\nu}(x)$ and $\hat{T}_2^{\mu\nu}(x)$ to 
obtain the expressions for the momentum operators $\hat{P}_1^\mu$ and $\hat{P}_2^\mu$. In the second case, an integration over $-\infty<\lambda<\infty$ is also necessary. 
The results are 
\begin{eqnarray}
\hat{P}_1^\mu&=&\intq Mq^\mu\ha^\dagger(\bs{q})\ha(\bs{q})\nonumber\\
\hat{P}_2^\mu&=&\int_{-\infty}^\infty d\ka\intq m(\ka)q^\mu\hB^\dagger(\bs{q},\ka)\hB(\bs{q},\ka)\label{2.2.16}
\end{eqnarray}
Again, these expressions are consistent with \eqref{2.1.39} and \eqref{2.2.7}, obtained directly from representations of the Poincar\'e group. 

\subsubsection{The interaction} 
 Let us now consider an interaction between the two systems, defined by the Lagrangian density
 \begin{equation}
 \hat{{\cal L}}_{\rm int}(x)=\frac{\beta}{2}\int_{-\infty}^\infty d\lambda\,f(\lambda)\hphi_2(x,\lambda)\hphi_1(x)\label{2.2.17}
 \end{equation}
 where $f(\lambda)$ is a real-valued form factor and $\beta$, a dimensionless coupling constant. 
 Since $\hphi_2$ is an even function of $\lambda$, we take $f$ to be an even function as well. 
 
Since this interaction does not involve the derivatives of the fields, according to \eqref{2.1.31}, the Lorentz generators 
remain interaction-free,  
\begin{equation}
\hat{J}^{\mu\nu}=\hat{J}_0^{\mu\nu},\label{2.2.17b}
\end{equation}
while according to \eqref{2.1.29}, all components of the four momentum operator become interaction-incorporating:  
\begin{equation}
\hat{P}^\mu_{\rm int}=-\frac{\beta}{2}\int d^4x\,\delta(x^2-\tau^2)\theta(x^0)x^\mu\hphi_1(x)\int_{-\infty}^\infty d\lambda\,f(\lambda)\hphi_2(x,\lambda)\label{2.2.18}
\end{equation}
Substituting \eqref{2.2.8} for the field operators in \eqref{2.2.18}, we obtain
\begin{eqnarray}
\hat{P}^\mu_{\rm int}&=&-\frac{\beta M}{2(2\pi)^3}\intqq\intq\int_{-\infty}^\infty d\ka\, m(\ka)\alpha(\ka)\nonumber\\
&&\quad\int d^4x\,\delta(x^2-\tau^2)\theta(x^0)x^\mu\nonumber\\
&&\left[ \ha(\bs{q}')\hB(\bs{q},\ka)e^{-i\left(Mq'+m(\ka)q\right)\cdot x}+\ha^\dagger(\bs{q}')\hB^\dagger(\bs{q},\ka)e^{i\left(Mq'+m(\ka)q\right)\cdot x}\right.\nonumber\\
 &&\left.+ \ha(\bs{q}')\hB^\dagger(\bs{q},\ka)e^{-i\left(Mq'-m(\ka)q\right)\cdot x}+\ha^\dagger(\bs{q}')\hB(\bs{q},\ka)e^{i\left(Mq'-m(\ka)q\right)\cdot x}\right]\nonumber\\
 \label{2.2.19}
 \end{eqnarray}
 where $\alpha(k)$, a real even function, is the Fourier transform of $f$: 
 \begin{equation}
 \alpha(\ka):=\int_{-\infty}^\infty d\lambda\, f(\lambda)\cos(\lambda\ka)\label{2.2.20}
 \end{equation}
 Using \eqref{2.1.37},
 \begin{eqnarray}
 \hat{P}^\mu_{\rm int}&=&-\frac{\beta M}{2(2\pi)^3}\intqq\intq\int_{-\infty}^\infty d\ka\, m(\ka)\alpha(\ka)\nonumber\\
&&\ \ha(\bs{q}')\hB(\bs{q},\ka){\cal I}^\mu(Mq'+m(\ka)q)+\ha^\dagger(\bs{q}')\hB^\dagger(\bs{q},\ka){\cal I}^\mu(-Mq'-m(\ka)q)\nonumber\\
&&\ + \ha(\bs{q}')\hB^\dagger(\bs{q},\ka){\cal I}^\mu(Mq'-m(\ka)q)+\ha^\dagger(\bs{q}')\hB(\bs{q},\ka){\cal I}^\mu(-Mq'+m(\ka)q)\nonumber\\
\label{2.2.21}
\end{eqnarray} 
 Therewith, we have the Poincar\'e generators for the interacting system: 
 \begin{eqnarray}
 \hat{P}^\mu&=&\hat{P}_0^\mu+\hat{P}^\mu_{\rm int}=\hat{P}_1^\mu\otimes\hat{I}_2+\hat{I}_1\otimes\hat{P}_2^\mu+\hat{P}_{\rm int}^\mu\label{2.2.22}\\
 \hat{J}^{\mu\nu}&=&\hat{J}_0^{\mu\nu}=\hat{J}_1^{\mu\nu}\otimes\hat{I}_2+\hat{I}_1\otimes\hat{J}_2^{\mu\nu}\label{2.2.23}
 \end{eqnarray}
The explicit expressions for $\hat{P}_1^\mu$, $\hat{P}_2^\mu$ and $\hat{P}^\mu_{\rm int}$ are as given by \eqref{2.2.16} and \eqref{2.2.21}. We have not computed 
the explicit expressions for $J^{\mu\nu}$ as we will not need these. That the $J^{\mu\nu}$ are the same as the free generators ensures that these operators do 
integrate to a unitary representation of the Lorentz group, with the operators ${\hat{U}}(\Lambda)$ being the same as the free operators ${\hat{U}}_0(\Lambda)$. Under the action 
of $\hat{U}_0(\Lambda)$, the momentum operator \eqref{2.2.22} transforms as a four vector, the only property that we will need in what follows. 

\section{Construction of the physical vacuum and Fock space}\label{sec3}
\subsection{The eigenvalue problem}
The task now is to diagonalize  the full momentum operators \eqref{2.2.22}. This is tantamount to finding a set of creation operators $\hc^\dagger(\bs{q},\ms)$ and annihilation operators
$\hc(\bs{q},\ms)$ such that
\begin{eqnarray}
\left[\hat{P}^\mu,\hc^\dagger(\bs{q},\ms)\right]&=&\sqrt{\ms}q^\mu\hc^\dagger(\bs{q},\ms)\label{3.1.1}\\
\left[\hat{P}^\mu,\hc(\bs{q},\ms)\right]&=&-\sqrt{\ms}q^\mu\hc(\bs{q},\ms)\label{3.1.2}
\end{eqnarray}
and constructing a vacuum state $\Omega$ annihilated by $\hc(\bs{q},\ms)$. Once the new vacuum $\Omega$ has been constructed, 
our task in section \ref{sec3.1}, the repeated application of the creation operator $\hc^\dagger(\bs{q},\ms)$ on $\Omega$ will generate 
the entire Fock space. 

From the general structure of the Bogolubov transformation, we anticipate $\hc^\dagger(\bs{q},\ms)$ to be a linear combination of the operators $\ha^\dagger(\bs{q})$, $\ha(\bs{q})$, 
$\hB^\dagger(\bs{q},\ka)$ and $\hB(\bs{q},\ka)$. Therefore, we let
\begin{eqnarray}
\hc^\dagger(\bs{q},\ms)&=&\intqq \int_{-\infty}^\infty d\ka \left(T(\ms,\ka; q,q')\hB^\dagger(\bs{q}',\ka)+R(\ms,\ka;q,q')\hB(\bs{q}',\ka)\right)\nonumber\\
&&+\intqq\left(t(\ms;q,q')\ha^\dagger(\bs{q}')+r(\ms;q,q')\ha(\bs{q}')\right)\label{3.1.3}
\end{eqnarray}
As mentioned at the end of section \ref{sec2}, in point-form dynamics, ${\hat{U}}(\Lambda)={\hat{U}}_0(\Lambda)$. That is to say that the transformation of \emph{all} the operators of \eqref{3.1.3} under 
Lorentz transformations is furnished by the same set of unitary operators. From \eqref{2.2.6} and the analogous equation for $\hc^\dagger(\bs{q},\ms)$, it then follows 
that the coefficient functions $T$, $R$, $t$, and $r$ are all Lorentz scalars. Therefore, they must be of the form
\begin{eqnarray}
T(\ms,\ka;q,q')&=&T'(\ms,\ka;q\cdot q')+T(\ms,\ka)2q^0\delta(\bs{q}-\bs{q}')\nonumber\\
R(\ms,\ka;q,q')&=&R'(\ms,\ka;q\cdot q')+R(\ms,\ka)2q^0\delta(\bs{q}-\bs{q}')\nonumber\\
t(\ms;q,q')&=&t'(\ms;q\cdot q')+t(\ms)2q^0\delta(\bs{q}-\bs{q}')\nonumber\\
r(\ms;q',q)&=&r'(\ms;q\cdot q')+r(\ms)2q^0\delta(\bs{q}-\bs{q}')\label{3.1.4}
\end{eqnarray}
Now, we will consider a special class of solutions by suppressing the terms $T'(\ms,\ka;q\cdot q')$, $R'(\ms,\ka;q\cdot q')$, $t'(\ms;q\cdot q')$ and 
$r'(\ms;q\cdot q')$ in \eqref{3.1.4}. 
This choice is equivalent to restricting ourselves to couplings that are only local in the momentum variables. Therefore, \eqref{3.1.3} becomes 
\begin{equation}
\hc^\dagger(\bs{q},\ms)= \int_{-\infty}^\infty d\ka \left(T(\ms,\ka)\hB^\dagger(\bs{q},\ka)+R(\ms,\ka)\hB(\bs{q},\ka)\right) +t(\ms)\ha^\dagger(\bs{q})+r(\ms)\ha(\bs{q})\label{3.1.5}
\end{equation}

Substituting \eqref{2.2.16}, \eqref{2.2.21} and \eqref{3.1.5} into \eqref{3.1.1}, and making use of the commutation relations \eqref{2.2.7}, we obtain the following four coupled 
equations for the coefficient functions of \eqref{3.1.5}: 
\begin{eqnarray}
(\rms-M) t(\ms)&=&-\frac{1}{2(2\pi)^3}\beta M\int_{-\infty}^\infty d\ka\, m(\ka)\alpha(\ka)\intqq\nonumber\\
&&\qquad\left[T(\ms,\ka){\cal D}\left(-Mq'+m(\ka)q\right)-R(\ms,\ka){\cal D}\left(-Mq'-m(\ka)q\right)\right]\nonumber\\
(\rms+M)r(\ms)&=&-\frac{1}{2(2\pi)^3}\beta M\int_{-\infty}^\infty d\ka\, m(\ka)\alpha(\ka)\intqq\nonumber\\
&&\qquad\left[T(\ms,\ka){\cal D}\left(Mq'+m(\ka)q\right)-R(\ms,\ka){\cal D}\left(Mq'-m(\ka)q\right)\right]\nonumber\\
 (\rms-m(\ka))T(\ms,\ka)&=&-\frac{1}{2(2\pi)^3}\beta M m(\ka)\alpha(\ka)\intqq\nonumber\\
 &&\qquad\quad\left[t(\ms){\cal D}\left(Mq-m(\ka)q'\right)-r(\ms){\cal D}\left(-Mq-m(\ka)q'\right)\right]\nonumber\\
  (\rms+m(\ka)) R(\ms,\ka)&=&-\frac{1}{2(2\pi)^3}\beta M m(\ka)\alpha(\ka)\intqq\nonumber\\
  &&\quad\quad\left[t(\ms){\cal D}\left(Mq+m(\ka)q'\right)-r(\ms){\cal D}\left(-Mq+m(\ka)q'\right)\right]\nonumber\\
  \label{3.1.7}
  \end{eqnarray}
 where 
 \begin{equation}
 {\cal D}\left(m_1q'+m_2q\right):=q_\mu{\cal I}^\mu(m_1q'+m_2q)\label{3.1.7b}
 \end{equation}
 and  ${\cal I}^\mu(m_1q'+m_2q)$ is defined by \eqref{2.1.37}. 
 \\
 
\noindent We solve these equations in Appendix \ref{appB} to obtain: 
 \begin{eqnarray}
t(\ms)&=&\left(\rms+M\right)H_1^{(1)}(M\tau)\rho(\ms)G(\ms)\nonumber\\
r(\ms)&=&-\left(\rms-M\right)H_1^{(2)}(M\tau)\rho(\ms)G(\ms)\nonumber\\
T(\ms,\ka)&=&C(\ms)\delta(\rms-m(\ka))+\frac{i\pi\beta}{8}\frac{\alpha(m(\ka))H_1^{(1)}(m(\ka)\tau)}{\rms-m(\ka)}{\cal H}_{1,2}(\rms\tau,M\tau)\rho(\ms)G(\ms)\nonumber\\
R(\ms,\ka)&=&-\frac{i\pi\beta}{8}\frac{\alpha(m(\ka))H_1^{(2)}(m(\ka)\tau)}{\rms+m(\ka)}{\cal H}_{1,2}(\rms\tau,M\tau)\rho(\ms)G(\ms)\
\label{3.1.20a}
\end{eqnarray}
where $H^{(1)}_\alpha$ and $H^{(2)}_\alpha$ are the first and second kind Hankel functions  of order $\alpha$, respectively, and 
$C(\ms)$ is an arbitrary function of $\ms$. The quantities $\rho(\ms)$, $\Pi(\ms)$, $G(\ms)$ and ${\cal H}_{\mu,\nu}(x,y)$ are defined by
\begin{eqnarray}
\rho(\ms)&:=&\frac{iC\pi\beta}{8}\frac{\alpha(\ms){\cal H}_{1,2}(\rms\tau,\rms\tau)}{2m\sqrt{\ms-4m^2}\,H_1^{(1)}(\rms\tau)}=\frac{iC\pi\beta}{4}\frac{\rms\alpha(\ms)H_2^{(2)}(\rms\tau)}{2m\sqrt{\ms-4m^2}}\nonumber\\
\label{3.1.18a}\\
\Pi(\ms)&:=&-\frac{(\pi\beta)^2}{32}{\cal H}_{1,2}(\rms\tau,M\tau)\int_{2m}^\infty d\mu(\ka)\,\frac{\alpha(m(\ka))^2{\cal H}_{1,2}(\rms\tau,m(\ka)\tau)}{\ms-m(\ka)^2}\nonumber\\
\label{3.1.16a}\\
G(\ms)&:=&\frac{1}{\ms-M^2-\Pi(\ms)}\label{3.1.19a}\\
{\cal H}_{\mu,\nu}(y,x)&:=&{y\left(H^{(1)}_\mu(x)H^{(2)}_\nu(x)+H^{(2)}_\mu(x)H^{(1)}_\nu(x)\right)}+\nonumber\\
&&\quad\quad{x\left(H^{(1)}_\mu(x)H^{(2)}_\nu(x)-H^{(2)}_\mu(x)H^{(1)}_\nu(x)\right)}\label{3.1.10c}
\end{eqnarray}

As shown in Appendix \ref{appB}, a non-trivial solution \eqref{3.1.7}, and therewith to the eigenvalue problem \eqref{3.1.1}, exists only for 
$4m^2\leq\ms<\infty$.  Therefore, the spectrum of the interacting square-mass operator ${\hat{M}}^2={\hat{P}}_\mu {\hat{P}}^\mu$ is 
\begin{equation}
4m^2\leq\ms<\infty\label{3.1.17a}
\end{equation}
\\
The $G(\ms)$ defined by \eqref{3.1.19a} is the Green's function for the model. Unstable states, which will occupy much of our attention in the next section, correspond to poles of
$G(\ms)$.  Furthermore, since the Green's function $G(\ms)$, as well as 
all of the coefficient functions \eqref{3.1.20a}, are Lorentz scalars, we are in a position to carry out the analysis of the resonance behavior in a manifestly 
Lorentz invariant manner. \emph{This would decidedly not have been the case if we had used instant-form dynamics}. 

With \eqref{3.1.20a}, we have obtained a formal solution to the eigenvalue problem \eqref{3.1.1}-- formal because we have not specified how to 
handle the singularities of the integrals that define $\rho(\ms)$ and $\Pi(\ms)$. For the former, we can make the integral well defined by 
 simply demanding that the form factor $\alpha(m(\ka))$ vanish 
at $m(\ka)=2m$ sufficiently fast. For the latter, on the other hand, it is necessary to consider the integral \eqref{3.1.16a} for complex values of $\ms$. For a suitable 
class of form factors $\alpha(m(\ka))$, the integral then defines a function that is analytic everywhere except for the branch cut $[2m,\infty)$. The function $\Pi(\ms)$
must be obtained as the limit of this analytic function for real $2m\leq\ms<\infty$. Corresponding to the limit from above and below, we will have two functions $\Pi_\pm(\ms)$, and 
therewith also two sets of solutions to \eqref{3.1.20a}. We will come back to these considerations in section \ref{sec4} where we will take up the question of resonance poles. 

In the remainder of this section, we will lay out the general program for the formal solution \eqref{3.1.20a}, including the construction of the physical vacuum. 
These considerations will carry over to the specific solutions treated in section~\ref{sec4}. 

As shown in Appendix \ref{appB}, the annihilation operator $\hc(\bs{q},\ms)$ can be obtained by solving \eqref{3.1.2} following same procedure used to determine $\hc^\dagger(\bs{q},\ms)$: 
\begin{equation}
\hc(\bs{q},\ms)=\int_{-\infty}^\infty d\ka \left(T^*(\ms,\ka)\hB(\bs{q},\ka)+R^*(\ms,\ka)\hB^\dagger(\bs{q},\ka)\right) +t^*(\ms)\ha(\bs{q})+r^*(\ms)\ha^\dagger(\bs{q})\label{3.1.5b}
\end{equation}
where $T^*(\ms,\ka)$, $R^*(\ms,\ka)$, $t^*(\ms)$ and $r^*(\ms)$ are the complex conjugates of the corresponding functions of \eqref{3.1.20a}. Hence, not surprisingly, we see that $\hc(\bs{q},\ms)$ is the 
formal adjoint of $\hc^\dagger(\bs{q},\ms)$. 

From the structure of \eqref{3.1.5b}, it is clear that $\hc(\bs{q},\ms)$ does not annihilate the original Fock vacuum. 
Therefore, we must construct a Poincar\'e invariant physical vacuum state vector $\Omega$ 
annihilated by $\hc(\bs{q},\ms)$. Once this is done, our task of section \ref{sec3.1}, we can define what would be the analogue of ``one-particle states" $|\bs{q},\ms\rangle$ as the image of 
$\Omega$ under $\hc^\dagger(\bs{q},\ms)$: $|\bs{q},\ms\rangle=\hc^\dagger(\bs{q},\ms)|\Omega\rangle$. Clearly, there is no obvious direct particle 
interpretation associated with the states $|\bs{q},\ms\rangle$ as $\ms$ varies over $2m\leq\ms<\infty$ and is 
not restricted to be a single number as would be required for an irreducible representation. We will see in section \ref{sec4} that they relate to scattering states. 

Let us normalize the states $|\bs{q},\ms\rangle$ as 
\begin{equation}
\langle\bs{q},\ms|\bs{q}',\ms'\rangle=2q^0\delta(\bs{q}-\bs{q}')\delta(\rms-\rrms)\label{3.1.20b}
\end{equation}
By way of the nuclear spectral theorem of Gelf'and and Maurin \cite{spectraltheorem}, the vectors $|\bs{q},\ms\rangle$ can be defined as the evaluation functionals
on a suitable space of test functions $\left\{\psi\right\}$ such that $\psi(\bs{q},\ms)=\langle\bs{q},\ms|\psi\rangle$. The completion of this test function space $\{\psi\}$ 
with respect to the norm topology gives what would be the analogue of the ``one-particle Hilbert space'', which we denote by 
\begin{equation}
\int_{4m^2}^\infty\,d\ms\,{\cal H}(\ms)\label{3.1.20c}
\end{equation} 
This Hilbert space is isomorphic to the $j=0$ subspace of \eqref{2.2.3}. 

The repeated application of $\hc^\dagger(\bs{q},\ms)$ on $\Omega$ would create 
what would be the analogue of ``multi-particle states"--for instance, $\hc^\dagger(\bs{q},\ms)\hc^\dagger(\bs{q}',\ms')\Omega=|\bs{q},\ms;\bs{q}',\ms'\rangle$. 
The completion of the linear span of these states with respect the norm topology gives rise the tensor products of the Hilbert space \eqref{3.1.20c}. We then take the 
direct sum of these tensor product spaces to obtain the full Fock space for the system: 
\begin{equation}
{\cal H}=\mathbb{C}\oplus\int_{4m^2}^\infty\,d\ms\,{\cal H}(\ms)\oplus\int_{4m^2}^\infty\,d\ms\int_{4m^2}^\infty\,d\ms'\,{\cal H}(\ms)\otimes{\cal H}(\ms')\oplus\cdots\label{3.1.20d}
\end{equation}
where $\mathbb{C}$ is the field of complex numbers, isomorphic to the one-dimensional Hilbert space spanned by the vacuum state $\Omega$.

With the creation operator $\hc^\dagger(\bs{q},\ms)$ defined as an operator in the Fock space \eqref{3.1.20d}, the annihilation operator \eqref{3.1.5b} 
fulfilling \eqref{3.1.2} and annihilating the vacuum $\Omega$ will be defined as the adjoint of $\hc^\dagger(\bs{q},\ms)$ in \eqref{3.1.20d}. The action 
of $\hc^\dagger(\bs{q},\ms)$ and $\hc(\bs{q},\ms)$ in \eqref{3.1.20d} would be completely fixed by the identity $\hc(\bs{q},\ms)|\Omega\rangle=0$, the definition
$\hc^\dagger(\bs{q},\ms)|\Omega\rangle=|\bs{q},\ms\rangle$ and the normalization condition \eqref{3.1.20b}. Therewith, we have the canonical commutation relations 
\begin{equation}
\left[\hc(\bs{q},\ms),\hc^\dagger(\bs{q}'\ms')\right]=2q^0\delta(\bs{q}-\bs{q}')\delta(\rms-\rrms)\label{3.1.20e}
\end{equation}
On the other hand, \eqref{3.1.1} and \eqref{3.1.2} determine the operators $\hc^\dagger(\bs{q},\ms)$ and $\hc(\bs{q},\ms)$ only up to an arbitrary scalar function that we denoted 
by $C(\ms)$ in the solution \eqref{3.1.20a}. A direct calculation using \eqref{3.1.5}, \eqref{3.1.5b} and \eqref{3.1.20a} gives the commutation relations 
\begin{equation}
\left[\hc(\bs{q},\ms),\hc^\dagger(\bs{q}',\ms')\right]=\frac{\rms}{m{\sqrt{\ms-4m^2}}}\left|C(\ms)\right|^2\,2q^0\delta(\bs{q}-\bs{q}')\delta(\rms-\rrms)\label{3.1.26}
\end{equation}
We can now fix the arbitrary function $C(\ms)$ by imposing the normalization condition \eqref{3.1.20b} so that \eqref{3.1.26} reduces to the canonical commutation relations \eqref{3.1.20e}.

\subsection{The vacuum}\label{sec3.1}
Let $|0\rangle$ be the vacuum state annihilated by  $\ha(\bs{q})$ and $\hB(\bs{q},\ka)$. Then, it is clear from \eqref{3.1.5b} that the operator $\hc(\bs{q},\ms)$ does not 
annihilate $|0\rangle$.  Therefore, in order to obtain the Fock space \eqref{3.1.20d} for the physical states by the repeated application of the creation 
operator $\hc^\dagger(\bs{q},\ms)$, we must first construct the physical vacuum state. Let $\Omega$ be that state, i.e., 
\begin{equation}
\hc(\bs{q},\ms)|\Omega\rangle=0\label{3.2.0}
\end{equation}
By the general structure of Bogolubov transformation, 
we expect 
\begin{equation}
\Omega=e^{\hV}|0\rangle\label{3.2.1}
\end{equation}
where $\hV$ is a quadratic function of the free creation operators $\ha^\dagger$ and $\hB^\dagger$.  Therefore, let 
\begin{eqnarray}
\hV&=&\int d\ka\, d\ka'\frac{d^3\bs{q}}{2\sqrt{1+\bs{q}^2}}\frac{d^3\bs{q}'}{2\sqrt{1+{\bs{q}'}^2}} f_3(\ka,\ka';q,q')\hB^\dagger(\bs{q},\ka)\hB^\dagger(\bs{q}',\ka')\nonumber\\
&&+\int d\ka\frac{d^3\bs{q}}{2\sqrt{1+\bs{q}^2}}\frac{d^3\bs{q}'}{2\sqrt{1+{\bs{q}'}^2}} f_2(\ka;q,q')\hB^\dagger(\bs{q},\ka)\ha^\dagger(\bs{q}')\nonumber\\
&&+\int\frac{d^3\bs{q}}{2\sqrt{1+\bs{q}^2}}\frac{d^3\bs{q}'}{2\sqrt{1+{\bs{q}'}^2}} f_1(q,q')\ha^\dagger(\bs{q})\ha^\dagger(\bs{q}')\label{3.2.2}
\end{eqnarray}
Furthermore, since the Lorentz transformations are kinematic in point-form dynamics, both 
$\Omega$ and $|0\rangle$ remain invariant under the same set of operators $\hat{U}(\Lambda)=\hat{U}_0(\Lambda)$. This fact, once again, greatly simplifies 
the matters. We obtain 
\begin{eqnarray}
\Omega&=&\hat{U}(\Lambda)\Omega=\hat{U}_0(\Lambda)\Omega\nonumber\\
&=&\hat{U}_0(\Lambda)e^{\hV}\hat{U}^{-1}_0(\Lambda)|0\rangle\nonumber\\
&=&e^{\hat{U}_0(\Lambda){\hV}\hat{U}_0^{-1}(\Lambda)}|0\rangle\label{3.2.3}
\end{eqnarray}
Hence, we take $\hV$ to be a Lorentz scalar. This  in turn means that the coefficient functions $f_1$, $f_2$ and $f_3$ must be Lorentz scalars:
\begin{eqnarray}
f_3(\ka,\ka';q,q')&=&f_3(\ka,\ka';q\cdot q')+f_3(\ka,\ka')\delta(q-q')\nonumber\\
f_2(\ka;q,q')&=&f_2(\ka;q\cdot q')+f_2(\ka)\delta(q-q')\nonumber\\
 f_1(q,q')&=&f_1(q\cdot q')+A\delta(q-q')\label{3.2.4}
 \end{eqnarray}
 where $A$ is a constant. However, 
 as in \eqref{3.1.4}, we consider couplings that are local in velocity variables and use \eqref{3.2.4} in \eqref{3.2.2} to obtain
 \begin{eqnarray}
\hV&=&\int\frac{d^3\bs{q}}{2\sqrt{1+\bs{q}^2}}\left\{\int d\ka\,d\ka'\,f_3(\ka,\ka')\hB^\dagger(\bs{q},\ka)\hB^\dagger(\bs{q},\ka')\right.\nonumber\\
 &&+\left.\int d\ka\, f_2(\ka)\hB^\dagger(\bs{q},\ka)\ha^\dagger(\bs{q})
 +A\ha^\dagger(\bs{q})\ha^\dagger(\bs{q})\right\}\label{3.2.5}
 \end{eqnarray}
 Note that $f_3(\ka,\ka')=f_3(\ka',\ka)$. Further, since $\hB^\dagger(\bs{q},\ka)$ is an even function of $\ka$, $f_2$ is an even function of $\ka$. 
 
 The coupling functions $f_3$, $f_2$ and the constant $A$ must be determined by the requirement that $\Omega$ be annihilated by $\hc(\bs{q},\ms)$:
 \begin{equation}
 \hc(\bs{q},\ms)|\Omega\rangle=\hc(\bs{q},\ms)e^{\hV}|0\rangle=0\label{3.2.6}
 \end{equation}
We can bring this equation into an equivalent, simpler form by considering the formal Taylor expansion of $e^V$ and the action of $\hc(\bs{q},\ms)$ on 
$\hV^n$ for each $n$. To that end, note since $\hV$ is a function of only the creation operators $\ha^\dagger$ and $\hB^\dagger$ and since 
$\hc$ is a linear function of $\ha$, $\ha^\dagger$, $\hB$ and $\hB^\dagger$, the commutator $[\hc(\bs{q},\ms),\hV]$ is a function of only 
creation operators $\ha^\dagger$ and $\hB^\dagger$. Therefore, $[\hc(\bs{q},\ms),\hV]$ commutes with $\hV$. With this observation, 
the following identity can be readily proved by induction: 
\begin{equation}
\frac{\hc(\bs{q},\ms)\hV^n}{n!}=\frac{\hV^n\hc(\bs{q},\ms)}{n!}+\frac{\hV^{n-1}}{(n-1)!}\left[\hc(\bs{q},\ms),\hV\right]\label{3.2.7}
 \end{equation}
Summing over $n=1,2,3,\cdots$, we then obtain 
\begin{equation}
\hc(\bs{q},\ms)e^{\hV}=e^{\hV}\left(\hc({\bs{q}},\ms)+\left[\hc(\bs{q},\ms),\hV\right]\right)\label{3.2.8}
\end{equation}
Therewith, the requirement \eqref{3.2.6} can be fulfilled by demanding that the following equation hold:
\begin{equation}
\left(\hc({\bs{q}},\ms)+\left[\hc(\bs{q},\ms),\hV\right]\right)|0\rangle=0\label{3.2.9}
\end{equation}
In other words, $\left(\hc({\bs{q}},\ms)+[\hc(\bs{q},\ms),\hV]\right)$ must not contain creation operators $\ha^\dagger(\bs{q})$ 
or $\hB^\dagger(\bs{q},\ka)$. Using \eqref{3.1.5b}  and \eqref{3.2.5} in \eqref{3.2.9} and setting the coefficients 
of $\ha^\dagger(\bs{q})$ and $\hB^\dagger(\bs{q},\ka)$ to be equal to zero, we obtain the following two singular integral equations:
\begin{eqnarray}
R^*(\ms,\ka)+2\int d\ka'\,T^*(\ms,\ka')f_3(\ka,\ka')+t^*(\ms)f_2(\ka)&=&0\label{3.2.10}\\
r^*(\ms)+2A t^*(\ms)+\int d\ka'\,T^*(\ms,\ka')f_2(\ka')&=&0\label{3.2.11}
\end{eqnarray}
In view of the relationship \eqref{3.1.8} between $r(\ms)$ and $t(\ms)$, 
we note that the coupling function $f_2(\ka)$ defines an integral operator that maps $T^*(\ms,\ka)$ to $t^*(\ms)$ (times a function of $\ms)$. 
Once $f_2$ is determined from \eqref{3.2.11}, 
the other function $f_3(\ka,\ka')$ can be determined from \eqref{3.2.10}.

Substituting \eqref{3.1.8}, \eqref{3.1.12} and \eqref{3.1.13}   in \eqref{3.2.10} and \eqref{3.2.11}, we obtain
\begin{eqnarray}
0&=&4C\frac{\rms}{2m\sqrt{\ms-4m^2}}f_3(m(\ka),\ms)+t^*(\ms)\Bigl\{f_2(\ka)+\Bigr.\nonumber\\
&&\quad+\frac{i\pi\beta}{8H_1^{(2)}(M\tau)}\frac{\alpha(m(\ka))H_1^{(1)}(m(\ka)\tau)}{\rms+m(\ka)}\frac{{\cal H}^*_{1,2}(\rms\tau,M\tau)}{\rms+M}\nonumber\\
&&\qquad\left.-\frac{i\pi\beta}{2H_1^{(2)}(M\tau)}\frac{{\cal H}^*_{1,2}(\rms\tau,M\tau)}{\rms+M}\int_{2m}^\infty d\mu(\ka')\,\frac{\alpha(m(\ka'))H_1^{(2)}(m(\ka')\tau)f_3(m(\ka),m(\ka'))}{\rms-m(\ka')}\right\}\nonumber\\
\label{3.2.12}\\
0&=&2C\frac{\rms}{2m\sqrt{\ms-4m^2}}f_2(\ms)+t^*(\ms)\Bigl\{2A-\frac{(\rms-M)H_1^{(1)}(M\tau)}{(\rms+M)H_1^{(2)}(M\tau)}\Bigr.\nonumber\\
&&\qquad-\left.\frac{i\pi\beta}{4H_1^{(2)}(M\tau)}\frac{{\cal H}^*_{1,2}(\rms\tau,M\tau)}{\rms+M}\int_{2m}^\infty d\mu(\ka')\,\frac{\alpha(m(\ka'))H_1^{(2)}(m(\ka')\tau)f_2(m(\ka'))}{\rms-m(\ka')}\right\}\nonumber\\
\label{3.2.13}
\end{eqnarray}
Substituting from \eqref{3.1.20a} for $t^*(\ms)$ and using \eqref{3.1.15} for $\rho(\ms)$, we finally get
\begin{eqnarray}
0&=&4f_3(\ms,m(\ka))-\frac{i\pi\beta}{4}\alpha(\ms)H_2^{(1)}(\rms\tau)G^*(\ms)\Bigl\{(\rms+M)H_1^{(2)}(M\tau)f_2(m(\ka))\Bigr.\nonumber\\
&&\qquad+\frac{i\pi\beta}{2}\frac{\alpha(m(\ka))H_1^{(1)}(m(\ka)\tau)}{\rms+m(\ka)}{\cal H}^*_{1,2}(\rms\tau,M\tau)\nonumber\\
&&\qquad-\left.\frac{i\pi\beta}{2}{\cal H}^*_{1,2}(\rms\tau,M\tau)\int_{2m}^\infty d\mu(\ka')\,\frac{\alpha(m(\ka'))H_1^{(2)}(m(\ka')\tau)f_3(m(\ka),m(\ka'))}{\rms-m(\ka')}\right\}\nonumber\\
\label{3.2.14}\\
0&=&2f_2(\ms)-\frac{i\pi\beta}{4}\alpha(\ms)H_2^{(1)}(\rms\tau)G^*(\ms)\left\{2A(\rms+M)H_1^{(2)}(M\tau)-(\rms-M)H_1^{(1)}(M\tau)\right.\nonumber\\
&&\quad\left.-\frac{i\pi\beta}{4}{\cal H}^*_{1,2}(\rms\tau,M\tau)\int_{2m}^\infty d\mu(\ka')\,\frac{\alpha(m(\ka'))H_1^{(2)}(m(\ka')\tau)f_2(m(\ka'))}{\rms-m(\ka')}\right\}\nonumber\\
\label{3.2.15}
\end{eqnarray}
The task now is to evaluate the singular integrals in \eqref{3.2.14} and \eqref{3.2.15} and solve these equations 
for the unknown functions $f_2(m(\ka))$ and $f_3(m(\ka),m(\ka'))$. These existence of singular integrals clearly depend on the form factor 
$\alpha(m(\ka))$. Therefore, we must consider \eqref{3.2.14} and \eqref{3.2.15} as a set of constraints that determine the permissible 
class of form factors such as those that fulfill the H\"older condition \cite{muskhelishvili}.  For such form factors, the existence of the physical vacuum $\Omega$ annihilated 
by $\hc(\bs{q},\ms)$ is ensured by the solutions of \eqref{3.2.14} and  \eqref{3.2.15}.

\section{Resonances and representations of the causal Poincar\'e semigroup}\label{sec4}
\subsection{Introduction and summary of results} 
As stated above, the pole singularities of the Green's function \eqref{3.1.19a} correspond to resonances, our main focus of this section. 
Therefore, let us consider the analyticity properties of the Green's function. These in turn follow from those of the function $\Pi(\ms)$ defined by 
\eqref{3.1.16a}: 
\begin{eqnarray}
\Pi(\ms)&=&-\frac{(\pi\beta)^2}{32}{\cal H}_{1,2}(\rms\tau,M\tau)\int_{2m}^\infty d\mu(\ka)\,\frac{\alpha(m(\ka))^2{\cal H}_{1,2}(\rms\tau,m(\ka)\tau)}{\ms-m(\ka)^2}\nonumber\\
&=&-\frac{(\pi\beta)^2}{32}{\cal H}_{1,2}(\rms\tau,M\tau)\int_{2m}^\infty \frac{m(\ka)dm(\ka)}{2m\sqrt{m(\ka)^2-4m^2}}\frac{\alpha(m(\ka))^2{\cal H}_{1,2}(\rms\tau,m(\ka)\tau)}{\ms-m(\ka)^2}\nonumber\\
\label{4.1}
\end{eqnarray}
where $2m\leq m(\ka)<\infty$ and $2m\leq\sqrt{\ms}<\infty$. 

The integral is clearly singular. The singularity at $m(\ka)=2m$ can be removed by  demanding that the form factor $\alpha(m(\ka))$ vanish sufficiently fast at this point. 
For a suitable class of form factors $\alpha(m(\ka))$ that fulfills this condition, taking into account the analyticity properties of Hankel functions, we see that 
the integral exists for complex valued $\ms$ and defines $\Pi(\ms)$ as a function analytic everywhere 
except for the branch line $[2m,\infty)$.  The discontinuity of $\Pi(\ms)$ across this branch cut can be computed the usual way: 
\begin{equation}
\Pi_+(\ms)-\Pi_-(\ms)= -\frac{i\pi^3\beta^2}{16}\frac{\rms}{2m\sqrt{\ms-4m^2}}\alpha(\ms)^2{\cal H}_{1,2}(\rms\tau,M\tau){\cal H}_{1,2}(\rms\tau,\rms\tau)\label{4.2}
\end{equation}
where 
\begin{equation}
\Pi_\pm(\ms)=\lim_{\epsilon\to0}\Pi(\ms\pm i\epsilon).\label{4.2b}
\end{equation}
It then follows from the definition \eqref{3.1.19a} that the Green's function $G(\ms)$ is also analytic everywhere on the complex $\ms$-plane, 
except for the branch cut $[2m,\infty)$ and the zeros of the denominator.  The limit functions from above and below, 
\begin{equation}
G_\pm(\ms)=\frac{1}{\ms-M^2-\Pi_\pm(\ms)},\label{4.3}
\end{equation}
can be analytically continued across the branch cut to a second Riemann sheet. If we denote the Green's function on the second sheet by $G^{II}(\ms)$,  then  $\lim_{\epsilon\to0}G^{II}(\ms\pm i\epsilon)=G_\mp(\ms)$. 
On the first sheet itself, the discontinuity of the Green's function across the branch cut follows from \eqref{4.2} and \eqref{4.3}:
\begin{eqnarray}
G_+(\ms)-G_-(\ms)&=&-\frac{i\pi^3\beta^2}{16}\frac{\rms}{2m\sqrt{\ms-4m^2}}\alpha(\ms)^2{\cal H}_{1,2}(\rms\tau,M\tau){\cal H}_{1,2}(\rms\tau,\rms\tau)\times\nonumber\\
&&\qquad\qquad G_-(\ms)G_+(\ms).\label{4.4}
\end{eqnarray}
From this relationship, we infer the dispersion relation
\begin{eqnarray}
G(\ms)&=&-\frac{(\pi\beta)^2}{32}{\cal H}_{1,2}(\rms\tau,M\tau)\times\nonumber\\
&&\qquad \int_{2m}^\infty d\mu(\ka)\,\frac{\alpha^2(m(\ka)){\cal H}_{1,2}(\rms\tau,m(\ka)\tau)G_-(m(\ka))G_+(m(\ka))}{\ms-m(\ka)^2}.\nonumber\\
\label{4.5}
\end{eqnarray}

The number and distribution of the poles of $G^{II}(\ms)$ on the second sheet are determined by the form factor $\alpha(m(\ka))$.  In general, there may be an infinite number of resonance poles, both simple 
and of higher order. Resonance poles appear in pairs, one in the lower half plane and the other in the upper half plane at the conjugate position. We typically consider the poles located on the lower half plane as these may be associated with the decaying part of the resonance. 

The standard practice of determining the positions of these poles is to determine the 
zeros of the real part of the inverse Green's function $\ms-M^2-\Pi(\ms)$, which gives the real-valued mass of the resonance $\bar{M}_R$, and evaluate the imaginary part of  $\ms-M^2-\Pi(\ms)$
at the value of the zero of the real part to obtain the resonance width $\bar{\Gamma}_R$. The complex square mass of the resonance is then given by 
\begin{equation}
{\bar{\ms}}_R=\bar{M}^2_R-i{\bar{M}_R}{\bar{\Gamma}}_R.\label{4.6}
\end{equation}
This definition of resonance mass and width is motivated by the fact that the scattering amplitude has a local maximum at $\ms=\bar{\ms}_R$. However,  the parameters 
$\bar{M}_R$ and $\bar{\Gamma}_R$ obtained by  the above procedure have been found to be gauge dependent \cite{consoli,bohm&harshman}. Therefore, we favor defining 
the resonance pole position $\ms_R$ by the zeros of the inverse Green's function on the second Riemann sheet:
\begin{equation}
\ms_R-M^2-\Pi_+(\ms_R)=0.\label{4.7}
\end{equation}
The mass and width extracted out of  the solution of \eqref{4.7} will be automatically gauge invariant (and coincide with the definition of resonances as poles of the analytic $S$-matrix).
Still, a given complex solution $\ms_R$ of \eqref{4.7} can be parametrized in great many ways to obtain a real-valued resonance mass and width. A highlight of the results obtained in this paper is, 
as seen from the analysis in 
section~\ref{sec4.5} below, that a resonance 
can be associated to an irreducible representation of the causal Poincar\'e semigroup, much the same way as Wigner's unitary, irreducible representations of the Poincar\'e group are associated 
with stable particles. This allows for a state-vector description for resonances and the transformation properties of resonance state vectors under the causal Poincar\'e semigroup 
single out the following \emph{unique} definitions of resonance mass and 
width: 
\begin{eqnarray}
M_R&=&\Re\,\left(\sqrt{\ms_R}\right)\nonumber\\
\Gamma_R&=&-2\Im\,\left(\sqrt{\ms_R}\right)\label{4.8}
\end{eqnarray}
so that $\ms_R=\left(M_R-\frac{i}{2}\Gamma_R\right)^2$. 

In particular, as seen from \eqref{4.5.4} and \eqref{4.5.6} below, only this definition of width satisfies the lifetime-with relation
\begin{equation}
\tau_R=\frac{1}{\Gamma_R}\label{4.9}
\end{equation}
as an exact and universal identity. The use of this lifetime-width relation as the fundamental criterion for defining the mass and width of a resonance was first advocated in \cite{bohm&harshman}. 
See \cite{bohm&harshman, bohm&sato} for further discussions on the conceptual and computational difficulties of various definitions 
of resonance mass and width. 
\\

When the coupling constant $\beta$ is small, an approximate solution to \eqref{4.7} can be obtained by ignoring the principal value part of the singular integral 
that defines $\Pi_+(\ms)$:
\begin{eqnarray}
\Pi_+(\ms)&=&-\frac{(\pi\beta)^2}{32}{\cal H}_{1,2}(\rms\tau,M\tau)\int_{2m}^\infty d\mu(\ka)\,\frac{\alpha(m(\ka))^2{\cal H}_{1,2}(\rms\tau,m(\ka)\tau)}{\ms-m(\ka)^2}\nonumber\\
&=&-\frac{(\pi\beta)^2}{32}{\cal H}_{1,2}(\rms\tau,M\tau)\left\{PV\int_{2m}^\infty d\mu(\ka)\,\frac{\alpha(m(\ka))^2{\cal H}_{1,2}(\rms\tau,m(\ka)\tau)}{\ms-m(\ka)^2}\right.\nonumber\\
&&\quad+\left.i\pi\int_{2m}^\infty d\mu(\ka)\,\delta(\rms-m(\ka))\alpha(m(\ka))^2{\cal H}_{1,2}(\rms\tau,m(\ka)\tau)\right\}\nonumber\\
&\approx&-\frac{i\pi^3\beta^2}{32}\frac{\rms}{2m\sqrt{\ms-4m^2}}\alpha(\ms)^2{\cal H}_{1,2}(\rms\tau,\rms\tau)\label{4.10}
\end{eqnarray}
From \eqref{4.10} and \eqref{4.7}, we then obtain the equation for the resonance pole position $\ms_R$:
\begin{equation}
\ms_R-M^2+\frac{i\pi^3\beta^2}{32}\frac{\sqrt{\ms_R}}{2m\sqrt{\ms_R-4m^2}}\alpha(\ms_R)^2{\cal H}_{1,2}(\sqrt{\ms_R}\,\tau,\sqrt{\ms_R}\,\tau)=0\label{4.11}
\end{equation}
In order to solve for $\ms_R$ explicitly, obviously we need the explicit form of $\alpha$. In general, it is clear that $M_R=\Re\left(\sqrt{\ms_R}\right)$ will be different from $M$. 
Hence, the resonance occurs at a mass value different from that of the stable particle that couples to the continuum mass $m(\ka)$. 

The state vector description for the resonances alluded to above will be obtained by constructing the state vectors $|\bs{q},\ms^\pm\rangle$ corresponding to 
$\Pi_\pm(\ms)$  and then taking the weak analytic extensions of $|\bs{q},\ms^\pm\rangle$ to the upper and lower half complex 
$\ms$-plane on the second Riemann sheet. The evaluation of the analytic extension of $|\bs{q},\ms^-\rangle$ at the 
pole position \eqref{4.7} defines the state vectors $|\bs{q},\ms_R^-\rangle$ that describe the decaying resonance. We call these vectors 
\emph{Gamow vectors} in honor of George Gamow who first gave a heuristic treatment of  
eigenvectors of the Hamiltonian with complex eigenvalues to describe decaying states. The mathematical structure that allows 
this construction rigorously is that of a rigged Fock space of Hardy-type. In the following subsections of this section, 
we will take on the task of constructing these rigged Fock spaces and representations of the causal Poincar\'e semigroup, and discuss the 
transformation properties of resonance state vectors under these representations. 

\subsection{In-states and out-states: the rigged Hilbert spaces of Hardy functions}\label{sec4.1}
As seen above, the specification of how to handle the singularities of the integral that defines $\Pi(\ms)$ gave us the boundary value functions $\Pi_\pm(\ms)$. Therewith, we
also have Green's functions $G_\pm(\ms)$. Replacing $G(\ms)$ of the formal solution \eqref{3.1.20a} to the eigenvalue problems \eqref{3.1.1} and \eqref{3.1.2} with 
$G_\pm(\ms)$, we can obtain two different solutions to \eqref{3.1.1} and \eqref{3.1.2}, which we denote by $\hc^\dagger_\pm(\bs{q},\ms)$ and 
$\hc_\pm(\bs{q},\ms)$:
\begin{eqnarray}
\hc^\dagger_\pm(\bs{q},\ms)&=&\int_{-\infty}^\infty d\ka\,\left(T_\pm(\ms,\ka)\hB^\dagger(\bs{q},\ka)+R_\pm(\ms,\ka)\hB(\bs{q},\ka)\right)+t_\pm(\ms)\ha^\dagger(\bs{q})+r_\pm(\ms)\ha(\bs{q})\nonumber\\
\label{4.1.1}\\
\hc_\pm(\bs{q},\ms)&=&\int_{-\infty}^\infty d\ka\,\left(R^*_\pm(\ms,\ka)\hB^\dagger(\bs{q},\ka)+T^*_\pm(\ms,\ka)\hB(\bs{q},\ka)\right)+r^*_\pm(\ms)\ha^\dagger(\bs{q})+t^*_\pm(\ms)\ha(\bs{q})\nonumber\\
\label{4.1.2}
\end{eqnarray}
where 
\begin{eqnarray}
t_\pm(\ms)&=&\left(\rms+M\right)H_1^{(1)}(M\tau)\rho(\ms)G_\pm(\ms)\nonumber\\
r_\pm(\ms)&=&-\left(\rms-M\right)H_1^{(2)}(M\tau)\rho(\ms)G_\pm(\ms)\nonumber\\
T_\pm(\ms,\ka)&=&C\delta(\rms-m(\ka))+\frac{i\pi\beta}{8}\frac{\alpha(m(\ka))H_1^{(1)}(m(\ka)\tau)}{\rms-m(\ka)\pm i\epsilon}{\cal H}_{1,2}(\rms\tau,M\tau)\rho(\ms)G_\pm(\ms)\nonumber\\
R_\pm(\ms,\ka)&=&-\frac{i\pi\beta}{8}\frac{\alpha(m(\ka))H_1^{(2)}(m(\ka)\tau)}{\rms+m(\ka)}{\cal H}_{1,2}(\rms\tau,M\tau)\rho(\ms)G_\pm(\ms)\
\label{4.1.3}
\end{eqnarray}
with $G_\pm(\ms)$ defined by \eqref{4.2b} and \eqref{4.3}.

From \eqref{3.2.10} and \eqref{3.2.11}, we see that the vacuum state annihilated by the operator $\hc_-(\bs{q},\ms)$ is also annihilated by $\hc_+(\bs{q},\ms)$.  Therefore, the action of the 
creation operators \eqref{4.1.1} on this vacuum  defines two sets of states: 
\begin{equation}
\left|\bs{q},\ms^\pm\right.\rangle:=\hc^\dagger_\pm(\bs{q},\ms)\left|\Omega\right\rangle\label{4.1.4}
\end{equation}
Let us normalize both sets of states as
\begin{equation}
\langle^\pm\bs{q},\ms|\bs{q}',\ms'^\pm\rangle=2q^0\delta(\bs{q}-\bs{q}')\delta(\rms-\rrms).\label{4.1.5}
\end{equation}
This normalization condition also fixes the arbitrary scaling constants of \eqref{4.1.1} and \eqref{4.1.2} so that $\hc(\bs{q},\ms)$ and $\hc^\dagger(\bs{q},\ms)$ fulfill the canonical 
commutation relations $[\hc_\pm(\bs{q},\ms),\hc^\dagger_\pm(\bs{q}',\ms')]=2q^0\delta(\bs{q}-\bs{q}')\delta(\rms-~\rrms)$.

The operators $\hc_+^\dagger(\bs{q},\ms)$ and $\hc_+(\bs{q},\ms)$ can be used to define a causal scalar field $\hat{\varphi}_+(x)$ which fulfills the in-going boundary conditions for $t\to{-\infty}$. 
Similarly, operators $\hc_-^\dagger(\bs{q},\ms)$ and $\hc_-(\bs{q},\ms)$ define a scalar field $\hat{\varphi}_-(x)$ which fulfills the out-going boundary conditions for $t\to\infty$. These limits 
are to be understood not in the strong operator sense but in the weak sense.\footnote{However, it is apparent  that, due to the continuous mass distribution, the fields obtained by these 
limits do not have a direct particle interpretation.  For such a particle interpretation, we must consider the inverse problem of the direct integral decomposition \eqref{2.2.3} and construct a tensor 
product space from the continuum. We will not deal with this problem in the paper.}

Therefore,  continua of vectors \eqref{4.1.4} represent the scattering states, with $|\bs{q},\ms^+\rangle$ (or, more precisely, their smooth superpositions) 
as having evolved from asymptotically free in-states and $|\bs{q},\ms^-\rangle$ (or, more precisely, their smooth superpositions) as evolving into asymptotically free
out-states. The overlap $\langle^-\bs{q},\ms\left|\right.\bs{q},\ms^+\rangle$, equal to the vacuum expectation value of the commutator 
$[\hc_-(\bs{q}',\ms'),\hc_+^\dagger(\bs{q},\ms)]$, defines the analytic $S$-matrix: 
\begin{eqnarray}
\langle^-\bs{q}',\ms'\left|\right.\bs{q},\ms^+\rangle&=&\langle\Omega\left|\right.\hc_-(\bs{q}',\ms')\hc^\dagger_+(\bs{q},\ms)\Omega\rangle\nonumber\\
&=&\langle\Omega\left|\right.[\hc_-(\bs{q}',\ms'),\hc^\dagger_+(\bs{q},\ms)]\Omega\rangle\nonumber\\
&=&2q^0\delta(\bs{q}-\bs{q}')\delta(\rms-\rrms)\,S(\ms)\label{4.1.6}
\end{eqnarray}
It is known that the both the analytic $S$-matrix  and the Green's function lead to the same lineshape for a resonance scattering process~\cite{bohm&harshman}.  In particular, 
this means that  the resonance poles of the Green's function on the second Riemann sheet correspond to the poles of the analytic $S$-matrix, also on the second Riemann sheet. 
This property allows us to carry out the analysis using the analyticity structure of \eqref{4.1.6} and arrive at a description of the resonances given by the solutions of \eqref{4.7}.  

Let us now turn to the problem of constructing the Fock space from the basis vectors \eqref{4.1.4}. 
As discussed in the context of the formal solution of section~\ref{sec3.1}, we can take all square integrable smooth superpositions of either $|\bs{q},\ms^-\rangle$ or 
$|\bs{q},\ms^+\rangle$ and complete this space with respect to the ensuing norm topology to obtain a Hilbert space $\int_{4m^2}^\infty d\ms\,{\cal H}_-(\ms)$ or 
$\int_{4m^2}^\infty d\ms\,{\cal H}_+(\ms)$. Let us further assume that these two Hilbert spaces are the same, i.e., that 
asymptotic completeness holds: 
\begin{equation}
\int_{4m^2}^\infty d\ms\,{\cal H}_-(\ms)=\int_{4m^2}^\infty d\ms\,{\cal H}_+(\ms)\equiv\int_{4m^2}^\infty d\ms\,{\cal H}(\ms)\equiv{\cal H}\label{4.1.7}
\end{equation}
The asymptotic completeness imposes further restrictions on the interaction, a very complicated and subtle mathematical problem that will 
take us far afield from our main focus.  

It is often claimed that \eqref{4.1.7} means that the two sets of vectors $|\bs{q},\ms^\pm\rangle$ furnish two bases for the same Hilbert space. 
However, since the vectors $|\bs{q},\ms^\pm\rangle$ are themselves not square integrable, they do not belong to the Hilbert space \eqref{4.1.7} but must be defined 
as functionals on a suitable subspace of test functions.  The vectors $|\bs{q},\ms^\pm\rangle$ do form bases for these test function spaces. The crucial point to emphasize 
is that \eqref{4.1.7} does not dictate that the two sets of vectors $|\bs{q},\ms^\pm\rangle$ be defined on the same test function space--it is entirely possible that they be 
defined on two different dense spaces of test functions so that the norm completion of each leads to the same Hilbert space \eqref{4.1.7}. In fact, it is equally possible 
to introduce topologies  stronger than the norm topology on these test function space so that upon completion the spaces remain distinct. Our entire analysis below hinges on 
this point and in this regard, we view the construction developed below as a topological refinement of the principle of asymptotic completeness. 

To construct the test function spaces, suppose there exist 
in-vectors $\phi^+\in{\cal H}$ and out-vectors $\psi^-\in{\cal H}$ so that the following expansions hold: 
\begin{eqnarray}
\phi^+&=&\intq\int_{4m^2}^\infty d\ms\,|\bs{q},\ms^+\rangle\langle^+\bs{q},\ms|\phi^+\rangle\nonumber\\
\psi^-&=&\intq\int_{4m^2}^\infty d\ms\, |\bs{q},\ms^-\rangle\langle^-\bs{q},\ms|\psi^-\rangle\label{4.1.8}
\end{eqnarray}
In order to determine the defining criteria for $\phi^+$ and $\psi^-$, let us consider their inner product.  Using \eqref{4.1.8} and \eqref{4.1.6}, we obtain
\begin{equation}
\langle\psi^-|\phi^+\rangle=\intq\int_{4m^2}^\infty d\ms\,\langle^-\psi|\bs{q},\ms^-\rangle S(\ms)\langle^+\bs{q},\ms|\phi^+\rangle\label{4.1.9}
\end{equation}
As pointed out above, the meromorphic  function $S(\ms)$ has poles corresponding to resonances at $\ms_{R_i}$
in the lower half plane of the second Riemann sheet. In order to ascertain the contribution of these poles to the inner product \eqref{4.1.9}, we 
 must require that the integral over the square mass variable be deformed to a contour integral in the lower half plane of the second sheet 
 of the $S$-matrix. Since $S(\ms)$ is a meromorphic  function, it is only necessary that we require that the functions $\langle^-\psi|\bs{q},\ms^-\rangle$ 
 and $\langle^+\bs{q},\ms|\phi^+\rangle$ admit analytic extensions into the lower half plane.  Further, if the analytic extensions of 
$\langle\psi^-|\bs{q},\ms^-\rangle$ and $\langle^+\bs{q},\ms|\phi^+\rangle$ 
decrease  
sufficiently fast for $|\ms|\to\infty$, then the contribution to the integral from the infinite semicircle on the lower half plane will vanish. 
Using Cauchy's theorem to evaluate the integral around the poles,   \eqref{4.1.9}  may then be written, on the second sheet,  as 
\begin{eqnarray}
\langle\psi^-|\phi^+\rangle&=&-4\pi i\sum_{i}R_i\intq\langle\psi^-|\bs{q},\ms_{R_i}^-\rangle\langle^+\bs{q},\ms_{R_i}|\phi^+\rangle
\nonumber\\
&&+\int_{-\infty}^{4m^2}d\ms \intq\langle\psi^-|\bs{q},\ms^-\rangle S(\ms)\langle^+ \bs{q},\ms\phi^+\rangle\nonumber\\
\label{4.1.10}
\end{eqnarray}
where $\ms_{R_i}$ are the solutions of \eqref{4.7} (with $\Im\left(\sqrt{\ms_{R_i}}\right)<0$) 
and $R_i$ is the residue of the Laurent expansion of $S$-matrix around $\ms_{R_i}$. (While the Laurent expansion 
itself can be generally carried out only in the neighborhood of any one of the pole positions $\ms_{R_i}$, the superposition 
of all the pole contributions of \eqref{4.1.10} holds.)

Motivated by the analyticity properties required for \eqref{4.1.10}, we choose the functions 
$\langle\psi^-|\bs{q},\ms^-\rangle$ and $\langle^+\bs{q},\ms|\phi^+\rangle$ to be of Hardy class 
\cite{koosis} from below in the square mass variable $\ms$.\footnote{In the non-relativistic case, the use 
of Hardy class functions in scattering theory was first proposed in \cite{bohm}. Further mathematical details of this construction, also 
in the non-relativistic setting, can be found in \cite{gadella-gomez}.} In addition, we also require these 
wavefunctions to be smooth as well as rapidly decreasing both at infinity and at the origin $\ms=0$. 
As shown in \cite{sw1,sw2}, these additional conditions guarantee that the integral representation \eqref{4.1.10} of 
$\langle\psi^-|\phi^+\rangle$ exists for $S$-matrix functions $S(\ms)$ 
that do not diverge for $|\ms|\to\infty$ 
faster than a polynomial of any order. Finally, we require that $\langle\psi^-|\bs{q},\ms^-\rangle$ and $\langle^+\bs{q},\ms|\phi^+\rangle$ be Schwartz functions 
in the velocity variable $\bs{q}$. These conditions also permit a realization of the interacting 
Poincar\'e algebra \eqref{2.2.22}  and \eqref{2.2.23} by everywhere defined, continuous operators in a 
countably normed vector space. For discussion of this point, see \cite{sw2}.

Specifically, let\footnote{The discrepancy in the signs is due the conventions in physics and mathematics. In scattering theory, the in vectors $\phi^+$ and 
out vectors $\psi^-$ have been defined by their behavior at $t\to-\infty$ and $t\to\infty$, respectively. On the other hand, the notation for 
Hardy spaces ${\cal H}_-^2$ and ${\cal H}_+^2$ has been used in mathematics based on the analyticity properties on the open lower 
half plane $\mathbb{C}_-$ and open upper half plane $\mathbb{C}_+$, respectively.}
\begin{subequations}
\label{4.1.11}
\begin{equation}
\tag{\ref{4.1.11}+}
\{\langle^-\bs{q},\ms|\psi^-\rangle\}\equiv{\cal K}_+:=\left.\mathcal{M}\cap\HS_+^2\right|_{{\mathbb{R}}_{\ms_0}}
\otimes\mathcal{S}({\mathbb{R}}^3)\label{4.1.11+}
\end{equation}
\begin{equation}
\tag{\ref{4.1.11}$-$}
\{\langle^+\bs{q},\ms|\phi^+\rangle\}\equiv{\cal K}_-:=\left.\mathcal{M}\cap\HS_-^2\right|_{{\mathbb{R}}_{\ms_0}}
\otimes\mathcal{S}({\mathbb{R}}^3)\label{4.1.11-}
\end{equation}
\end{subequations}
Here,
\begin{enumerate}
\item [{\rm (4.2.a)}] $\mathcal{S}({\mathbb{R}}^3)$ is the Schwartz space on $\mathbb{R}^3$, i.e., smooth, complex valued functions over ${\mathbb{R}}^3$ 
that decrease at infinity faster than any inverse polynomial. As functions of $\bs{q}$, we take $\langle\psi^-|\bs{q},\ms^-\rangle$ and $\langle^+\bs{q},\ms|\phi^+\rangle$ to belong 
to ${\mathcal S}(\mathbb{R}^3)$. 
\item [{\rm (4.2.b)}] $\HS_\pm^2$ are Hardy class functions on ${\mathbb{C}}_\pm$. 
\item [{\rm (4.2.c)}] $\mathcal{M}=\{f:\ f \in{\mathcal{S}}({\mathbb{R}}); \left.\frac{1}{x^m}\frac{d^nf}{dx^n}\right|_{x=0}=0,\ n,m=0,1,2,\cdots\}$.  
\item [{\rm (4.2.d)}] The support of every function 
in $\HS_\pm^2$ is the entire line $\mathbb{R}$ and the symbol $\left.{}\right|_{{\mathbb{R}_{\ms_0}}}$ indicates space of functions 
obtained by the restricting the domain of $\mathcal{M}\cap\HS_\pm^2$ to the spectrum of $M^2$, ${\mathbb{R}}_{\ms_0}=[4m^2,\infty)$. As functions of $\ms$, 
we take $\langle^-\bs{q},\ms|\psi^-\rangle$ and $\langle^+\bs{q},\ms|\phi^+\rangle$ to belong to $\left.\mathcal{M}\cap\HS_+^2\right|_{{\mathbb{R}}_{\ms_0}}$ and 
$\left.\mathcal{M}\cap\HS_-^2\right|_{{\mathbb{R}}_{\ms_0}}$, respectively. 
\end{enumerate}

It is a property of Hardy class functions that if $f\in \HS_\pm^2$, then the complex conjugate function ${f}^*\in\HS_\mp^2$. Therefore, \eqref{4.1.11} implies that 
$\langle^-\psi|\bs{q},\ms^-\rangle={\langle^-\bs{q},\ms|\psi^-\rangle}^*$ has the required analyticity properties on the lower half complex $\ms$-plane. 

The complementary analyticity properties imposed on the wave functions of $\phi^+$ and $\psi^-$ make them mathematically distinct objects (in contrast to their 
Hilbert space representation where both types of vectors are represented by square integrable functions in the same Hilbert space \eqref{4.1.7}). 
They are distinct conceptually and operationally as well. In a typical scattering experiment, the vector $\phi^+$ is taken to have evolved from 
an asymptotic in-state $\phi^{\rm in}$, prepared by a device such as a particle accelerator. On the other hand, $\psi^-$ evolves into an out- vector 
$\psi^{\rm out}$ which represents properties measured by means of a device such as a detector. In discussions on the foundations of quantum physics, 
a philosophical distinction is often made between prepared states and measured observables \cite{states-observables}. The differences between  
out-states and in-states  in terms of the degree of correlations have been noted in \cite{tdlee}.  Somewhat along the same lines, in the setting of quantum 
entanglement theory, it has been argued that in and out-states are different in terms of their entanglement content, namely scattering 
in-state vectors $\phi^{\rm in}$ are separable, while out vectors $\psi^{\rm out}$ 
representing observables are not \cite{harshman}. The characterization of $\phi^+$ and $\psi^-$ vectors in terms of Hardy class functions 
from below and above, respectively, entails a distinction in the same spirit between the two types of wavefunctions. 
For further discussions on this point, albeit in connection with non-relativistic scattering theory, see \cite{hardy-postulate}. 

 The spectral theorem of von Neumann ensures that there exists a realization of the Hilbert space \eqref{4.1.7} by $L^2$-functions defined 
 on the spectra of $\bs{\hat{P}}$ and $\hat{M}^2=\hat{P}_\mu\hat{P}^\mu$, a complete system of commuting operators (CSCO) for our system, such 
 that these operators all act as multiplication operator in the $L^2$-space. Let $L^2\left(\mathbb{R}_{\ms_0},\mathbb{R}^3\right)$ denote this 
 realization of the Hilbert space. Then, specifically, the spectral theorem ensures the existence of a unitary operator $\hat{U}$, not 
 necessarily unique, that establishes the equivalence of $\int_{4m^2}^\infty d\ms\,\HS(\ms)$ and $L^2\left(\mathbb{R}_{\ms_0},\mathbb{R}^3\right)$:
 \begin{equation}
 \hat{U}\left(\int_{4m^2}^\infty d\ms\,\HS(\ms)\right)=L^2\left(\mathbb{R}_{\ms_0},\mathbb{R}^3\right)\label{4.1.12}
 \end{equation}
 
 It has been shown in \cite{sw2} that the function spaces \eqref{4.1.11+} and \eqref{4.1.11-} are dense in the Hilbert space 
 $L^2\left(\mathbb{R}_{\ms_0},\mathbb{R}^3\right)$. Therefore, the completion of either space ${\cal K}_\pm$ under the usual norm topology leads to the 
 same Hilbert space $L^2\left(\mathbb{R}_{\ms_0},\mathbb{R}^3\right)$. What this  means is that the topological 
structure of the Hilbert space is rather coarse so that any 
distinctions between in and out vectors $\phi^+$ and $\psi^-$ encoded in the analyticity requirements of \eqref{4.1.11} will wash out  upon completion. 
This is certainly not surprising since asymptotic completeness was assumed at the outset. Obversely, if we want to sustain the 
analytic and regularity structures of \eqref{4.1.11} in the setting of a complete vector space, it is necessary to find a topology finer than the Hilbert space topology. 
In \cite{sw2}, it has also been shown that each space ${\cal K}_\pm$ can in fact be equipped with a nuclear Fr\'echet topology. 
It is in this sense, as stated above,  that the theory we are developing may be viewed as a topological refinement of the principle of asymptotic completeness.

From the existence of this topology and the denseness of \eqref{4.1.11} in $L^2\left(\mathbb{R}_{\ms_0},\mathbb{R}^3\right)$, we conclude that 
\begin{subequations}
\label{4.1.13}
\begin{equation}
\tag{\ref{4.1.13}+}
{\mathcal{K}}_+\subset L^2\left({\mathbb{R}}_{\ms_0},{\mathbb{R}}^3\right)\subset{\mathcal K}_+^\times
\label{4.1.13+}
\end{equation}
\begin{equation}
\tag{\ref{4.1.13}$-$}
{\mathcal{K}}_-\subset L^2\left({\mathbb{R}}_{\ms_0}, {\mathbb{R}}^3\right)\subset{\mathcal K}_-^\times
\label{4.1.13-}
\end{equation}
\end{subequations}
are a pair of rigged Hilbert spaces. (See \cite{sw1,sw2} for the detailed construction of these rigged Hilbert spaces.) 
Here,   ${\cal K}_\pm^\times$ are the  spaces of continuous antilinear functionals  equipped with the usual weak-* topology. 
The interacting Poincar\'e algebra \eqref{2.2.22} and \eqref{2.2.23} acts the same way on the spaces ${\cal K}_\pm$, a crucial requirement of 
a relativistic quantum theory (see \cite{weinberg}, p.~119). While the Lie algebra acts the same way, the operators representing 
finite Poincar\'e transformations \emph{do not} have the same action on ${\cal K}_\pm$. 
 
Taking the inverse images of rigged Hilbert spaces \eqref{4.1.13} under the unitary operator $\hat{U}$ of \eqref{4.1.12} and transporting the topologies of 
${\cal K}_\pm$ to these inverse images (so that $\hat{U}$ is a homeomorphism), we obtain a pair of 
abstract rigged Hilbert spaces that are unitarily equivalent to
the Hardy-type rigged Hilbert spaces \eqref{4.1.13}:
\begin{equation}
\Phi_\pm\subset \int_{4m^2}^\infty d\ms\,\HS(\ms)\subset \Phi_\pm^\times\label{4.1.14}
\end{equation}
Within this mathematical structure, the heuristic basis vector expansions \eqref{4.1.8}, originally conceived of by Dirac, hold as a rigorous result, 
known as the Nuclear Spectral Theorem \cite{spectraltheorem}, provided $\phi^+\in\Phi_-,\ |\bs{q},\ms^+\rangle\in\Phi_-^\times$ and 
$\psi^-\in\Phi_+,\ |\bs{q},\ms^-\rangle\in\Phi_+^\times$.
 
\subsection{Gamow vectors}\label{sec4.2}
The pole terms of \eqref{4.1.10} suggests that we identify $|\bs{q},\ms_{R_i}^-\rangle$ as the vectors associated with the resonance defined by
the pole at $\ms_{R_i}$. We call $|\bs{q},\ms_{R_i}^-\rangle$ Gamow vectors. 
According to hypothesis \eqref{4.1.11}, the wavefunctions of out vectors $\psi^-$ are of Hardy class from above. Therefore, their complex conjugates ${\psi^-}^*$
are of Hardy class from below, having analytic extensions into the lower half plane. It then follows from \eqref{4.1.10} that the Gamow vectors $|\bs{q},\ms_{R_i}^-\rangle$ 
are defined as the 
functionals that furnish the 
evaluation of these analytic extensions in the lower half plane at the 
pole position $\ms=\ms_{R_i}$\footnote{Since our focus is on decay processes, we only consider poles on the lower half plane. 
An entirely similar analysis can be carried out for the poles on the upper half plane.  }:
\begin{equation}
|\bs{q},\ms_{R_i}^-\rangle:\quad \psi^-\to{\psi^-}^*(\bs{q},\ms_{R_i})=
\langle^-\psi|\bs{q},\ms_{R_i}^-\rangle\label{4.2.1}
\end{equation}
 Similarly, the scattering vectors $|\bs{q},\ms^\pm\rangle$ are the  evaluation of functionals of  $\phi^+$ and $\psi^-$ at real $\ms$: 
\begin{eqnarray}
|\bs{q},\ms^-\rangle:\quad \psi^-\to{\psi^-}(\bs{q},\ms)&=&
\langle\bs{q},\ms^-|\psi^-\rangle\nonumber\\
|\bs{q},\ms^+\rangle:\quad \phi^+\to{\phi^+}(\bs{q},\ms)&=&
\langle\bs{q},\ms^+|\phi^+\rangle
\end{eqnarray}
Both the out-scattering vectors $|\bs{q},\ms^-\rangle$ and Gamow vectors 
$|\bs{q},\ms_{R_i}^-\rangle$ exist as elements of the dual space $\Phi_+^\times$.  The in-scattering vectors 
$|\bs{q},\ms^+\rangle$ exist as elements of the dual space $\Phi_-^\times$. The proofs of these statement are essentially the same as the proof given 
in \cite{bohm-gadella, civitarese-gadella} for non-relativistic resonance. 

The Gamow vectors are also generalized eigenvectors of the operators 
$\hat{M}$ and $\hat{P}_\mu$ with complex eigenvalues:
\begin{eqnarray}
\hat{M}|\bs{q},\ms_{R_i}^-\rangle&=&\sqrt{\ms_{R_i}}|\bs{q},\ms_{R_i}^-\rangle\nonumber\\
\hat{P}_\mu|\bs{q},\ms_{R_i}^-\rangle&=&\sqrt{\ms_{R_i}}q_\mu|\bs{q},\ms_{R_i}^-\rangle\label{4.2.2}
\end{eqnarray}
In the event the system under consideration has non-zero spin, Gamow vectors will also be eigenvectors of the second Casimir operator 
$\hat{W}$ of the Poincar\'e group and the spin projection $\hat{S}_3$. 

Recall that complex-valued solutions to the eigenvalue problem do not exist 
in the Hilbert space by the self-adjointness of observables. On the other hand, as is the case here, 
 complex solutions to the eigenvalue problem are possible in the dual space of a 
suitably constructed rigged Hilbert space. It is important to emphasize that all the observables are still self-adjoint 
as operators defined in the Hilbert space of the triplets \eqref{4.1.14}, albeit their extensions to the dual spaces $\Phi_\pm^\times$
 may have complex eigenvalues.  
 
 It is also important to emphasize that, although the Green's function and therewith the $S$-matrix may have pole singularities, the wavefunctions 
 $\langle^-\psi|\bs{q},\ms^-\rangle$ and $\langle^+\bs{q},\ms|\phi^+\rangle$ themselves \emph{do not} have any singularities  on the lower-half plane; they 
 are Hardy class functions and, as such, analytic everywhere in the open lower-half plane. 
 Similarly, functions $\langle^+\phi|\bs{q},\ms^+\rangle$ and $\langle^-\bs{q},\ms|\psi^-\rangle$ are analytic 
 everywhere in the open upper-half plane. In particular, the Gamow vectors should not be understood as singularities of the 
 wave function $\langle^-\psi|\bs{q},\ms^-\rangle$ at $\ms=\ms_{R_i}$. Rather, they are the evaluation functionals 
 of (analytic) wave functions  at the pole positions of the $S$-matrix. The analytic extensions of wavefunctions developed 
 here is quite a different matter from the possible analytic extensions (beyond the infinitesimals $\pm i\epsilon$ employed for the evaluation of singular integrals) 
 of the operators $\hc^\dagger_\pm(\bs{q},\ms)$ and $\hc_\pm(\bs{q},\ms)$, a  problem  that we have not dealt with in this paper. 

As seen from section \ref{4.3} below,  the linear span of Gamow vectors over  $\bs{q}$ 
 (and the spin projection, for a resonance with spin)  defines the vector space that provides a state vector description for the resonance.

\subsection{Rigged Fock spaces of Hardy-type}\label{sec4.3}
 The repeated application of the creation operators $\hc_\pm^\dagger(\bs{q},\ms)$ 
on the vacuum gives state vectors such as $|\bs{q}_1,\ms_1;\bs{q}_2,\ms_2;\cdots;\bs{q}_n,\ms_n^\pm\rangle$, the analogue of an $n$-particle state in our problem.  These 
vectors can be defined as functionals on an $n$-fold $\pi$-tensor product of the topological vector spaces $\Phi_\pm$ of \eqref{4.1.14}. Specifically, let 
\begin{equation}
\left(\Phi_\pm\right)_n=\Pi_{i=1}^n\Phi^{(i)}_\pm,\quad \Phi^{(i)}_\pm=\Phi_\pm;\quad\left(\Phi_\pm\right)_0\equiv\mathbb{C}\label{4.3.1}
\end{equation}
and define
\begin{equation}
Y_\pm=\sum_{n=0}^\infty\oplus \left(\Phi_\pm\right)_n\label{4.3.2}
\end{equation}
The vector spaces $\left(\Phi_\pm\right)_n$ and $Y_\pm$ have the following properties: 
\begin{enumerate}
\item [{(4.4.1)}] It is proved in \cite{sw2} that the spaces $\Phi_\pm$, homeomorphic to the function spaces ${\cal K}_\pm$ defined in \eqref{4.1.11}, 
are nuclear Fr\'echet spaces. Being $\pi$-tensor products of nuclear spaces, the $\left(\Phi_\pm\right)_n$ are also nuclear Fr\'echet spaces for each $n$. 
\item [{(4.4.2)}] Since $\Phi_\pm$ are Fr\'echet spaces, the strong dual $\left(\Phi_\pm\right)^\times_n=\Pi_{i=0}^n \left(\Phi_\pm^{(i)}\right)^\times$ and 
$|\bs{q}_1,\ms_1;\bs{q}_2,\ms_2;\cdots;\bs{q}_n,\ms_n^\pm\rangle\in\left(\Phi_\pm\right)_n^\times$. 
\item [{(4.4.3)}] For each $n$, the spaces $\left(\Phi_\pm\right)_n$ are algebraically isomorphic and topologically homeomorphic to a proper subspace of 
$\left(\Phi_\pm\right)_{n+1}$. In particular, the original $\pi$-product topology of $\left(\Phi_\pm\right)_n$ coincides 
with that which it inherits as a subspace of $\left(\Phi_\pm\right)_{n+1}$. Therefore, we have a continuous embedding 
$\left(\Phi_\pm\right)_n\to\left(\Phi_\pm\right)_n\subset\left(\Phi_\pm\right)_{n+1}$ for each $n$.  
\item [{(4.4.4)}] The preceding property together with the definition \eqref{4.3.2}  implies that the spaces $Y_\pm$ are the strict countable inductive limits of $\left(\Phi_\pm\right)_n$:
\begin{equation}
Y_\pm=\bigcup_{n}\left(\Phi_\pm\right)_n\label{4.3.3}
\end{equation}
Since $\left(\Phi_\pm\right)_n$ are nuclear Fr\'echet spaces for each $n$, $Y_\pm$ are nuclear LF spaces. 
\item [{(4.4.5)}] Let $Y_\pm^\times$ be the topological (anti)dual of $Y_\pm$, i.e., the space of continuous antilinear functionals on $Y_\pm$, endowed with its usual 
weak$^*$-topology. It follows that for each $n$, the dual space $\left(\Phi_\pm\right)^\times_n$ is a proper subspace of $Y_\pm^\times$. 
\end{enumerate}

Now, let $\HS_n$ be the $n$-fold $\pi$-tensor product of the Hilbert space \eqref{4.1.7} and let ${\cal H}$ be the strict countable inductive limit of $\HS_n$. 
Then, we have the triplets of spaces, 
\begin{equation}
Y_\pm\subset\HS\subset Y_\pm^\times,\label{4.3.4}
\end{equation}
which we call \emph{rigged Fock spaces of Hardy-type}. 

The Gamow vectors $|\bs{q},\ms_{R_i}^-\rangle$ introduced above are elements of $\Phi_+^\times$ and therefore also elements of $Y_+^\times$. Its null space 
consists of $Y_+-\Phi_+$. 

\subsection{Representations of the causal Poincar\'e semigroup}\label{sec4.4}
It remains to investigate the properties of $|\bs{q},\ms^\pm\rangle$ and $|\bs{q},\ms_{R_i}^-\rangle$ under Poincar\'e transformations.  This problem has been studied in 
\cite{sw1,sw2} within the context of the integrability of a Poincar\'e algebra obtained by means of the Bakamjian-Thomas construction. Just as in the model developed here, 
in \cite{sw1,sw2} the construction of the interaction-incorporting Poincar\'e algebra is done in the point-form. However, unlike here, the interaction-incorporating momentum 
operators are all obtained in \cite{sw1,sw2} from a single interaction-incorporating mass operator and 
interaction-free velocity operators: $\hat{P}^\mu=\hat{P}_0^\mu+\Delta\hat{P}^\mu=\hat{M}\hat{Q}_0^\mu,\ 
\hat{M}=\hat{M}_0+\Delta\hat{M}$. The choice of the CSCO $\left\{\bs{\hat{Q}},\ \hat{M}\right\}$ (for the spinless case) then leads to 
generalized eigenvectors $|\bs{q},\ms^\pm\rangle$, defined as elements of the dual spaces of the same pair of rigged Hilbert spaces as in this present case, given 
by \eqref{4.1.14} or its $L^2$-realization \eqref{4.1.13}. 

While, as demonstrated in this paper, the point-form dynamics is possible in a field theoretical construction, it is not possible, 
at least within the Lagrangian formalism,  to factorize the interacting momentum operators \eqref{2.2.22} into a mass operator that completely encompasses 
interactions and velocity operators that are interaction free. Notwithstanding this, observe that all of the dynamics of the model developed here 
are encapsulated in functions of the square mass variable alone--for 
instance, recall that the Green's function is a function of $\ms$  and does not contain momentum variables.  From this point of view, the construction presented 
here bears a strong parallel to that given in \cite{sw1,sw2}. The crux of the matter is the point-form dynamics, the present paper and \cite{sw1,sw2} providing two
different realizations thereof. 
 
Since the rigged Hilbert spaces \eqref{4.1.13} and \eqref{4.1.14} are identical to those utilized in \cite{sw1,sw2}, 
the construction of the representations of Poincar\'e transformations in these spaces can be done exactly as in \cite{sw1,sw2}.  Therefore, here we will briefly 
outline the procedure and some of the main results, referring the reader to \cite{sw1,sw2} for details. 

Now that we have constructed the Hilbert space \eqref{4.1.7} in which interacting momentum four vector \eqref{2.2.22} and Lorentz tensor \eqref{2.2.23}  are defined 
as self-adjoint operators furnishing a basis for a representation of the Poincar\'e algebra, the problem at hand is to integrate this operator Lie algebra to obtain a unitary 
representation of the Poincar\'e group in \eqref{4.1.7}. In view of the spectrum of the mass operator $\hat{M}$, note that such a representation will be clearly reducible. 
Once the unitary representation in \eqref{4.1.7} has been obtained, the representation in the Fock space can be constructed by taking tensor products of the representation 
in \eqref{4.1.7}. 

Recall that we constructed the creation and annihilation operators $\hc_\pm^\dagger(\bs{q},\ms)$ and $\hc_\pm(\bs{q},\ms)$ by demanding that they 
be solutions to the eigenvalue problems \eqref{3.1.1} and \eqref{3.1.2}. From this, it immediately follows that 
\begin{equation}
\hat{P}^\mu|\bs{q},\ms^\pm\rangle=\sqrt{\ms}q^\mu|\bs{q},\ms^\pm\rangle.\label{4.4.1}
\end{equation}
In view of \eqref{2.2.23}, the operators $\hc_\pm^\dagger(\bs{q},\ms)$ and $\hc_\pm(\bs{q},\ms)$ 
transform just the same way as the free creation and annihilation operators \eqref{2.2.6} 
under the direct product representation $\hat{U}(\Lambda)=\hat{U}_0(\Lambda)=\hat{U}_1(\Lambda)\otimes I_2+I_1\otimes\hat{U}_2(\Lambda)$ 
of the Lorentz group,
where $\hat{U}_1(\Lambda)$ and $\hat{U}_2(\Lambda)$ are given by \eqref{2.2.1} and \eqref{2.2.2}. It then follows
\begin{equation}
\hat{U}(\Lambda)|\bs{q},\ms^\pm\rangle=|\bs{\Lambda q},\ms^\pm\rangle\label{4.4.2}
\end{equation}
In particular, it follows from \eqref{4.4.2} that 
the action of $\hat{U}(\Lambda)$ in the $L^2$-realization \eqref{4.1.12} has the form
\begin{equation}
\left(\hat{U}(\Lambda)\varphi\right)(\bs{q},\ms)=\varphi\left(\bs{{\Lambda^{-1}q}},\ms\right),\qquad\varphi\in L^2\left(\mathbb{R}_{\ms_0},\mathbb{R}^3\right)\label{4.4.3}
\end{equation}
This means that the integrability of the Lorentz algebra spanned by operators \eqref{2.2.23} is automatic in the Hilbert space generated 
from the new vacuum state \eqref{3.2.1} and creation operators $\hc^\dagger_+(\bs{q},\ms)$ or $\hc^\dagger_-(\bs{q},\ms)$. 
This is a particular strength of formulating dynamics in the point-form. In the more conventionally used instant-form, the integrability of the Lorentz 
algebra is a much more difficult, if not an intractable, problem.  

The eigenvalue equations \eqref{4.4.1} show that the Abelian subalgebra generated by the interaction-incorporating momentum 
operators also integrate in the Hilbert space \eqref{4.1.7} to furnish a unitary representation of spacetime translations: 
\begin{equation}
\left(\hat{U}(a)\varphi\right)(\bs{q},\ms)=e^{-i\sqrt{\ms}q\cdot a}\varphi(\bs{q},\ms),\qquad\varphi\in 
L^2\left(\mathbb{R}_{\ms_0},\mathbb{R}^3\right),\quad a\in\mathbb{R}^4\label{4.4.4}
\end{equation}
From the group composition $U(\Lambda,a)=U(a)U(\Lambda)$ and expressions \eqref{4.4.3} and \eqref{4.4.4}, we then obtain
\begin{equation}
\left(\hat{U}(\Lambda,a)\varphi\right)(\bs{q},\ms)=e^{-i\sqrt{\ms}q\cdot\Lambda a}\varphi(\bs{\Lambda^{-1}q},\ms)\label{4.4.5}
\end{equation}\\

Let us now take for $\varphi$ a function a function $\phi^+\in{\cal K}_-\subset L^2\left(\mathbb{R}_{\ms_0},\mathbb{R}^3\right)$. 
Note that in the tensor product decomposition $L^2\left(\mathbb{R}_{\ms_0},\mathbb{R}^3\right)=
L^2\left(\mathbb{R}_{\ms_0}\right)\otimes L^2\left(\mathbb{R}^3\right)$, the unitary operators $\hat{U}(\Lambda)$ are local in the 
subspace $L^2\left(\mathbb{R}^3\right)$. In other words, the $\ms$-dependence of functions $\varphi$ is not affected by 
Lorentz transformations. This is in fact the main reason for our using the velocity operators, rather than the more common choice of momentum 
operators, for our CSCO. Furthermore, from \eqref{4.4.2} and 
the continuity of $\Lambda$ on $\mathbb{R}^4/[1,\infty)$ it follows that $\hat{U}(\Lambda)$ is reduced by the Schwartz space ${\cal S}(\mathbb{R}^3)$. 
Hence, $\hat{U}(\Lambda)\phi^+\in{\cal K}_-$ for every $\phi^+\in{\cal K}_-$.

On the other hand, the operators $\hat{U}(a)$ do not separate with respect to the tensor product decomposition $L^2\left(\mathbb{R}_{\ms_0},\mathbb{R}^3\right)=
L^2\left(\mathbb{R}_{\ms_0}\right)\otimes L^2\left(\mathbb{R}^3\right)$ and they may affect the properties of  $\phi^+$ as a function of $\ms$. 
Specifically, recall that $\phi^+\in{\cal K}_-$ has an analytic extension into the lower-half 
$\ms$-plane and that this analytic function decreases faster than the inverse of any polynomial for $\ms\to\infty$. From \eqref{4.4.4} and 
for a suitable branch that ensures analyticity of $\sqrt{\ms}$ in the open lower half plane, say 
\begin{equation}
-\pi< {\rm Arg}\ \ms\leq\pi,\label{4.4.6}
\end{equation}
we see that $\hat{U}(a)$ preserves the analyticity and regularity properties of $\phi^+\in{\cal K}_-$ 
for any four-vector $a$ if $a\cdot q\geq0$ for all $q=\left(\sqrt{1+\bs{q}^2},\bs{q}\right),\ 
\bs{q}\in\mathbb{R}^3$. Hence, $\cal{K}_-$ reduces $\hat{U}(a)$ for any $a$ with $a^0\geq0$ and $a^2\geq0$. This property is what is at the heart of the 
 representations of the causal Poincar\'e semigroup. 
 
To make these observations more precise, let ${\cal P}_+$ be the semigroup defined by \eqref{0.1}. 
As noted before, ${\cal P}_+$ is closed under proper orthochronous Lorentz transformations $\Lambda$, but  not under the inverse 
operation $(\Lambda, a)\to (\Lambda, a)^{-1}$. Hence, ${\cal P}_+$ is a subsemigeoup 
of the Poincar\'e group ${\cal P}$. It is the semidirect product of the semigroup of spacetime translations into the closed 
forward lightcone and the group of proper orthochronous Lorentz transformations. 

Next, define a set of operators $\hat{\cal T}_-(\Lambda,a)$ on ${\cal K}_-$ by
\begin{eqnarray}
\left(\hat{{\cal T}}_-(\Lambda,a)\phi^+\right)(\bs{q},\ms)&:=&\left(\hat{U}(\Lambda,a)\phi^+\right)(\bs{q},\ms),\qquad (\Lambda,a)\in\calP_+,\ \phi^+\in{\cal K}_-\nonumber\\
&=&e^{-i\sqrt{\ms}q\cdot \Lambda a}\phi^+(\bs{\Lambda^{-1}\bs{q}},\ms)\label{4.4.8}
\end{eqnarray}
where the second equality follows from \eqref{4.4.5}. 
Recall that ${\cal K}_-$ is endowed with a nuclear Fr\'echet topology. The semigroup $\calP_+$ inherits the locally 
Euclidean topology of $\calP$, i.e., it is a Lie subsemigroup of $\calP$.
With respect to these topological and algebraic structures, the following properties have been established in \cite{sw2}: 
\begin{enumerate}
\item [{\rm (4.5.a)}] 
For any $(\Lambda,a)\not\in\calP_+$, there exists at least one function $\phi^+\in{\cal K}_-$ 
such that 
$\hat{U}(\Lambda,a)\phi^+\not\in{\mathcal K}_-$. Hence, $\hat{{\cal T}}_-(\Lambda,a)$ are defined only for 
$(\Lambda,a)\in\calP_+$.  Obviously, from \eqref{4.4.8}  
$\hat{{\cal T}}_-(\Lambda_2,a_2)\hat{{\cal T}}_-(\Lambda_1,a_1)=\hat{{\cal T}}_-(\Lambda_2\Lambda_1,a_2+\Lambda_2a_1)$. 
\item [{\rm (4.5.b)}]  The operators $\hat{{\cal T}}_-(\Lambda,a)$ are continuous for all $(\Lambda,a)\in\calP_+$. 
\item [{\rm (4.5.c)}] The mapping ${\cal K}_-\times\calP_+\to {\cal K}_-$ is differentiable. That is, the map 
$(\phi^+,(\Lambda,a))\to\hat{{\cal T}}_-(\Lambda,a)\phi^+$ 
is differentiable for every $\phi^+\in{\cal K}_-$. 
\end{enumerate}
 
Therefore, \emph{the function space ${\mathcal K}_-$ furnishes a differentiable representation 
$\hat{{\mathcal T}}_-:\ \calP_+\times{\mathcal K}_-\to{\mathcal K}_-$ 
of the semigroup 
$\calP_+$. This representation of ${\cal P}_+$ does {\rm{not}} extend to a representation of the whole Poincar\'e group $\calP$}. 
Furthermore, from \eqref{4.4.8} we see that this differentiable representation 
of the semigroup $\calP_+$ in ${\cal K}_-$ coincides with the restriction of the unitary representation 
$\hat{U}(\Lambda,a)$ of $\calP$ to ${\cal K}_-$. 

Similarly, a differentiable representation $\hat{{\mathcal T}}_+:\ \calP_+\times{\mathcal K}_+\to{\mathcal K}_+$ exists on the vector space ${\mathcal K}_+$. 
By appealing to the duality  between $\phi^+$ and $\psi^-$ implied by \eqref{4.1.9} or \eqref{4.1.10}, we  define  operators $\hat{{\cal T}}_+(\Lambda,a)$ by
\begin{equation}
\hat{\mathcal T}_+(\Lambda,a)\psi^-:=\Bigl(\hat{U}\left(\Lambda^{-1},-\Lambda^{-1}a\right)\Bigr)^\dagger\psi^-,\quad (\Lambda,a)\in\calP_+,\  \psi^-\in{\cal K}_+
\label{4.4.9}
\end{equation}
On an arbitrary function $\psi^-\in{\mathcal K}_+$, these operators  
have the action
\begin{equation}
\Bigl(\hat{\mathcal{T}}_+(\Lambda,a)\psi^-\Bigr)\left(\bs{q},\ms\right)=e^{i\sqrt{\ms}q\cdot a}
\psi^-\left(\bs{\Lambda^{-1}q},\ms\right)\label{4.4.10}
\end{equation}

Since the abstract rigged Hilbert spaces \eqref{4.1.14} are homeomorphic to the $L^2$-realizations \eqref{4.1.13}, there exist differentiable representations of 
$\calP_+$ in  $\Phi_-$ and $\Phi_+$ that are unitarily equivalent to \eqref{4.4.8} and \eqref{4.4.9}. For notational simplicity, we will denote these abstract representations also by $\hat{\mathcal T}_\pm$. 

By duality, the representations $\hat{\cal T}_\pm$ in $\Phi_\pm$ induce 
representations ${\hat{\mathcal T}}_\pm^\times$  in the dual spaces 
$\Phi_\pm^\times$.  For any $(\Lambda,a)\in\calP_+$
\begin{subequations}
\label{4.4.11}
\begin{equation} 
\tag{\ref{4.4.11}+}
\langle\hat{\mathcal T}_+(\Lambda,a)\psi^-|F^-\rangle=\langle\psi^-|\hat{\mathcal T}_+^\times\left(\Lambda^{-1},-\Lambda^{-1}a\right) F^-\rangle,\  \psi^-\in\Phi_+,\ F^-\in\Phi_+^\times\label{4.4.11+}
\end{equation}
\begin{equation}
\tag{\ref{4.4.11}$-$}
\langle\hat{\mathcal T}_-(\Lambda,a)\phi^+|F^+\rangle=\langle\phi^+|\hat{\mathcal T}_-^\times\left(\Lambda^{1},-\Lambda^{-1}a\right)F^+\rangle,\  \phi^+\in\Phi_-,\ F^+\in\Phi_-^\times~\label{4.4.11-}
\end{equation}
\end{subequations}
The action of operators $\hat{\mathcal{T}}_\pm^\times$ on the generalized eigenvectors $|\bs{q},\ms^\mp\rangle$ follows from the definitions \eqref{4.4.11} 
and the explicit expression \eqref{4.4.5}:
\begin{equation}
\hat{{\mathcal T}}_\mp^\times(\Lambda,a)|\bs{q},\ms^\pm\rangle=e^{\pm i\sqrt{\ms}q\cdot a} |\bs{q},\ms^\pm\rangle,\ (\Lambda,a)\in\calP_+\label{4.4.12}
\end{equation}

Equations \eqref{4.4.12} show that for each value of $\ms\in[4m^2,\infty)$, there exists an irreducible representation of the semigroup $\calP_+$ in each of the two rigged 
Hilbert spaces \eqref{4.1.14}. These two representations are different but connected by the duality relation \eqref{4.4.9}.  In view of \eqref{4.4.8} and \eqref{4.4.10}, we note 
that nuclear Fr\'echet spaces $\Phi_\pm$ have a direct integral decomposition
\begin{equation}
\Phi_\pm=\int_{4m^2}^\infty d\ms\,\Phi_\pm(\ms)\label{4.4.13}
\end{equation}
such that each $\Phi_\pm(\ms)$ carries an irreducible representation of the causal Poincar\'e semigroup $\calP_+$.  The representations of $\calP_+$ in the rigged Fock spaces \eqref{4.3.4} can be obtained 
by taking the direct sums of tensor products of the representations \eqref{4.4.8} and \eqref{4.4.9}. 

\subsection{Characterization of resonances by irreducible representations of $\calP_+$: Resonance mass, width and lifetime}\label{sec4.5}
That the in-states $\phi^+$  and out-states $\psi^-$ are, respectively, of Hardy class from below and above in the square mass variable $\ms$ is the key requirement 
that led to the  differentiable representations of the $\calP_+$ furnished by \eqref{4.4.8} and \eqref{4.4.10}. In particular, the invariance of ${\cal K}_+$ under 
$\hat{\cal T}_+(\Lambda,a)$ implies that the function $\left(\hat{\mathcal{T}}_+(\Lambda,a)\psi^-\right)(\bs{q},\ms)$ has a unique analytic extension into the 
open upper half complex $\ms$-plane for every $\psi^-\in{\cal K}_-$ and $(\Lambda,a)\in\calP_+$. 
Likewise, its complex conjugate has a unique analytic extension into the lower half plane. Evaluating this analytic extension at a resonance pole position $\ms_{R_i}$ in
 the lower half plane, we obtain
  \begin{equation}
\left(\hat{\mathcal T}_+(\Lambda,a)\psi^-\right)^*\left(\bs{q},\ms_{R_i}\right)=e^{i\sqrt{\ms_{R_i}}\,q\cdot a}
{\psi^-}^*\left(\bs{\Lambda^{-1}q},\ms_{R_i}\right)\label{4.5.1}
\end{equation}
The Gamow vectors \eqref{4.2.1} are 
the evaluation functionals of the analytic extensions of the complex conjugates of out-state wave functions at the resonance pole positions $\ms_{R_i}$. 
With this definition of Gamow vectors and the definition \eqref{4.4.11+} of the dual representation $\hat{\cal T}_+$, it follows that Gamow vectors transform 
under $\calP_+$ as 
\begin{equation}
\hat{\mathcal T}_+^\times(\Lambda,a)|\bs{q},\ms_{R_i}^-\rangle=e^{-i\sqrt{\ms_{R_i}}\,q\cdot a}
  |\bs{\Lambda q},\ms_{R_i}^-\rangle\label{4.5.2}
\end{equation}
That the Gamow vectors are generalized eigenvectors of the momentum operators $\hat{P}^\mu$ and mass operator $\hat{M}$, as claimed 
in \eqref{4.2.2}, follows from the transformation formula \eqref{4.5.2}. It should be noted that the differentiations with respect to the 
translation semigroup parameters $a^\mu$ needed to derive \eqref{4.2.2} from \eqref{4.5.2} must be done with respect to the weak$^*$-topology 
of the dual space $\Phi_+^\times$. 

The conclusion to be drawn from \eqref{4.5.1} and \eqref{4.5.2} is that \emph{there exists an irreducible representation the causal Poincar\'e semigroup characterized by the complex square mass $\ms_{R_i}$}.
In order to obtain the rigged Hilbert space for this irreducible representation of ${\cal P}_+$, 
note that \eqref{4.5.1} is defined for the set of functions $\psi^-(\bs{q},\ms_{R_i})$, where $\ms_{R_i}$ is a \emph{fixed} pole position. By the construction of the rigged Hilbert spaces \eqref{4.1.13} and the definitions \eqref{4.1.11} 
of the spaces ${\cal K}_\pm$, recall that as functions of the velocity variable $\bs{q}$, the 
$\psi^-(\bs{q},\ms)$ (as well as $\phi^+(\bs{q},\ms)$) are Schwartz functions. Hence, it follows that the vector space of functions for which \eqref{4.5.1} holds is isomorphic to the Schwartz space: $\{\psi^-(\bs{q},\ms_{R_i})\}={\cal S}(\mathbb{R}^3)$. To indicate that the 
functions  $\psi^-(\bs{q},\ms_{R_i})$ are obtained by analytic extension of the (complex conjugates of the) functions of ${\cal K}_+$ to the resonance pole position $\ms=\ms_{R_i}$, let us introduce the notation 
\begin{equation}
{\mathcal K}_{\ms_{R_i}}:=\{\psi^-(\bs{q},\ms_{R_i})\}={\cal S}(\mathbb{R}^3)
\end{equation}
It is clear that the completion of ${\mathcal K}_{\ms_{R_i}}$ with respect to the norm topology gives the Hilbert space $L^2(\mathbb{R})$. Therewith 
we have the rigged Hilbert space 
\begin{equation}
{\mathcal K}_{\ms_{R_i}}\subset L^2\left({\mathbb{R}}^3\right)
\subset \left({\mathcal K}_{\ms_{R_i}}\right)^\times\label{4.5.3}
\end{equation}
That this triplet  is really a rigged Hilbert space immediately follows from the fact that ${\mathcal K}_{\ms_{R_i}}$ is isomorphic to the Schwartz space ${\mathcal S}(\mathbb{R}^3)$. In particular, 
the dual space $\left({\mathcal K}_{\ms_{R_i}}\right)^\times$ is the space of tempered distributions. This is the rigged Hilbert space that furnishes the irreducible representation of ${\cal P}_+$ that describes the 
resonance at $\ms=\ms_{R_i}$. If the resonance has a non-zero spin value $j$, then the product of \eqref{4.5.3} with the $2j+1$ dimensional vector space $\mathbb{C}^{(2j+1)}$ furnishes 
the relevant irreducible representation of ${\cal P}_+$. 

A comparison of \eqref{4.5.2} and \eqref{2.1.1} shows us that, in much the same way 
a stable particle is characterized by a unitary irreducible representation of the Poincar\'e group, a resonance is characterized by an 
\emph{irreducible} representation of the causal Poincar\'e semigroup. In both cases, the irreducible representation is defined by 
eigenvalues of the two Casimir operators $\hat{M}^2=\hat{P}^\mu\hat{P}_\mu$ and $\hat{W}=\frac{1}{\hat{M}^2}\hat{w}_\mu\hat{w}^\mu$, where 
$\hat{w}^\mu$ is the Pauli-Lubanski vector. For stable particles, $\hat{M}^2$ has a real eigenvalue whereas 
for resonances $\hat{M}^2$ has a complex eigenvalue.  
In the construction, this complex eigenvalue naturally arises as  the resonance pole position of the $S$-matrix. 
In this sense, \eqref{4.5.2} and \eqref{4.5.3} establish 
a synthesis between the description of resonances by $S$-matrix poles and the description of particles by 
its state vector space which furnishes an irreducible representation of Poincar\'e transformations.

The time evolution of the resonance state can now be obtained by evaluating \eqref{4.5.1} for $(\Lambda,a)=(I,t)$: 
\begin{eqnarray}
\left(\hat{\mathcal T}_+(I,t)\psi^-\right)^*\left(\bs{q},\ms_{R_i}\right)&=&e^{-i\sqrt{\ms_{R_i}}q^0t}
{\psi^-}^*\left(\bs{q},\ms_{R_i},j\right)\nonumber\\
&=&e^{-\frac{\Gamma{R_i}}{2}q_0t}e^{-im_{R_i}q_0t}{\psi^-}^*\left(\bs{q},\ms_{R_i}\right)\label{4.5.4}
\end{eqnarray}
where 
\begin{equation}
m_{R_i}=\Re{\sqrt{\ms_R}}\quad\text{and}\quad\Gamma_{R_i}=-2\Im{\sqrt{\ms_R}}\label{4.5.5}
\end{equation}
The equation \eqref{4.5.4} shows that Gamow vectors have the exponential decay behavior. Furthermore, the decay rate 
in the rest frame ($q_0=1$) is related to the imaginary part of the resonance pole position: 
\begin{equation}
\tau=\frac{-1}{2\Im{\sqrt{\ms_R}}}\label{4.5.6}
\end{equation}
These considerations justify our choice of defining a resonance as associated with the zeros of the denominator of the Green's function \eqref{4.7}. 
Furthermore, the transformation properties of the Gamow vector states under the irreducible representation of $\calP_+$ that characterizes 
them give a unique and unambiguous criterion for extracting real valued mass and width parameters from the resonance pole position. This definition 
of width in turn establishes the lifetime-width relation as an exact identity. 

As a final remark, we recall that there are roughly two kinds of unstable particle: resonances and decaying states. Resonances are associated with the $S$-matrix poles and they 
display  a Breit-Wigner type mass distribution. Decaying states, on the other hand, are associated with  the exponential decay and are characterized by their lifetime. In this regard, 
the theoretical significance of  Gamow vectors is quite noteworthy as they encapsulate the properties of both resonances and decaying states.  The Gamow vectors are eigenvectors of the interacting mass operator with the complex eigenvalue $\sqrt{\ms_{R_i}}$, where $\ms_{R_i}$ is the $S$-matrix pole position of the resonance, and they furnish irreducible representations of the semigroup $\calP_+$, leading to an exact exponential decay for all $t\geq0$. From these considerations  we infer that the irreducible representations of the causal Poincar\'e semigroup defined by \eqref{4.5.1} and \eqref{4.5.2} provide a unifying mathematical image for both 
resonances and decaying states, suggesting the phrase \emph{quasistable states} used in the title as a collective  term for both.

\section{Concluding remarks}\label{sec5}
In this paper, we have constructed a field theoretical model that builds upon and advances the theory of relativistic quasistable states developed in \cite{sw1,sw2}. 
The key features of this theory  are the use of point-form dynamics and the introduction of topological and analytical structures, which are suggested by 
the properties of the analytic $S$-matrix, for the spaces of scattering in-states 
and out-states that lead to a pair of rigged Hilbert spaces \eqref{4.1.13}.   
In point-form dynamics, 
all four components of the momentum operator becomes interaction-dependent while the Lorentz generators remain interaction free: $\hat{P}^\mu=\hat{P}_0^\mu+\Delta\hat{P}^\mu$ and 
$\hat{J}^{\mu\nu}=\hat{J}_0^{\mu\nu}$ where the subscript $0$ indicates the free generators.
An important consequence of the synthesis of these two key features is  that the Poincar\'e algebra, while realized by everywhere defined continuous operators in each of the two rigged Hilbert spaces, does not 
integrate to furnish a representation of the Poincar\'e group in either rigged Hilbert space. Instead, a subset of the algebra integrates to furnish two different differentiable 
representations of the causal Poincar\'e semigroup \eqref{0.1} in the two rigged Hilbert spaces. These representations are not irreducible, but they have a direct integral decomposition into irreducible 
representations that involves an integration over spectrum of the interaction-incorporating square mass operator. The evaluation of the analytic extensions of the wavefunctions in the square mass variable (the Hardy class property of \eqref{4.1.13}) at the resonance pole position of the $S$-matrix gives the Gamow vectors and therewith an \emph{irreducible} representation of the causal Poincar\'e semigroup. These representations 
provide a unified description of resonances and decaying states along the same lines of Wigner's description of stable particles by unitary irreducible representations of the Poincar\'e group. 

The main theoretical advantage of point-form dynamics is that it provides the cleanest separation between interaction-incorporating and interaction-free (kinematic) generators. While in all three forms of Dirac's dynamics 
the kinematic generators close to form a subalgebra of the Poincar\'e algebra, it is only in the point-form that the interaction-incorporating generators also close to form a subalgebra. Furthermore, since 
the Lorentz generators are unaffected by interactions, the integrability of the Lorentz subalgebra is automatic and the Lorentz transformations in the interacting system are furnished by the same operators 
$\hat{U}_0(\Lambda)$ that implement Lorentz transformations in the non-interacting system. This allows additional requirements, such as the Hardy class condition used in \eqref{4.1.13},  to be 
imposed on the interacting system in a manner consistent with Lorentz covariance. 

In this paper, we have worked within the main theoretical framework developed in \cite{sw1,sw2}, but used quantum fields to construct a particular realization of the point-form generators.  
In particular, we obtained the point-form Poincar\'e generators for the interacting system by integrating the tensor densities $\hat{T}^{\mu\nu}$ and $\hat{M}^{\rho\mu\nu}$ on a forward hyperboloid 
$x_\mu x^\mu=\tau^2,\ x^0\geq0$, as opposed to the more common choice $x^0=constant$ that leads to instant-form dynamics. Whenever the interaction Lagrangian density does not contain derivative couplings, the integration of these tensor densities on a forward hyperboloid in 
spacetime always leads to point-form dynamics. By contrast, the point-form generators were obtained in \cite{sw1,sw2} by means of a Bakamjian-Thomas construction, i.e., by introducing a 
perturbation to the mass operator, $\hat{M}=\hat{M}_0+\Delta\hat{M}$, and defining the momentum operators $\hat{P}^\mu=\hat{M}\hat{Q}_0^\mu$ so that the velocity operators 
$\hat{Q}_0^\mu$, as well as the Lorentz generators, remained interaction free. Hence, all of the dynamics were encoded in the mass operator. This is clearly not the case for a Lagrangian field theory, including the 
one constructed here. However, while a decomposition of the momentum operators, such as $\hat{P}^\mu=\hat{M}\hat{Q}_0^\mu$, into an interacting mass operator and interaction-free velocity operators is not possible for the model developed here, 
all of the key constructions, such as the creation and annihilation operators, the vacuum state and the Green's function for the interacting system, reduce to calculations involving 
functions of only the spectral values of the the square mass operator. It is this remarkable feature, which in turn is grounded in the choice of point-form dynamics, that allows us to carry out the construction of the 
rigged Hilbert spaces, Gamow vectors and representations of the causal Poincar\'e semigroup exactly same way as in \cite{sw1,sw2}.

\section*{Acknowledgments} 
SW is grateful for financial support from Research Corporation. He is grateful to M.~Gadella and the Department of Theoretical Physics, Atomic Physics and Optics of the University of Valladolid where 
he was a visitor while a part of this work was done.

%---------------------------
\appendix
\section{Singular Integrals}\label{appA}
%---------------------------
In this Appendix, we calculate some of the singular integrals that appear in the text. 
\begin{enumerate}
\item {\bf Integral $\bs{I(p,\tau)}$ of Eq.~\eqref{2.1.35}}\\
Observe that this is really the integral that defines the 
Pauli-Jordan function. Normally, the integration is over the momentum variables constrained to the forward hyperboloid $p_\mu p^\mu=m^2,\ p^0>0$. For instance, 
see \cite{bogolubov}. In contrast, the integration in our case is over spacetime points on the forward hyperboloid $x_\mu x^\mu=\tau^2,\ x^0>0$. Therefore, 
the integral $I(p,\tau)$ can be evaluated quite simply by interchanging the $x$ and $p$ variables in the standard calculation. 

\begin{eqnarray}
I({p},\tau)&=&\int 2d^4x\,\delta(x^2-\tau^2)\theta(x^0)e^{-i{p}\cdot x}\nonumber\\
&=&\int\frac{d^3{\bs{x}}}{x^0}\,e^{-i({p}^0x^0-\bs{{p}}\cdot\bs{x})}\nonumber\\
&=&-2\pi\int_0^\infty\frac{r^2dr}{x^0}\,e^{-i{ p}^0x^0}\int_1^{-1}d\cos\theta\,e^{i|\bs{{p}}|r\cos\theta}\nonumber\\
&=&\frac{2\pi i}{|\bs{{p}}|} \int_{-\infty}^\infty \frac{rdr}{x^0}e^{-i({ p}^0x^0+|\bs{{p}}|r)}\nonumber\\
&=&-\frac{2\pi }{|\bs{{p}}|}\frac{\partial}{\partial|\bs{{p}}|}\int_{-\infty}^\infty \frac{dr}{x^0}\,e^{-i({p}^0x^0+|\bs{{p}}|r)}\label{a.1}
\end{eqnarray}
Since $I(p,\tau)$ is Lorentz invariant, it can only depend on $p_\mu p^\mu$ and $x_\mu x^\mu$ and, when $p$ or  $x$ is a timelike vector, 
the sign of $p^0$ or ${ x}^0$, also Lorentz invariant quantities.  For spacelike $x$ or $p$, we do not expect $I(p,\tau)$ to 
depend on the sign of $x^0$ or $p^0$. The derivative of \eqref{a.1} should be understood 
in the distribution sense. As evident from the calculation below,  if either $x$ or ${ p}$ is a lightlike vector, then $I(p,\tau)$ contains a 
delta-type singularity. 
%If both $x$ and $p$ are timelike,  integral is undefined and \eqref{a.1} should be defined as the weak limit of 
%a suitable sequence of regularized functions. 

In our case, since we are restricted to the forward hyperboloid $x^\mu x_\mu=\tau^2,\ x^0>0$, $x$ is always a  
timelike vector with a positive time component. The momentum  vector ${p}$, on the other hand, 
can be a timelike vector with $p^0>0$ and a continuous or discrete mass distribution, a spacelike vector with a continuous mass distribution, 
or the null vector which appears as the limit of spacelike vector ${\sf p}=m(q-q')$ of \eqref{2.1.34}.  
We consider these cases separately:
%Therefore, we need not worry about the lightcone singularities arising from either momentum or position variables. 
%case where ${\sf p}$ is a 
%lightlike vector will be treated separately. 
%The integral Next, introduce the change of variables 
%\begin{eqnarray}
%x^0=\tau\cosh\phi&&r=\tau\sinh\phi\nonumber\\
%{\sf p}^0=\sqrt{{\tt s}}\cosh\phi_0&&|\bs{\sf{p}}|=\sqrt{{\tt s}}\sinh\phi_0\nonumber\\
%\varphi=\phi+\phi_0&&\label{a.2}
%\end{eqnarray}
%Then, \eqref{a.1} becomes 
%\begin{eqnarray}
%I({\sf p},\tau)&=&-\frac{2\pi }{|\bs{\sf{p}}|}\frac{\partial}{\partial|\bs{\sf{p}}|}\int_{-\infty}^\infty d\phi\,e^{-i\sqrt{{\tt s}}\tau\cosh(\phi+\phi_0)}\nonumber\\
%&=&-\frac{2\pi }{|\bs{\sf{p}}|}\left(-\frac{|\bs{\sf{p}}|}{\sqrt{{\tt s}}}\frac{\partial}{\partial\sqrt{\tt s}}+\frac{{\sf p}^0}{\tt s}\frac{\partial}{\partial\phi_0}\right)\int_{-\infty}^\infty d\varphi\,e^{-i\sqrt{\tt s}\tau\cosh\varphi}\nonumber\\
%&=&\frac{2\pi\tau}{\sqrt{\tt s}}\frac{\partial}{\partial(\sqrt{\tt s}\tau)}\int_{-\infty}^\infty d\varphi\,e^{-i\sqrt{\tt s}\tau\cosh\varphi}\label{a.3}
%\end{eqnarray}
\begin{enumerate}
\item ${p}^2={\tt s}>0,\ \ {p}^0>0$\\ 
Introducing the change of variables 
\begin{eqnarray}
x^0=\tau\cosh\phi,&&r=\tau\sinh\phi,\nonumber\\
{\sf p}^0=\sqrt{{\tt s}}\cosh\phi_0,&&|\bs{{p}}|=\sqrt{{\tt s}}\sinh\phi_0,\nonumber\\
\varphi=\phi+\phi_0,&&\label{a.2}
\end{eqnarray}
\begin{equation}
\int_{-\infty}^\infty \frac{dr}{x^0}\,e^{-i({p}^0x^0+|\bs{{p}}|r)}=\int_{-\infty}^\infty d\phi\,e^{-i\sqrt{{\tt s}}\tau\cosh(\phi+\phi_0)}\nonumber
\end{equation}
This is a well-known integral representation of cylinder functions \cite{}. Therewith, 
\begin{eqnarray}
\int_{-\infty}^\infty d\phi\,e^{-i\sqrt{{\tt s}}\tau\cosh(\phi+\phi_0)}&=&-\pi\left(N_0(\sqrt{\tt s}\tau)+iJ_0(\sqrt{\tt s}\tau)\right)\nonumber\\
&=&-i\pi H_0^{(2)}(\sqrt{\tt s}\tau)
\label{a.3}
\end{eqnarray}
where $J_0$ and $N_0$ are, respectively, the Bessel and Neumann functions of zeroth order and 
$H_0^{(2)}$ is the zeroth-order Hankel function of the second kind. 
Using the change of variables \eqref{a.2} to rewrite  the derivatives of \eqref{a.1}, 
\begin{eqnarray}
\frac{\partial}{\partial\left|{\bs{p}}\right|}&=&\frac{\partial\sqrt{\tt s}}{\partial\left|{\bs{p}}\right|}\frac{\partial}{\partial\sqrt{\tt s}}+\frac{\partial\phi_0}{\partial\left|{\bs{p}}\right|}\frac{\partial}{\partial\phi_0}\nonumber\\
&=&-\frac{\left|{\bs{p}}\right|}{\sqrt{\tt s}}\frac{\partial}{\partial\sqrt{\tt s}}+\frac{{p}^0}{{\tt s}}\frac{\partial}{\partial\phi_0}\label{a.4}
\end{eqnarray}
we then obtain
\begin{equation}
I({p},\tau)=-i\frac{2\pi^2\tau}{\sqrt{\tt s}}\frac{\partial}{\partial\left(\sqrt{\tt s}\tau\right)}\Bigl(J_0(\sqrt{\tt s}\tau)-iN_0(\sqrt{\tt s}\tau)\Bigr)\
\label{a.5}
\end{equation}
The derivatives here  can be computed by using the well-known identity 
\begin{equation}
Y_n(x)=(-1)^nx^n\left(\frac{1}{x}\frac{d}{dx}\right)^nY_0(x)\label{a.6}
\end{equation}
where $Y_n$ can be a Bessel, Neumann or Hankel function of order $n$. In particular, \eqref{a.9} gives $Y_1(x)=-\frac{d}{dx}Y_0(x)$. 
Therewith, 
\begin{equation}
I({p},\tau)=i\frac{2\pi^2\tau }{\sqrt{\tt s}}\Bigl(J_1(\sqrt{\tt s}\tau)-iN_1(\sqrt{\tt s}\tau)\Bigr)=i\frac{2\pi^2\tau }{\sqrt{\tt s}}H_1^{(2)}(\sqrt{\tt s}\tau)
\label{a.7}
\end{equation}

\item ${p}^2={\tt s}\leq0$\\
Introducing the change of variables 
\begin{eqnarray}
x^0=\tau\cosh\phi,&&r=\tau\sinh\phi,\nonumber\\
{p}^0=\sqrt{-{\tt s}}\sinh\phi_0,&&|\bs{{p}}|=\sqrt{-{\tt s}}\cosh\phi_0,\nonumber\\
\varphi=\phi-\phi_0,&&\label{a.8}
\end{eqnarray}
\begin{equation}
\int_{-\infty}^\infty \frac{dr}{x^0}\,e^{-i({ p}^0x^0+|\bs{{p}}|r)}=\int_{-\infty}^\infty d\phi\,e^{-i\sqrt{-{\tt s}}\tau\sinh(\phi+\phi_0)}\nonumber
\end{equation}
Again, this integral furnishes a representation of the Hankel function \cite{watson}: 
\begin{equation}
\int_{-\infty}^\infty d\phi\,e^{-i\sqrt{-{\tt s}}\tau\sinh(\phi+\phi_0)}=i\pi H^{(1)}_0(i\sqrt{-{\tt s}}\tau)=-i\pi H_0^{(2)}(-i\sqrt{-{\tt s}}\tau)\label{a.9}
\end{equation}
where $H^{(1)}_0$ and $H^{(2)}_0$ are the zeroth-order Hankel function of the first and second kind, respectively. 

As in the previous case, we use the change of variables \eqref{a.8} to express the derivative of \eqref{a.1}. Therewith, 
\begin{eqnarray}
I({p},\tau)&=&i2\pi^2 \Bigl(\frac{1}{\sqrt{-{\tt s}}}\frac{\partial}{\partial{\sqrt{-{\tt s}}}}+\frac{\tanh\phi_0}{{\tt s}}\frac{\partial}{\partial\phi_0}\Bigr)H^{(1)}_0(i\sqrt{-\tt s}\tau)\nonumber\\
&=&-\frac{2\pi^2\tau}{\sqrt{-{\tt s}}}\frac{\partial}{\partial\left(i{\sqrt{-{\tt s}}}\tau\right)}H^{(1)}_0(i\sqrt{-{\tt s}}\tau)\label{a.10}
\end{eqnarray}
This derivative can be simply determined by means of  the identity \eqref{a.6} for cylinder functions. However, if ${\tt s}=0$ is in the spectrum of $p_\mu p^\mu$, then 
due to the discontinuity of $H_0^(1)$ at the origin, there arises a delta-function singularity. Hence, 
\begin{eqnarray}
I({p},\tau)=-i\frac{2\pi^2\tau}{\sqrt{\tt s}}\epsilon(p^0)\delta(\sqrt{\tt s}\tau)&+&\frac{2\pi^2\tau }{\sqrt{-{\tt s}}}H^{(1)}_1(i\sqrt{-{\tt s}}\,\tau),\nonumber\\
&& \qquad p_\mu p^\mu={\tt s}\leq0,\ p\not=0\nonumber\\
\label{a.11}
\end{eqnarray}
where $\epsilon(p^0)=\theta(p^0)-\theta(-p^0)$ is the sign of $p^0$. (See \cite{bogolubov}.)  If, on the other hand,  the ${\tt s}=0$ point appears in the spectrum of $p_\mu p^\mu$ 
as a result of the vanishing $p^\mu$, as is the case with ${\sf p}=m(q-q')$ defined by \eqref{2.1.34}, then $\epsilon(p^0)$ is undefined. However, if we set $\theta(0)=\frac{1}{2}$, then we obtain 
\begin{equation}
I({p},\tau)=\frac{2\pi^2\tau }{\sqrt{-{\tt s}}}H^{(1)}_1(i\sqrt{-{\tt s}}\,\tau),\ p_\mu p^\mu={\tt s}\leq0,\ {\tt s}=0\Rightarrow p=0\label{a.12}
\end{equation}

\end{enumerate}
\item {\bf Distribution ${\cal I}^\mu({p},\tau)$ and the evaluation of \eqref{2.1.38}}\\
As defined in \eqref{2.1.37}, 
\begin{equation}
{\cal I}^\mu({ p},\tau)\equiv i\frac{\partial}{\partial {{p}}_\mu}I({p},\tau)=\int 2d^4x\,\delta(x^2-\tau^2)\theta(x^0)x^\mu e^{-ix\cdot{p}}\label{a.2.1}
\end{equation}
From \eqref{a.7} and \eqref{a.11}, we note that $I({p},\tau)$ depends only on the Lorentz invariant parameters $\sqrt{\tt s}$ and $\tau$. 
Therefore, using ${ p}_\mu{p}^\mu={\tt s}$ and the identity \eqref{a.6}, we obtain
\begin{eqnarray}
{\cal I}^\mu({p},\tau)&=& i2{{p}}^\mu\frac{\partial}{\partial {\tt s}}I({p},\tau)=i\frac{p^\mu\tau}{\sqrt{\tt s}}\frac{\partial}{\partial{\sqrt{\tt s}}\tau}I(p,\tau)\nonumber\\
&=&\frac{2\pi^2\tau^2}{{\tt s}}{p}^\mu H_2^{(2)}(\sqrt{\tt s}\tau)\quad \text{for}\quad {\tt s}>0\label{a.2.2}\\
&=&\frac{2\pi^2\tau^2}{{\tt s}}{p}^\mu H_2^{(1)}(i\sqrt{-{\tt s}}\tau)\quad\text{for}\quad {\tt s}<0\label{a.2.3}
\end{eqnarray}

With these results, we can compute the distributions that appear in \eqref{2.1.38}. First observe 
that $P=m(q+q')$ defined by \eqref{2.1.34} is a timelike vector with $P^0>0$, while ${\sf p}=m(q-q')$ is a 
spacelike vector. Moreover, $P$ and ${\sf p}$ are orthogonal. Therefore, from \eqref{a.2.2} and \eqref{a.2.3}, 
we obtain
\begin{eqnarray}
P^\mu P\cdot{\cal I}(P,\tau)&=&{2\pi^2\tau^2}{P}^\mu H_2^{(2)}(\sqrt{\tt s}\tau)\nonumber\\
P\cdot P\,{\cal I}^\mu(P,\tau)&=&{2\pi^2\tau^2}{P}^\mu H_2^{(2)}(\sqrt{\tt s}\tau)\nonumber\\
{\sf p}^\mu{\sf p}\cdot{\cal I}(P,\tau)&=&\frac{2\pi^2\tau^2}{{\tt s}}{\sf p}^\mu {\sf p}\cdot{P} H_2^{(2)}(\sqrt{\tt s}\tau)=0\label{a.2.4}
\end{eqnarray}
The last equality follows from the orthogonality of $P$ and ${\sf p}$. 
These identities readily show that  the first term of \eqref{2.1.38} vanishes. That the second term of \eqref{2.1.38} also vanishes 
follows from \eqref{a.2.4} and the observation that ${\cal I}^\mu(-P,\tau)$ is the complex conjugate of ${\cal I}^\mu(P,\tau)$. 

The last two terms of \eqref{2.1.38} involves the distribution ${\cal I}^\mu({\sf p})$, where ${\sf p}=m(q-q')$. Since the ${\tt t}=0$ point is in the 
spectrum of ${\tt t}={\sf p}_\mu{\sf p}^\mu$ and corresponds to the null vector ${\sf p}=0$, we can't simply use \eqref{a.2.2} and \eqref{a.2.3} to determine 
the coefficient terms of $\ha^\dagger(\bs{q}',m)\ha(\bs{q},m)$ and $\hat{b}^\dagger(\bs{q}',m)\hat{b}(\bs{q},m)$. In particular, while we might expect from \eqref{a.2.4}
${\cal I}^\mu({\sf p},\tau)$ to be proportional to ${\sf p}^\mu$, 
\begin{equation}
{\cal I}^\mu({\sf p},\tau)=\frac{2\pi^2\tau^2}{{\tt t}}{\sf p}^\mu H_2^{(1)}(i\sqrt{-{\tt t}}\tau),\quad {\sf p}_\mu{\sf p}^\mu={\tt t},\label{a.2.5}
\end{equation}
and therefore ${\cal I}({\sf p},\tau)$ to be orthogonal to $P$, 
the term $P\cdot{\cal I}({\sf p},\tau)$ is in fact non-zero due to the contribution of the single point 
${\sf p}=0$ to ${\cal I}^\mu({\sf p},\tau)$.  To see this, let us directly compute the integral $P\cdot{\cal I}({\sf p},\tau)
=\int 2d^4x\,\delta(x^2-\tau^2)\theta(x^0)P_\mu x^\mu e^{-ix\cdot{\sf p}}$. 
Since the expression is Lorentz invariant, we may evaluate the integral in the frame of reference in which  the spatial component of 
$P$ vanishes so that $P^0=\sqrt{\tt s}$. Since ${\sf p}$ is orthogonal to $P$, in such a reference frame
the temporal component of ${\sf p}$ must vanish. Let us denote the spatial component of ${\sf p}$ in this frame by $m\bs{q}$. 
Therewith, 
\begin{eqnarray}
P\cdot{\cal I}({\sf p},\tau)&=&\int 2d^4x\,\delta(x^2-\tau^2)\theta(x^0)\sqrt{\tt s}x^0e^{-im\bs{x}\cdot\bs{q}}\nonumber\\
&=&\sqrt{\tt s}\int d^3x\,e^{-im\bs{x}\cdot\bs{q}}\nonumber\\
&=&(2\pi)^3\sqrt{\tt s}\delta(m\bs{q})\nonumber
\end{eqnarray}
Transforming back to the original reference frame, we obtain
\begin{equation}
P\cdot{\cal I}({\sf p},\tau)=(2\pi)^3P^0\delta(\bs{\sf p})\label{a.2.6}
\end{equation}
In addition, from \eqref{a.2.5}, we also have 
\begin{eqnarray}
{\sf p}^\mu {\sf p}\cdot{\cal I}({\sf p},\tau)&=&{2\pi^2\tau^2}{\sf p}^\mu H_2^{(2)}(i\sqrt{-{\tt t}}\tau)\nonumber\\
{\sf p}\cdot{\sf p}\,{\cal I}^\mu({\sf p},\tau)&=&{2\pi^2\tau^2}{\sf p}^\mu H_2^{(2)}(i\sqrt{-{\tt t}}\tau)\label{a.2.7}
\end{eqnarray}
where ${\tt t}={\sf p}_\mu{\sf p}^\mu$. The distribution coefficients 
$P^\mu P\cdot{\cal I}(-{\sf p},\tau)$,  ${\sf p}^\mu {\sf p}\cdot{\cal I}(-{\sf p},\tau)$ and  ${\sf p}\cdot{\sf p}\,{\cal I}^\mu(-{\sf p},\tau)$ of the 
$\hat{b}^\dagger(\bs{q}',m)\hat{b}(\bs{q},m)$ term can also be easily 
obtained from \eqref{a.2.6} and \eqref{a.2.7} and the observation that ${\cal I}^\mu(-{\sf p},\tau)$ is the 
complex conjugate of ${\cal I }^\mu({\sf p},\tau)$. 

Substituting from \eqref{a.2.6} and \eqref{a.2.7} into  \eqref{2.1.38} and \eqref{2.1.39} and carrying out the integration with respect 
to the $q'$-variable, we obtain the standard result \eqref{2.1.39}.

\item {\bf Distribution $\bs{{\cal D}(m_1q'\pm m_2q,\tau)}$ of Eq.~{\bf\eqref{3.1.7b}}}\\
By definition \eqref{3.1.7b}, 
\begin{equation}
{\cal D}(m_1q'\pm m_2q,\tau)=q\cdot{\cal I}(m_1q'\pm m_2q,\tau)\label{a.3.1}
\end{equation}
where $q'$ and $q$ are the velocity four-vectors with ${q'}^2=q^2=1$ and $m_1$ and $m_2$ are positive real masses. 
They need not be constants, as is the case with $m(\ka)$. 
The vector distribution ${\cal I}^\mu(p,\tau)$ is as defined by \eqref{a.2.1}.  Therefore, 
\begin{eqnarray}
{\cal D}(m_1q'\pm m_2q,\tau)&=&\int 2d^4x\,\delta(x^2-\tau^2)\theta(x^0)\,q\cdot x\,e^{-i(m_1q'\pm m_2q)\cdot x}\nonumber\\
&=&\pm i\frac{\partial}{\partial m_2}\int \frac{d^3x}{\sqrt{\tau^2+\bs{x}^2}}\,e^{-i(m_1q'\pm m_2q)\cdot x}\nonumber\\
\label{a.3.2}
\end{eqnarray}
It is the integral of ${\cal D}(m_1q'\pm m_2q,\tau)$ with respect to the $q'$-variable that appears in \eqref{3.1.7}. Note that ${\cal D}(m_1q'\pm m_2q)$ does 
not have the same functional dependence on $q'$ and $q$, although our notation doesn't make this fact manifest. It is much easier to directly 
evaluate the integral $\intqq\,{\cal D}(m_1q'\pm m_2q,\tau)$:
\begin{eqnarray}
\intqq\,{\cal D}(m_1q'\pm m_2q,\tau)&=&\pm i\frac{\partial}{\partial m_2}\int \frac{d^3x}{\sqrt{\tau^2+\bs{x}^2}}\,e^{\mp im_2q\cdot x}\times\nonumber\\
&&\quad\intqq\,e^{-im_1q'\cdot x}\nonumber\\
\label{a.3.3}
\end{eqnarray}
We note that the second integral of \eqref{a.3.3} is really the same as the distribution $I(p,\tau)$ computed above, with only the integration variable changed to velocity rather 
than position. Since both $mq'$ and $x$ are timelike with positive time-components, from \eqref{a.7} we have
\begin{equation}
\intqq\,e^{-im_1q'\cdot x}=\frac{i\pi^2}{m_1\tau}H_1^{(2)}(m_1\tau)\label{a.3.4}
\end{equation}
Since this integral is independent of $x$, \eqref{a.3.3} separates: 
\begin{eqnarray}
\intqq\,{\cal D}(m_1q'\pm m_2q,\tau)&=&\mp\frac{\pi^2}{m_1\tau}H_1^{(2)}(m_1\tau)\frac{\partial}{\partial m_2}\int\frac{d^3x}{\sqrt{\tau^2+\bs{x}^2}}e^{\mp im_2q\cdot x}\nonumber\\
&=&\mp\frac{\pi^2}{m_1\tau}H_1^{(2)}(m_1\tau)\frac{\partial}{\partial m_2}\left(\pm i\frac{2\pi^2\tau}{m_2}H_1^{(2,1)}(m_2\tau)\right)\nonumber\\
&=&\frac{i2\pi^4\tau}{m_1m_2}H_1^{(2)}(m_1\tau)H_2^{(2,1)}(m_2\tau)\label{a.3.5}
\end{eqnarray}
where $(2,1)$ in the superscript of the Hankel function corresponds to the $\pm$ sign of the argument of ${\cal D}(m_1q'\pm m_2q)$, respectively. 
\end{enumerate}
%---------------------------
\section{The solution of \eqref{3.1.7}}\label{appB}
%---------------------------
 The first step in solving \eqref{3.1.7} is to explicitly determine the integrals $\intqq{\cal D}\left(m_1q'\pm m_2q\right)$ and $\intqq{\cal D}\left(-m_1q'\pm m_2q\right)$. 
 This is done in Appendix \ref{appA}. From \eqref{a.3.1}-\eqref{a.3.5}, 
 \begin{eqnarray}
 \intqq{\cal D}\left(m_1q'\pm m_2q\right)=i\frac{2\pi^4\tau}{m_1m_2}H^{(2)}_1(m_1\tau)H^{(2,1)}_2(m_2\tau)\nonumber\\
 \intqq{\cal D}\left(-m_1q'\pm m_2q\right)=-i\frac{2\pi^4\tau}{m_1m_2}H^{(1)}_1(m_1\tau)H^{(2,1)}_2(m_2\tau)\label{3.1.7c}
 \end{eqnarray}
 where $H^{(1)}_\alpha$ and $H^{(2)}_\alpha$ are first and second kind Hankel functions  of order $\alpha$, respectively. 
 The integral \eqref{3.1.7c} depends on the fact that $m_1$ and $m_2$ are both 
 positive Lorentz scalars and that $q$ and $q'$ are both timelike four vectors.  The symbol $(2,1)$ in the superscript of the 
 second Hankel functions in \eqref{3.1.7c} corresponds to the $\pm$ sign of $m_2$. 
 
With the aid of \eqref{3.1.7c}, the set of equations  \eqref{3.1.7} becomes 
\begin{eqnarray}
(\rms-M) t(\ms)&=&\frac{i\pi\beta \tau}{8}H_1^{(1)}(M\tau)\int_{-\infty}^\infty d\ka\,\alpha(\ka)\left[T(\ms,\ka)H_2^{(2)}(m(\ka)\tau)\right.\nonumber\\
&&\qquad\qquad\qquad\qquad\left.-R(\ms,\ka)H_2^{(1)}(m(\ka)\tau)\right]\nonumber\\
(\rms+M)r(\ms)&=&-\frac{i\pi\beta \tau}{8}H_1^{(2)}(M\tau)\int_{-\infty}^\infty d\ka\,\alpha(\ka)\left[T(\ms,\ka)H_2^{(2)}(m(\ka)\tau)\right.\nonumber\\
&&\qquad\qquad\qquad\qquad\left.-R(\ms,\ka)H_2^{(1)}(m(\ka)\tau)\right]\nonumber\\
 (\rms-m(\ka))T(\ms,\ka)&=&\frac{i\pi\beta \tau}{8}\alpha(\ka)H_1^{(1)}\left(m(\ka)\tau\right)\left[t(\ms)H_2^{(2)}(M\tau)-r(\ms)H_2^{(1)}(M\tau)\right]\nonumber\\
 (\rms+m(\ka)) R(\ms,\ka)&=&-\frac{i\pi\beta \tau}{8}\alpha(\ka)H_1^{(2)}\left(m(\ka)\tau\right)\left[t(\ms)H_2^{(2)}(M\tau)-r(\ms)H_2^{(1)}(M\tau)\right]\nonumber\\
  \label{3.1.6}
  \end{eqnarray}

From the first two equations, we obtain
\begin{equation}
-\frac{\rms+M}{H_1^{(2)}(M\tau)}r(\ms)=\frac{\rms-M}{H_1^{(1)}(M\tau)}t(\ms)\label{3.1.8}
\end{equation}
while from the second two, 
\begin{equation}
-\frac{\rms+m(\ka)}{H_1^{(2)}(m(\ka)\tau)}R(\ms,\ka)=\frac{\rms-m(\ka)}{H_1^{(1)}(m(\ka)\tau)}T(\ms,\ka)\label{3.1.9}
\end{equation}
Anticipating a positive definite spectrum for the square mass operator $\hat{M}$ (our ultimate  goal is to understand the resonance scattering in the model) 
we may divide \eqref{3.1.8} by $(\rms+M)$ and 
\eqref{3.1.9} by $(\rms+m(\ka))$.  Then, substituting \eqref{3.1.8} into the third and fourth equalities of \eqref{3.1.6}, we obtain
\begin{eqnarray}
T(\ms,\ka)&=&C\delta(\rms-m(\ka))\nonumber\\
&&\quad+\frac{i\pi\beta}{8H_1^{(1)}(M\tau)}\frac{\alpha(m(\ka))H_1^{(1)}(m(\ka)\tau)}{\rms-m(\ka)}\frac{{\cal H}_{1,2}(\rms\tau,M\tau)}{\rms+M}t(\ms)\nonumber\\
\label{3.1.12}\\
R(\ms,\ka)&=&-\frac{i\pi\beta}{8H_1^{(1)}(M\tau)}\frac{\alpha(m(\ka))H_1^{(2)}(m(\ka)\tau)}{\rms+m(\ka)}\frac{{\cal H}_{1,2}(\rms\tau,M\tau)}{\rms+M}t(\ms)\nonumber\\
\label{3.1.13}
\end{eqnarray}
where $C$ is a Lorentz scalar. In general, it can be a well-behaved function of $\ms$. The function ${\cal H}_{\mu,\nu}(y,x)$ is defined by 
\begin{eqnarray}
{\cal H}_{\mu,\nu}(y,x)&:=&{y\left(H^{(1)}_\mu(x)H^{(2)}_\nu(x)+H^{(2)}_\mu(x)H^{(1)}_\nu(x)\right)}+\nonumber\\
&&\quad\quad{x\left(H^{(1)}_\mu(x)H^{(2)}_\nu(x)-H^{(2)}_\mu(x)H^{(1)}_\nu(x)\right)}\label{3.1.10b}
\end{eqnarray}
By virtue of the Wronskian formula $H^{(1)}_\mu(x)H^{(2)}_{\mu+1}(x)-H^{(2)}_\mu(x)H^{(1)}_{\mu+1}(x)=\frac{4i}{\pi x}$,  for 
$\nu=\mu+1$, 
\eqref{3.1.10b} becomes
\begin{equation}
{\cal H}_{\mu,\mu+1}(x,y)={2y\left\{J_\mu(x)J_{\mu+1}(x)+N_\mu(x)N_{\mu+1}(x)\right\}-\frac{4i}{\pi}}\label{3.1.10c}
\end{equation}

Now, substituting \eqref{3.1.12} and \eqref{3.1.13} into the first equation of \eqref{3.1.6} and changing the integration measure from $d\ka$ to $d\mu(\ka)$, 
\begin{equation}
d\mu(\ka)=\frac{m(\ka)dm(\ka)}{2m\sqrt{m(\ka)^2-4m^2}}\label{3.1.14a}
\end{equation}
where $dm(\ka)$ is the Lebesgue measure on $[2m,\infty)$, 
we obtain the expression for $t(\ms)$:
\begin{equation}
(\ms-M^2-\Pi(\ms))t(\ms)=\left(\rms+M\right)H_1^{(1)}(M\tau)\rho(\ms)\label{3.1.14}
\end{equation}
where 
\begin{eqnarray}
\rho(\ms)&=&\frac{i\pi\beta C}{4}\int_{2m}^\infty d\mu(\ka)\,\delta(\rms-m(\ka))
\frac{\alpha(m(\ka)){\cal H}_{1,2}(\rms\tau,m(\ka)\tau)}{H_1^{(1)}(m(\ka)\tau)(\rms+m(\ka))}\nonumber\\
\label{3.1.15}\\
\Pi(\ms)&=&-\frac{(\pi\beta)^2}{32}{\cal H}_{1,2}(\rms\tau,M\tau)\int_{2m}^\infty d\mu(\ka)\,\frac{\alpha(m(\ka))^2{\cal H}_{1,2}(\rms\tau,m(\ka)\tau)}{\ms-m(\ka)^2}\nonumber\\
\label{3.1.16}
\end{eqnarray}
From \eqref{3.1.6}-\eqref{3.1.13}, we see that a non-trivial solution to the eigenvalue problem \eqref{3.1.1} exists only for 
$4m^2\leq\ms<\infty$.  Therefore, the spectrum of the interacting square-mass operator ${\hat{M}}^2={\hat{P}}_\mu {\hat{P}}^\mu$ is 
\begin{equation}
4m^2\leq\ms<\infty\label{3.1.17}
\end{equation}
For such $\ms$, integral \eqref{3.1.15} gives
\begin{equation}
\rho(\ms)=\frac{iC\pi\beta}{8}\frac{\alpha(\ms){\cal H}_{1,2}(\rms\tau,\rms\tau)}{2m\sqrt{\ms-4m^2}\,H_1^{(1)}(\rms\tau)}=\frac{iC\pi\beta}{4}\frac{\rms\alpha(\ms)H_2^{(2)}(\rms\tau)}{2m\sqrt{\ms-4m^2}}
\label{3.1.18}
\end{equation}
\\

Defining a Green's function,
\begin{equation}
G(\ms):=\frac{1}{\ms-M^2-\Pi(\ms)}\label{3.1.19}
\end{equation}
we obtain the following solution to \eqref{3.1.6}:
\begin{eqnarray}
t(\ms)&=&\left(\rms+M\right)H_1^{(1)}(M\tau)\rho(\ms)G(\ms)\nonumber\\
r(\ms)&=&-\left(\rms-M\right)H_1^{(2)}(M\tau)\rho(\ms)G(\ms)\nonumber\\
T(\ms,\ka)&=&C\delta(\rms-m(\ka))+\frac{i\pi\beta}{8}\frac{\alpha(m(\ka))H_1^{(1)}(m(\ka)\tau)}{\rms-m(\ka)}{\cal H}_{1,2}(\rms\tau,M\tau)\rho(\ms)G(\ms)\nonumber\\
R(\ms,\ka)&=&-\frac{i\pi\beta}{8}\frac{\alpha(m(\ka))H_1^{(2)}(m(\ka)\tau)}{\rms+m(\ka)}{\cal H}_{1,2}(\rms\tau,M\tau)\rho(\ms)G(\ms)\
\label{3.1.20}
\end{eqnarray}
where $\rho(\ms)$ is given by \eqref{3.1.18}. Green's function $G(\ms)$ is defined by \eqref{3.1.19}, with $\Pi(\ms)$ given by \eqref{3.1.16}. 
\\

\noindent{\bf The annihilation operator $\hc(\bs{q},\ms)$}\\
We can repeat the steps of the above calculation to determine the operator $\hc(\bs{q},\ms)$, defined  as the solution to the eigenvalue problem \eqref{3.1.2}, 
in terms of the free creation and annihilation operators $\ha^\dagger(\bs{q})$, $\ha(\bs{q})$, $\hB^\dagger(\bs{q},\ka)$ and $\hB(\bs{q},\ka)$. 
To that end, let 
\begin{equation}
\hc(\bs{q},\ms)=\int_{-\infty}^\infty d\ka \left(\tilde{T}(\ms,\ka)\hB^\dagger(\bs{q},\ka)+\tilde{R}(\ms,\ka)\hB(\bs{q},\ka)\right) +\tilde{t}(\ms)\ha^\dagger(\bs{q})+\tilde{r}(\ms)\ha(\bs{q})\label{3.1.21}
\end{equation}
Substituting this and the expression for the total momentum $\hat{P}^\mu$ into \eqref{3.1.2}, we obtain the same set of equations as \eqref{3.1.7}, but with $\rms$ 
replaced by $-\rms$ on the left hand side: 
\begin{eqnarray}
(-\rms-M)\tilde{t}(\ms)&=&-\frac{1}{2(2\pi)^3}\beta M\int_{-\infty}^\infty d\ka\,m(\ka)\alpha(\ka)\intqq\nonumber\\
&&\qquad\left[\tilde{T}(\ms,\ka){\cal D}(-Mq'+m(\ka)q)-\tilde{R}(\ms,\ka){\cal D}(-Mq'-m(\ka)q)\right]\nonumber\\
(-\rms+M)\tilde{r}(\ms)&=&-\frac{1}{2(2\pi)^3}\beta M\int_{-\infty}^\infty d\ka\, m(\ka)\alpha(\ka)\intqq\nonumber\\
&&\qquad\left[\tilde{T}(\ms,\ka){\cal D}(Mq'+m(\ka)q)-\tilde{R}(\ms,\ka){\cal D}(Mq'-m(\ka)q)\right]\nonumber\\
 (-\rms-m(\ka))\tilde{T}(\ms,\ka)&=&-\frac{1}{2(2\pi)^3}\beta M m(\ka)\alpha(\ka)\intqq\nonumber\\
 &&\qquad\left[\tilde{t}(\ms){\cal D}(Mq-m(\ka)q')-\tilde{r}(\ms){\cal D}(-Mq-m(\ka)q')\right]\nonumber\\
  (-\rms+m(\ka))\tilde{R}(\ms,\ka)&=&-\frac{1}{2(2\pi)^3}\beta M m(\ka)\alpha(\ka)\intqq\nonumber\\
  &&\qquad\left[\tilde{t}(\ms){\cal D}(Mq+m(\ka)q')-\tilde{r}(\ms){\cal D}(-Mq+m(\ka)q')\right]\nonumber\\
  \label{3.1.22}
  \end{eqnarray}

Using \eqref{3.1.7c}, we can rewrite these equations to obtain
\begin{eqnarray}
(\rms+M)\tilde{t}(\ms)&=&-\frac{i\pi\beta\tau}{8}H_1^{(1)}(M\tau)\int_{-\infty}^\infty d\ka\alpha(\ka)\left[\tilde{T}(\ms,\ka)H_2^{(2)}(m(\ka)\tau)\right.\nonumber\\
&&\qquad\qquad\qquad\qquad\quad -\left.\tilde{R}(\ms,\ka)H_2^{(1)}(m(\ka)\tau)\right]\nonumber\\
(\rms-M)\tilde{r}(\ms)&=&\frac{i\pi\beta\tau}{8}H_1^{(2)}(M\tau)\int_{-\infty}^\infty d\ka\alpha(\ka)\left[\tilde{T}(\ms,\ka)H_2^{(2)}(m(\ka)\tau)\right.\nonumber\\
&&\qquad\qquad\qquad\qquad\quad-\left.\tilde{R}(\ms,\ka)H_2^{(1)}(m(\ka)\tau)\right]\nonumber\\
(\rms+m(\ka))\tilde{T}(\ms,\ka)&=&-\frac{i\pi\beta\tau}{8}\alpha(\ka)H_1^{(1)}(m(\ka)\tau)\left[\tilde{t}(\ms)H_2^{(2)}(M\tau)-\tilde{r}(\ms)H_2^{(1)}(M\tau)\right]\nonumber\\
(\rms-m(\ka))\tilde{R}(\ms,\ka)&=&\frac{i\pi\beta \tau}{8}\alpha(\ka)H_1^{(2)}(m(\ka)\tau)\left[\tilde{t}(\ms)H_2^{(2)}(M\tau)-\tilde{r}(\ms)H_2^{(1)}(M\tau)\right]\nonumber\\
\label{3.1.24}
\end{eqnarray}
A comparison of \eqref{3.1.24} with the complex-conjugates of Eqs.~\eqref{3.1.6} shows, not surprisingly, that 
\begin{eqnarray}
\tilde{t}(\ms)&=&r^*(\ms)\nonumber\\
\tilde{r}(\ms)&=&t^*(\ms)\nonumber\\
\tilde{T}(\ms,\ka)&=&R^*(\ms,\ka)\nonumber\\
\tilde{R}(\ms,\ka)&=&T^*(\ms,\ka),\label{3.1.25}
\end{eqnarray}
where $r(\ms)$, $t(\ms)$, $R(\ms,\ka)$ and $T(\ms,\ka)$ are the functions given by \eqref{3.1.20},  furnish a solution to \eqref{3.1.24}. 
Hence, we see that $\hc(\bs{q},\ms)$ is the formal adjoint of $\hc^\dagger(\bs{q},\ms)$. After the Fock space \eqref{3.1.20d} has been constructed by the repeated 
application of $\hc^\dagger(\bs{q},\ms)$ on the vacuum state $\Omega$, the $\hc(\bs{q},\ms)$ in fact becomes the 
adjoint of $\hc^\dagger(\bs{q},\ms)$ in \eqref{3.1.20d}. This construction of \eqref{3.1.20d}, in particular the normalization condition \eqref{3.1.20b}, also 
fixes the arbitrary scalar function $C(\ms)$ to within which the operators $\hc^\dagger(\bs{q},\ms)$ and $\hc(\bs{q},\ms)$ are determined by 
the eigenvalue equations \eqref{3.1.1} and \eqref{3.1.2} (see \eqref{3.1.20e} and \eqref{3.1.26}).

\end{document}